\documentclass[preprint,amsmath,amssymb,tightenlines,nofootinbib]{revtex4}

\usepackage{graphicx}
\usepackage{hyperref}

\newcommand{\slp}{ {p\mspace{-7.5mu}/\mspace{-1.5mu}} }
\newcommand{\slP}{ {P\mspace{-11mu}/\mspace{2mu}} }
\newcommand{\gslash}[1]{#1\mspace{-9mu}/}
\newcommand{\g}{\gamma}
\newcommand{\e}{\epsilon}
\newcommand{\eps}{\varepsilon}
\newcommand{\la}{\lambda}
\newcommand{\ka}{\kappa}
\newcommand{\w}{\omega}

\newcommand{\G}{\Gamma}
\newcommand{\D}{\Delta}
\newcommand{\La}{\Lambda}

\newcommand{\Img}{\mathrm{Im}\,}
\newcommand{\img}{\mathrm{i}}
\newcommand{\df}{\mathrm{d}}
\newcommand{\GeV}{\,\mathrm{GeV}}
\newcommand{\MeV}{\,\mathrm{MeV}}

\newcommand{\bb}{\bar{b}}
\newcommand{\bn}{\bar{n}}
\newcommand{\bu}{\bar{u}}
\newcommand{\cD}{\mathcal{D}}

\newcommand{\cK}{\mathcal{K}}
\newcommand{\cO}{\mathcal{O}}
\newcommand{\cP}{\mathcal{P}}
\newcommand{\cQ}{\mathcal{Q}}
\newcommand{\cR}{\mathcal{R}}

\newcommand{\ord}[1]{\mathcal{O}(#1)}
\newcommand{\abs}[1]{\lvert#1\rvert}
\newcommand{\mae}[3]{\langle#1\lvert#2\rvert#3\rangle}
\newcommand{\vevB}[1]{\langle #1 \rangle_{\! B}}
\newcommand{\LQCD}{\La_\mathrm{QCD}}

\newcommand{\q}{,\quad}

\begin{document}

\preprint{\vbox{ \hbox{hep-ph/0503095} \hbox{LBNL-57264} \hbox{UCB-PTH-05/06} \hbox{March 10, 2005} } }

\title{\vspace*{2cm}\boldmath
Full-Phase-Space Twist Expansion in Semileptonic and Radiative $B$-Meson Decays}

\author{\vspace*{0.2cm}Frank J. Tackmann}
\affiliation{\vspace*{0.2cm}Department of Physics, University of California at Berkeley,
Ernest Orlando Lawrence Berkeley National Laboratory, Berkeley, CA 94720, USA
\vspace*{1cm}}

\begin{abstract}
\vspace*{0.5cm}
We study the $\LQCD/M_B$ corrections from subleading shape functions in inclusive $B$-meson decays. We propose a natural and smooth interpolation from the endpoint region to the full phase space, and derive expressions for the triple differential decay rate in $B \to X_u \ell \bar{\nu}_\ell$ and the photon energy spectrum in $B\to X_s \g$. Our results are valid to order $\LQCD/M_B$ for hadronic invariant masses of order $\LQCD M_B$ and to order $\LQCD^2/M_B^2$ for larger hadronic masses. They allow a systematic investigation of the transition between the separate regimes of the local and nonlocal expansions, and can be used to study decay distributions in any kinematic variables. We consider several examples of interest and point out that a combined measurement of hadronic energy and invariant mass provides an alternative to the extraction of $\abs{V_{ub}}$ which is largely independent of shape-function effects and in principle allows a higher accuracy than the combined measurement of leptonic and hadronic invariant masses.
We perform the expansion directly in QCD light-cone operators, and give a discussion of the general basis of light-cone operators. Reparametrization invariance under the change of the light-cone direction reduces the number of independent shape functions. We show that differing previous results for the lepton energy spectrum obtained from different choices of light-cone coordinates are in agreement.
\end{abstract}


\maketitle

\section{Introduction}

Inclusive decays of $B$ mesons offer a rich environment to explore the flavor sector of the Standard Model and to search for New Physics in radiative decays. Moment analyses of $B\to X_c \ell \bar{\nu}_\ell$ decay distributions have provided a precision measurement of the Cabibbo-Kobayashi-Maskawa (CKM) matrix element $\abs{V_{cb}}$ at the two-percent level, along with an extraction of the $b$-quark and $c$-quark masses, and higher order hadronic parameters \cite{Aubert:2004aw,Luth:2004gs,Bauer:2004ve}. Similarly, the study of inclusive decays $B\to X_u \ell \bar{\nu}_\ell$ allows for an extraction of $\abs{V_{ub}}$ with current errors of $10\%$--$15\%$ \cite{Gibbons:2004dg,Battaglia:2004ti,Aubert:2004tw,Aubert:2004bq,Aubert:2004bv,Abe:2004sc}.

The conventional treatment of inclusive $B$ decays relies on a local operator product expansion (OPE) in inverse powers of the large momentum $Q$ transferred to the hadronic system \cite{Chay:1990da,Bigi:1993fe,Blok:1994va,Manohar:1994qn,Mannel:1994su,Falk:1994dh}. However, for $b\to u$ transitions, tight experimental cuts are needed to suppress the overwhelming charm background. They usually put the kinematics close to the boundary of phase space where the final hadronic system has large energy in the $B$-meson rest frame but small invariant mass. That is, $Q$ lies close to the light-cone with $Q^2$ being much smaller then $p_B\cdot Q$. Consequently, the OPE in local operators breaks down. The large and small components of $Q$ can be separated by employing light-cone coordinates. The local OPE can then be replaced by an OPE in nonlocal light-cone operators \cite{Bauer:2001mh}, which only expands in inverse powers of the remaining large components of $Q$.

The leading term in this so-called twist expansion has been known for some time now \cite{Neubert:1994ch,Neubert:1994um,Bigi:1994ex,Mannel:1994pm}. When going beyond the tree-level approximation the separation of short- and long-distance contributions becomes important and schematically leads to a decay rate of the form \cite{Korchemsky:1994jb}
\begin{equation*}
\df\G = H \times J \otimes S
,\end{equation*}
which is factorized into a hard contribution multiplying the convolution of a jet and soft contribution. The hard, jet, and soft contributions are associated with the different scales $m_b^2 \gg \LQCD m_b \gg \LQCD^2$. The matrix elements parametrizing the soft contributions are usually referred to as shape functions.

With the current experimental reach of precision, the investigation of subleading twist corrections in $\LQCD/m_b$ has become important. They were considered at tree level for the photon energy spectrum in $B\to X_s \g$ \cite{Bauer:2001mh} and the lepton energy spectrum \cite{Leibovich:2002ys,Bauer:2002yu} and hadronic invariant mass spectrum \cite{Burrell:2003cf} in $B\to X_u \ell \bar{\nu}_\ell$. Baryonic decays have also been considered \cite{Kraetz:2002rv}. The first investigation of subleading twist corrections in $B\to X_c \ell \bar{\nu}_\ell$ was given recently in Ref.~\cite{Mannel:2004as}, and it was discovered that the matching of some subleading contributions in the earlier $B\to X_u \ell \bar{\nu}_\ell$ result \cite{Bauer:2002yu} are incorrect.

Beyond the tree-level approximation, the factorization into hard, jet, and soft contributions at subleading order in $\LQCD/m_b$ was first worked out by Lee and Stewart \cite{Lee:2004ja} within the framework of soft collinear effective theory (SCET). They investigated the general structure of subleading corrections, and gave results for decay rates to $\ord{\LQCD/m_b}$, including the full triple differential rate in $B\to X_u \ell \bar{\nu_\ell}$. The latter was also derived by Bosch, Neubert, and Paz \cite{Bosch:2004cb}. An analysis similar to Ref.~\cite{Lee:2004ja} was carried out by Beneke et al.\ \cite{Beneke:2004in}, too. In all cases, the subleading twist corrections to the differential decay rates were still given at tree level, mainly due to the increased complexity of the contributions arising beyond that.

In addition, it is usually difficult to assess how far away from their literal expansion region the twist or local expansions are still valid, and where the transition between them occurs. Having a single description for the entire phase space improves this situation. By comparing it with the predictions of the twist or local expansion, one can systematically investigate where the corrections to the local or twist expansion results become large.

In Ref.~\cite{Mannel:2004as} it was shown for the lepton energy spectrum in $B\to X_{u,c} \ell \bar{\nu}_\ell$ that the standard twist expansion can be modified, such that it becomes valid over the entire phase space. In the present paper we extend this approach to the triple differential decay rate in $B\to X_u \ell \bar{\nu}_\ell$ and the photon energy spectrum in $B\to X_s \g$. Our results are exact to $\ord{\LQCD/M_B}$ for hadronic invariant masses $s_H \sim \ord{\LQCD M_B}$ and to $\ord{\LQCD^2/M_B^2}$ away from this region. In particular, they contain the complete known result for the rate to $\ord{\LQCD^2/M_B^2}$ in the local OPE region, plus some higher order corrections.

The results in Refs.~\cite{Lee:2004ja,Bosch:2004cb,Beneke:2004in} were obtained by first matching QCD onto SCET, which acts at the intermediate scale $\mu^2 = \LQCD M_B$ and allows one to take into account perturbative corrections at this scale and sum logarithms between the hard and intermediate scales. In the second step SCET is matched onto heavy-quark effective theory (HQET). However, since we only work at tree level, there is no need to introduce an intermediate scale and go through this two-step matching procedure. Instead, we directly perform an expansion in QCD light-cone operators. The advantage of using QCD rather then HQET light-cone operators is, that it preserves the structure of the light-cone OPE, not mixing it with the separate expansion of QCD in HQET. It allows us to define shape functions in QCD, which automatically combine all higher order HQET shape functions that would normally arise from expanding QCD in HQET.

In the following section we give the basic ingredients to our calculation and discuss the power counting. In Sec.~\ref{sec:lc-ops} we discuss the general basis of light-cone operators and their parametrization in terms of shape functions. In Sec.~\ref{subsec:RPI} we include a discussion of reparametrization invariance under the change of the light-cone direction, which reduces the number of independent shape functions. In particular, we show that the results for the $B\to X_u \ell \bar{\nu}_\ell$ lepton energy spectrum derived in Ref.~\cite{Mannel:2004as} and with a different choice of light-cone direction in Refs.~\cite{Lee:2004ja,Bosch:2004cb} are in agreement. Sec.~\ref{sec:light-coneOPE} contains the matching calculation and the results for the light-cone OPE. The results for the differential decay distributions are presented and discussed in Sec.~\ref{sec:decayrates}, and we conclude in Sec.~\ref{sec:conclusions}.

\section{Basic Ingredients and Power Counting}

\subsection{Hadronic Tensor and Decay Rates}

We are interested in the semileptonic decay $B\to X_u \ell \bar{\nu}_\ell$ and the radiative decay $B\to X_s \g$. The effective weak Hamiltonian for the semileptonic decay is
\begin{equation}
\label{Heff_u}
H_W^u = \frac{4G_F}{\sqrt{2}}V_{ub}(\bar{u} \g_\alpha P_L b)(\bar{\ell}\g^\alpha P_L \nu_\ell)
,\end{equation}
with $P_L = (1 - \g_5)/2$, from which one obtains the triple differential decay rate in the rest frame of the $B$ meson \cite{manoharwise}
\begin{equation}
\label{dGu}
\frac{\df^3\G^u}{\df E_\ell \df E_\nu \df q^2}
= 48 \G^u_{0}
L^{\alpha\beta} W^u_{\alpha\beta} \theta(E_\ell) \theta(q^2) \theta(4E_\ell E_\nu- q^2)
,\end{equation}
where
\begin{equation}
\G^u_{0} = \frac{G_F^2 \abs{V_{ub}}^2}{192\pi^3}
.\end{equation}
Note, that we do not include the usual factor of $m_b^5$ in $\G_0$. The momentum $q = p_\ell + p_\nu$ is the total leptonic momentum, $E_\ell$ and $E_\nu$ are the charged lepton and neutrino energies in the rest frame of the decaying $B$ meson, and we explicitly kept all phase space limits. The leptonic tensor is
$L^{\alpha\beta} = \mathrm{Tr}[\slp_\nu \g^\alpha \slp_\ell \g^\beta P_L ]$ and $W^u_{\alpha\beta}$ denotes the hadronic tensor.

For the radiative decay $B \to X_s \g$ the effective weak Hamiltonian has the form
\begin{equation}
\label{Heff_s}
H^s_W = - \frac{4G_F}{\sqrt{2}} V_{tb} V^*_{ts} C_7^\mathrm{eff}(m_b) O_7
\quad \text{with} \quad
O_7 = \frac{e}{16\pi^2} \bar{s} \sigma_{\alpha\beta} F^{\alpha\beta} (\bar{m}_b P_R + \bar{m}_s P_L) b
.\end{equation}
Here, $F^{\alpha\beta}$ denotes the electromagnetic field strength, $P_{R,L} = (1 \pm \g_5)/2$. We restrict our discussion to the dipole operator $O_7$ and neglect the tiny $s$-quark mass. The photon energy spectrum in the $B$ rest frame is
\begin{equation}
\label{dGs}
\frac{\df\G^s}{\df E_\g}
= 8\G^s_0 E_\g^3 \theta(E_\g) \eps^\alpha \eps^{*\beta} W^s_{\alpha\beta}
.\end{equation}
In this case
\begin{equation}
\G^s_0 = \frac{G_F^2 \abs{V_{tb} V^*_{ts}}^2 \abs{C^\mathrm{eff}_7(m_b)}^2 \alpha_\mathrm{em} \bar{m}_b^2(m_b)}{32\pi^4}
,\end{equation}
where we only included the $\bar{m}_b^2$ from the effective weak Hamiltonian.
Summing over the photon polarization in Eq.~\eqref{dGs} yields $\sum \eps^\alpha \eps^{*\beta} = - \eta^{\alpha\beta}$.

The optical theorem allows one to express the hadronic tensor $W^f_{\alpha\beta}$ as the forward scattering matrix element
\begin{equation}
\label{Wf_def}
W^f_{\alpha\beta} = \frac{\mae{B}{T^f_{\alpha\beta}}{B}}{2M_B}
= \vevB{T^f_{\alpha\beta}}
,\end{equation}
where $f$ is either $u$ or $s$. We will use the shorthand $\vevB{\cO} = \mae{B(p_B)}{\cO}{B(p_B)}/2M_B$ to denote the $B$ expectation value of some operator $\cO$ between physical $B$-meson states. The operator $T^f_{\alpha\beta}$ is defined as the imaginary part of the time-ordered product of two effective weak currents,
\begin{equation}
\label{Tf_def}
T^f_{\alpha\beta} = - \frac{1}{\pi} \Img \biggl( -\img\int\df^4 x e^{-\img q\cdot x}
T \bigl[ J^{f\dagger}_\alpha(x) J^f_\beta(0) \bigr] \biggr)
.\end{equation}
The momentum $q$ is the momentum transferred away from the hadronic system, i.e.,
\begin{equation}
q = p_\ell + p_\nu \quad (f=u)
\quad\text{and}\quad
q = p_\g \quad (f = s)
.\end{equation}
The currents for $f = u,s$ following from Eqs.~\eqref{Heff_u} and \eqref{Heff_s} are
\begin{equation}
\label{currents}
J^u_\alpha = \bar{u} \g_\alpha P_L b
\q
J^s_\alpha = \bar{s} \gslash{\bn}\g_{\perp\alpha} P_R b
.\end{equation}
To write $J^s_\alpha$ we already used the definitions from Eqs.~\eqref{lc_comp} and \eqref{lc_fix} below. We will skip the flavor label $f$ in the following when unnecessary.

\subsection{Light-Cone Coordinates and Hadronic Variables}

As usual, we denote the $B$-meson velocity by $v$, and define light-cone coordinates by specifying two light-cone vectors $n$ and $\bn$, satisfying
\begin{equation}
\label{lc_def}
n^2 = \bn^2 = 0 \q n\cdot\bn = 2 ,\quad\text{and}\quad  2v = n + \bn
.\end{equation}
We use round or square brackets on indices to denote complete symmetrization or antisymmetrization, e.g.,
\begin{equation}
n^{(\mu} \bn^{\nu)} = \frac{1}{2} (n^\mu \bn^\nu + n^\nu \bn^\mu)
\q
n^{[\mu} \bn^{\nu]} = \frac{1}{2} (n^\mu \bn^\nu - n^\nu \bn^\mu)
.\end{equation}
The metric and Levi-Civita tensors are decomposed as (we use $\eps_{0123} = 1$)
\begin{equation}
\eta^{\mu\nu} = n^{(\mu} \bn^{\nu)} + \eta_\perp^{\mu\nu}
\quad \text{and} \quad
\eps^{\mu\nu\la\ka} = 6 \eps_\perp^{[\mu\nu} \bn^\la n^{\ka]}
.\end{equation}
The second relation defines $\eps_\perp^{\mu\nu} = \eps^{\mu\nu\la\ka} n_\la \bn_\ka/2$.
A generic four-vector $p$ can be written as
\begin{equation}
\label{lc_comp}
p^\mu = \frac{1}{2} p_- n^\mu + \frac{1}{2} p_+ \bn^\mu + p_\perp^\mu
,\end{equation}
with $p_+ = n \cdot p$, $p_- = \bn \cdot p$, and $p_\perp^\mu = \eta_\perp^{\mu\nu} p_\nu$.

The direction of the light-cone is fixed by setting $\mathbf{n} = - \mathbf{q}/\abs{\mathbf{q}}$, i.e.,
\begin{equation}
\label{lc_fix}
q_\perp = 0 \q q_\pm = q^0 \pm \abs{\mathbf{q}}
\quad\text{and thus}\quad q_- \leq q_+
.\end{equation}
For $f = s$ this means $q_+ = 2E_\g$ and $q_- = 0$. For $f = u$ we have $q^2 = q_+ q_-$ and $E_\nu = (q_+ + q_-)/2 - E_\ell$, and we choose $E_\ell$ and $q_\pm$ as our independent variables.

Usually, the hadronic tensor is decomposed into five scalar structure functions. For our purposes it will be most convenient to decompose it according to its light-cone structure,
\begin{equation}
\label{W_param}
W^{\alpha\beta}
= - \frac{1}{2}(\eta + \img\eps)_\perp^{\alpha\beta} W_1
- \frac{1}{2} (\eta - \img \eps)_\perp^{\alpha\beta} W_2
- n^{(\alpha} \bn^{\beta)} W_3 + n^\alpha n^\beta W_4 + \bn^\alpha \bn^\beta W_5
.\end{equation}
The structure functions $W_i$ are scalar functions of $q_+$ and $q_-$. In terms of these and changing variables from $q^2$ and $E_\nu$ to $q_\pm$ the triple differential rate \eqref{dGu} takes the form
\begin{equation}
\begin{split}
\label{dG3}
\frac{\df^3\G^u}{\df E_\ell \df q_+ \df q_-}
&= \frac{48 \G^u_0}{q_+ - q_-} \Bigl(
q_+ q_- \bigl(\bar{q}_-^2 W^u_1 + \bar{q}_+^2 W^u_2 \bigr) - 2 \bar{q}_+ \bar{q}_- \bigl(q_+ q_- W^u_3 + q_+^2 W^u_4 + q_-^2 W^u_5
\bigr) \Bigr)
\\ & \quad
\times \theta(q_-) \theta(2 E_\ell - q_-) \theta(q_+ - 2E_\ell)
,\end{split}
\end{equation}
where we defined $\bar{q}_\pm = q_\pm - 2E_\ell$. Integrating over $E_\ell$, the double differential rate becomes
\begin{equation}
\label{dG2}
\frac{\df^2\G^u}{\df q_+ \df q_-}
= 8 \G^u_0 (q_+ - q_-)^2  \bigl(q_+q_-(W^u_1 + W^u_2 + W^u_3) + q_+^2 W^u_4 + q_-^2 W^u_5 \bigr)
\theta(q_-) \theta(q_+ - q_-)
.\end{equation}
For $B \to X_s \g$, the photon spectrum \eqref{dGs} takes the form
\begin{equation}
\label{dGs2}
\frac{\df\G^s}{\df E_\g}
= 8\G^s_0 E_\g^3 \theta(E_\g) (W^s_1 + W^s_2 + 2W^s_3)
.\end{equation}

Usually, the hadronic tensor is computed in terms of partonic variables. To express the decay rates in terms of hadronic variables, the total parton momentum $m_b v - q$ is reexpressed in terms of the total hadronic momentum $P = M_B v - q$. For example, the light-cone component $m_b - q_+$ is shifted to $P_+ = M_B - q_+ = m_b + \la - q_+$, where $\la = M_B - m_b$ denotes the difference between the physical $B$-meson and $b$-quark masses. Since $\la \sim\ord{\LQCD}$, this change of variables yields an additional source of power corrections, which has to be taken care of when working to subleading order.

We follow a different approach and directly incorporate the hadronic variables in the OPE, because as discussed in Sec.~\ref{subsec:hadvspart}, they are better suited for an exact treatment of the phase space in the twist expansion. Usually, the OPE is constructed by splitting the $b$-quark momentum as $p_b = m_b v + k_b$\footnote{We use the label $b$ to distinguish the conventional definition $k_b = p_b - m_b v$ from ours.} and expanding in $k_b$. Instead, we use
\begin{equation}
p_b = M_B v - \la v + k_b \equiv M_B v + k
\quad\text{with}\quad k = p_b - M_B v = k_b - \la v
,\end{equation}
and expand in $k$. That is, we shift the residual momentum $k$ by $\la v$ compared to the conventional choice $k_b = p_b - m_b v$. This is allowed, because $k$ is only defined up to $\ord{\LQCD}$, and corresponds to constructing HQET with a residual mass term $\delta m = \la$ \cite{Falk:1992fm}. On the operator level $k_b$ and $k$ turn into
\begin{equation}
\label{cov_def}
\img\cD_b = \img D - m_b v \q \img\cD = \img D - M_B v = \img\cD_b - \la v
,\end{equation}
where $\img D$ is the full QCD covariant derivative corresponding to $p_b$.

The momentum $Q$ transferred to the hadronic system becomes
\begin{equation}
Q = p_b - q =  M_B v - q + k = P + k \quad\text{with}\quad P = M_B v - q
.\end{equation}
In light-cone coordinates,
\begin{equation}
P_\pm = M_B - q_\pm = P^0 \mp \abs{\mathbf{P}} \q P_\perp = 0
\q\text{and}\quad k_\pm = k_{b\pm} - \la \q k_\perp = k_{b\perp}
.\end{equation}
The decay rates will now explicitly contain only $M_B$, while all dependence on $m_b$ or $\la$ enters as higher order corrections through $k_\pm$, i.e., through the shape functions. In the local OPE this corresponds to reexpanding $m_b$ as $M_B - \la$, which normally does not yield a very good approximation to the local result, since it introduces sizable $1/m_b$ corrections, which are otherwise absent. However, this is not an issue in our case, since we are going to treat the complete $k_+$ dependence exactly, as described below. Concerning $k_-$, the contributions proportional to $k_-^n$ vanish at tree level for a massless final-state quark, the first nonzero term involving $k_-$ being of the form $k_- k_\perp = (k_{b-} - \la) k_\perp$. Therefore, when expanded in the local OPE, we effectively only expand the $m_b$ dependence of a certain class of higher order $1/m_b^2$ corrections, which should yield a very good approximation.

\subsection{Definition of Power Counting}
\label{subsec:powercounting}

For the purpose of our discussion we formulate both local and twist expansion in terms of hadronic variables. In light-cone coordinates, the local OPE is obtained by writing
\begin{equation*}
\label{QmuQ2_local}
Q^\mu = P_- \biggl(1 + \frac{k_-}{P_-} \biggr) \frac{n^\mu}{2}
 + P_+ \biggl(1 + \frac{k_+}{P_+} \biggr) \frac{\bn^\mu}{2} + k_\perp^\mu
\q
Q^2 = P_+P_-\biggl(1 + \frac{k_+}{P_+} + \frac{k_-}{P_-} + \frac{k_+ k_- + k_\perp^2}{P_+ P_-}\biggr)
,\end{equation*}
and expanding in powers of $\La \equiv \LQCD/M_B$, where $P_\pm$ and $k$ are treated as
\begin{equation}
\label{pc_local}
P_\pm \sim \ord{M_B} \q  k_\pm, k_\perp \sim \ord{\LQCD}
.\end{equation}
We will refer to this as ``local power counting''. The components of $k$ are always $\ord{\LQCD}$, but the size of $P_\pm$ varies over the phase space. When $P_+$ becomes $\ord{\LQCD}$, $k_+/P_+ \sim \ord{1}$ is not a valid expansion parameter anymore, and the local OPE breaks down.

The twist expansion avoids this breakdown by not expanding the $k_+$ dependence of $Q$ in $k_+/P_+$. Usually, the formal way to achieve this is to assign the power counting
\begin{equation}
\label{pc_twist}
P_- \sim \ord{M_B} \q P_+ \sim \ord{\LQCD} \q k_\pm, k_\perp \sim \ord{\LQCD}
,\end{equation}
which we refer to as ``twist power counting''. As $P_+$ is explicitly counted as $\ord{\LQCD}$, an expansion in $k_+/P_+$ is forbidden. However, at the same time the validity of the expansion is restricted to the phase space region where $P_+$ is small, which is called the shape-function region. In particular, the strict application of Eq.~\eqref{pc_twist} leads to an expansion in powers of $P_+/P_-$, including leptonic tensor and phase space, which introduces sizable errors due to neglected higher order terms.

However, we can choose a different approach, such that the twist expansion becomes valid over the entire phase space. The basic idea is to treat $P_+$ as an exact kinematic variable, i.e., to not count it as $\ord{\LQCD}$. At the same time we still do not expand in $k_+$ to avoid the breakdown of the local OPE. In other words, we only expand in $k_-$ and $k_\perp$ from the very beginning. To formalize this approach we define the power counting
\begin{subequations}
\label{pc_mod}
\begin{equation}
\label{pc_ks}
\frac{k_-}{M_B} \sim \ord{\eps} \q \frac{k_\perp}{M_B} \sim \ord{\eps}
.\end{equation}
Here, $\eps$ is meant to be a formal expansion parameter that counts powers of $k_-$ and $k_\perp$. When expanding in $\eps$, we treat all other quantities, including $k_+$ and $P_+$, as exact. In particular, we do not expand in $P_+/P_-$, as is done in the standard twist expansion. This modification of the usual twist expansion was applied in Ref.~\cite{Mannel:2004as} to the lepton energy spectrum in $B\to X_{u,c} \ell \bar{\nu}_\ell$, where the energy release $M_B - 2E_\ell$ plays the role of $P_+$.

On the operator level Eq.~\eqref{pc_ks} turns into
\begin{equation}
\label{pc_ops}
\frac{\img\cD_-}{M_B} \sim \ord{\eps} \q \frac{\img\cD_\perp}{M_B} \sim \ord{\eps}
,\end{equation}
\end{subequations}
where $\eps$ now counts the number of explicit covariant derivatives of a given light-cone operator. This already implies that expanding to $\ord{\eps^n}$ automatically contains the full result to $\ord{\La^n}$ in the local power counting. We perform the light-cone OPE to $\ord{\eps^2}$, that is, we obtain the full OPE coefficients of any appearing operator with up to two explicit covariant derivatives, which includes all corrections of subleading order in the twist power counting.

The size of an actual term in the expansion, for instance $k_\perp^2/P_+ P_-$, still depends on the region of phase space, i.e. the size of $P_\pm$. Since we do not count $P_\pm$ in any way, powers of $\eps$ do not correspond to powers of $\La \equiv \LQCD/M_B$, which is why we use $\eps$ rather then $\La$ to define the power counting. It also means that the accuracy in $\La$ of our expansion varies over the phase space.

\begin{figure}%
\centering
\includegraphics[width=0.3\columnwidth]{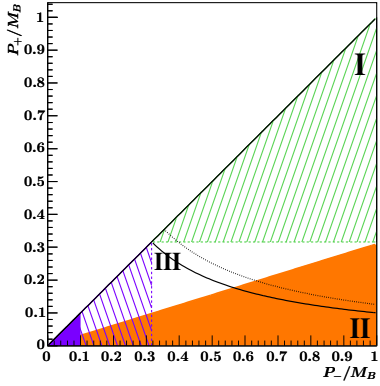}%
\caption{\label{fig:phsp}(color online) Phase space regions in the $P_\pm$ plane as discussed in the text. The solid line shows $s_H/M_B^2 = \La$ and the dashed one $s_H = M_D^2$. The light (orange) filled region is $P_+/P_- < \sqrt{\La}$, the light (green) and dark (violet) hatched regions are $P_+/M_B > \sqrt{\La}$ and $P_-/M_B < \sqrt{\La}$, and the dark (violet) filled region is the resonance region with $P_-/M_B < \La$. We take $\La = 0.1$ in all cases.}
\end{figure}

The phase space regions where the standard local and twist expansions are valid are pictured in Fig.~\ref{fig:phsp}. For illustration, we take $\La = 0.1$ and show the regions where the respective expansion parameters are less then $\sqrt{\La}$. Region I with $P_+ \sim \ord{M_B}$ is the region of the standard local OPE, and the light (green) hatched area corresponds to $\LQCD/P_+ < \sqrt{\La}$. Region II is the domain of the standard twist expansion $P_+ \sim \ord{\LQCD}$, $P_+/P_- \sim \ord{\La}$, and the light (orange) filled area shows $P_+/P_- < \sqrt{\La}$. The dark (violet) edge is the resonance region $P_-/M_B < \La$, where the inclusive treatment is invalid, and the expansions necessarily break down. The dark (violet) hatched area $\LQCD/P_- > \sqrt{\La}$ shows the transition into the resonance region. In region III, the vicinity of $P_+/M_B\sim P_-/M_B \sim \sqrt{\La}$, the local OPE is also applicable, except that the expansion is only in powers of $\sqrt{\La}$. The expansion in $\eps$ is valid anywhere away from the resonance region, and therefore provides a natural and smooth interpolation between the separate regimes of standard local and twist expansion.

To investigate the accuracy of our expansion, we write $Q^\mu$ as
\begin{equation}
\label{Qmu}
Q^\mu = \frac{1}{2} P_- \biggl(n^\mu + \frac{P_+ + k_+}{P_-} \bn^\mu + \frac{k_-}{P_-} n^\mu
+ 2\frac{k_\perp^\mu}{P_-} \biggr)
.\end{equation}
We can see that $Q^\mu$ itself contains an $\ord{\La^0}$ piece proportional to $n$. Taking the square,
\begin{equation}
\label{Q2}
Q^2
= P_-^2 \biggl(0 + \frac{P_+ + k_+}{P_-}  + \frac{P_+ + k_+}{P_-} \frac{k_-}{P_-} + \frac{k^2_\perp}{P_-^2} \biggr)
= (P_+ + k_+)P_-\biggl(1  + \frac{k_-}{P_-} + \frac{k^2_\perp}{(P_+ + k_+) P_-} \biggr)
,\end{equation}
the leading term $n^2 = 0$ vanishes, and the next largest term is $(P_+ + k_+)/P_-$. The scaling in $\La$ for the various terms in Eqs.~\eqref{Qmu} and \eqref{Q2} in regions I, II, and III is summarized in Table~\ref{scaling}. It shows that an expansion to $\ord{\eps^2}$ is exact to $\ord{\La^2}$ in region I, i.e., for $s_H \sim \ord{M_B^2}$, and to $\ord{\La}$ in regions II and III, i.e., for $s_H \sim \ord{\LQCD M_B}$. In particular, it includes all standard twist corrections of $\ord{\La}$, as well as the complete local $\ord{\La^2}$ result. The largest corrections occur in region III, where they are only suppressed by powers of $\sqrt{\La}$. In regions I and II the higher order corrections are suppressed by $\La$.

\begin{table}
\renewcommand{\arraystretch}{1.15}
\centering
\begin{tabular}{ccccc|cccc}
\hline\hline
\rule[-1ex]{0ex}{4ex} region & standard OPE & $s_H/M_B^2$ & $P_+/M_B$ & $P_-/M_B$ & $\frac{k_+}{P_+}$ & $\frac{P_+ + k_+}{P_-}$ & $\frac{k_-,k_\perp}{P_-}$ & $\frac{k_\perp^2}{(P_+ + k_+)P_-}$
\\ \hline
I   & local & $1$   & $1$   & $1$ & $\La$ & $1$ & $\La$ & $\La^2$ \\
II  & twist & $\La$ & $\La$ & $1$ & $1$ &$\La$ & $\La$ & $\La$ \\
III & $\sqrt{\mathrm{local}}$ & $\La$ & $\sqrt{\La}$ & $\sqrt{\La}$ & $\sqrt{\La}$ & $1$ & $\sqrt{\La}$ & $\La$
\\ \hline
\multicolumn{5}{r|}{assigned power in $\eps$:} & 1 & 1 & $\eps$ & $\eps^2$ \\
\hline\hline
\end{tabular}
\caption{\label{scaling}Scaling of expansion parameters in the construction of the OPE for different regions of phase space. The regions are shown in Fig.~\ref{fig:phsp} and discussed in the text.}
\end{table}

To explicitly see the difference to the standard twist expansion, we take a closer look at Eqs.~\eqref{Qmu} and \eqref{Q2}. The $k_-$ and $k_\perp^2$ terms in $Q^2$ are both twist $\ord{\La}$, but $\ord{\eps}$ and $\ord{\eps^2}$. In $Q^\mu$ itself, e.g. when multiplied by $\g_\mu$, the $k_-$ and $k_\perp$ terms are either twist $\ord{\La}$ or $\ord{\eps}$. These are the only terms, which have a power of $\eps$. Therefore, expanding to $\ord{\eps^2}$ includes all corrections of subleading twist. In addition it includes the $\ord{\La^2}$ twist terms $k_-^2/P_-^2$ and $k_- k_\perp/P_-^2$. These are precisely the $\ord{\La^2}$ twist contributions whose local expansions contain a local $\ord{\La^2}$ term. Their inclusion achieves the accuracy to local $\ord{\La^2}$. Note that we do not claim to include all $\ord{\La^2}$ twist contributions, which would require to include the terms $(k_\perp^2)^2$, $k_- k_\perp^2$, and $k_\perp k_\perp^2$. But the expansion to $\ord{\eps^2}$ is correct to $\ord{\La^2}$ in region I, and hence to $\ord{\La}$ in region III.

The second type of terms are those proportional to $P_+/P_-$, which are twist $\ord{\La}$, but local $\ord{1}$. Expanding these ``kinematic'' twist terms restricts the standard twist expansion to small $P_+$. In $Q^\mu$, the standard twist expansion also expands the term $(P_+ + k_+)/P_- \sim \ord{\La}$. In contrast, we do not assign a power counting to it and treat it exactly. As mentioned before, all kinematic factors from leptonic tensor and phase space are usually also treated as kinematic twist terms and expanded in $P_+/P_-$. Since they are unrelated to the OPE, we can treat them exactly, too. In summary, our expansion keeps all kinematic twist contributions, which makes it valid over the entire phase space.

\subsection{Phase Space and Hadronic vs Partonic Variables}
\label{subsec:hadvspart}

In this section, we point out a subtlety in the treatment of the phase space. Fundamentally, the hadronic tensor itself only contains the overall momentum conservation, while the remaining phase space limits are contained as $\theta$ functions in the decay rates. Taking the double differential rate \eqref{dG2} as example, these are $\theta(m_b - p_-)\theta(p_- - p_+)$, where $p = m_b v - q$ is the total parton momentum.

As defined in Eqs.~\eqref{Wf_def} and \eqref{Tf_def}, the hadronic tensor $W^f$ has support for positive and negative values of $P_-$, where the negative values correspond to different physical processes. In particular, its support is a priori not restricted to $0 \leq p_-$. Rather, when evaluating the imaginary part in $T^f$ one has to pick out the cut corresponding to $0 \leq P_-$. In the local OPE the hadronic tensor contains at tree level the partonic momentum conservation
\begin{equation}
\delta(p^2) = \delta(p_+ p_-) = \frac{1}{\lvert p_- \rvert} \delta(p_+)
= \frac{\theta(p_-) - \theta(-p_-)}{p_-} \delta(p_+)
.\end{equation}
The two terms in the last expression correspond to the two different cuts. The $\delta$ function sets $p_+ = 0$, with which the $\theta$ functions become $\theta(m_b - p_-)\theta(p_-)$. Therefore, the hadronic tensor is only evaluated for $0 \leq p_- \leq m_b$, which automatically picks out the correct cut.

Using the partonic variable $p$ in the twist expansion, the momentum conservation will be
\begin{equation}
\label{tw_part}
\delta\bigl((p_+ - \w) p_- \bigr) = \frac{\theta(p_-) - \theta(-p_-)}{p_-} \delta(p_+ - \w)
,\end{equation}
where $\w$ is the argument of the shape functions with support $-\la \leq \w$. With $p_+ = \w$ the phase space limits are $\w \leq p_- \leq m_b$, and $p_-$ can become negative. To pick the first term in Eq.~\eqref{tw_part} we must require $0\leq p_-$ by hand. This results in different limits on $p_-$ depending on the sign of $\w$, namely, $0\leq \w \leq p_- \leq m_b$ or $\w \leq 0 \leq p_- \leq m_b$, which is rather cumbersome. One could argue that this is irrelevant, since the twist expansion is only valid for large $p_-\sim\ord{m_b}$ anyway. Also, upon integration over $p_-$, the difference between $\w \leq p_-$ and $0\leq p_-$ will be of higher order. However, since $0\leq p_-$ restricts $P_-$ to $\la \leq P_-$, this seems to disallow an exact treatment of the phase space, which is what we aim for.

The problem is that we want hadronic phase space boundaries, while the partonic variables force us to expand around partonic phase space. In hadronic variables, the $\theta$ functions are $\theta(M_B - P_-)\theta(P_- - P_+)$. Using $P$ instead of $p$, the momentum conservation will be
\begin{equation}
\label{tw_hadr}
\delta\bigl((P_+ - \w) P_- \bigr) = \frac{\theta(P_-) - \theta(-P_-)}{P_-} \delta(P_+ - \w)
,\end{equation}
giving the limits $\w \leq P_- \leq M_B$. Since the support of the shape functions is now $0\leq \w$, this again picks out the correct cut by itself. In particular, $P_-$ can now extend into the region $0 \leq P_- \leq \la$. To summarize, we obtain $0 \leq P_+ \leq P_- \leq M_B$.
These are the physical phase space boundaries, provided we neglect the mass of the $\pi$ meson, which is much smaller then $\LQCD$. At present there is no way to consistently include the effects of $m_\pi$, because the twist expansion can only account for the nonperturbative effects due to the initial $B$ meson.

\section{QCD Light-Cone Operators}
\label{sec:lc-ops}

\subsection{General Operator Basis}
\label{subsec:operatorbasis}

All light-cone operators to $\ord{\eps^2}$ can be derived from the three kernels
\begin{equation}
\begin{split}
\cK_0^\G(\w) &= \bb \delta(\img\cD_+ + \w) \G b
,\\
\cK_1^{\G\mu}(\w_1, \w_2)
&= \bb \delta(\img\cD_+ + \w_1) \img \cD^\mu \delta(\img \cD_+ + \w_2) \G b
,\\
\cK_2^{\G\mu\nu}(\w_1, \w_2, \w_3)
&= \bb \delta(\img \cD_+ + \w_1) \img\cD^\mu \delta(\img\cD_+ + \w_2) \img\cD^\nu \delta(\img \cD_+ + \w_3) \G b
,\end{split}
\end{equation}
where $\G$ is some generic Dirac structure, and the $b$-quark fields are full QCD fields. When parametrizing the operators in Sec.~\ref{subsec:shapefunctions} we take $\img\cD = \img D - M_B v$, according to Eq.~\eqref{cov_def}, but the general discussion in this section is independent of the specific definition of $\cD$ and $\w$. Since $\cK_1^\G(\w_1,\w_2)$ and $\cK_2^\G(\w_1,\w_2,\w_3)$ depend on more then one variable, their parametrizations yield shape functions of two and three variables. In Refs.~\cite{Lee:2004ja,Beneke:2004in} such parametrizations are given for the SCET equivalents of these kernels. At tree level only bi-local operators appear, and therefores only integrals of the above operator kernels are needed. In the same way, the shape functions appearing at tree level only depend on one variable, and are effective combinations of the multivariable functions \cite{Lee:2004ja,Beneke:2004in}.

To save some writing, we abbreviate the Wilson lines as $\delta_+(\w) \equiv \delta(\img \cD_+ + \w)$.
The complete set of bi-local operators to $\ord{\eps^2}$ is
\begin{equation}
\label{lc-ops_basis}
\begin{split}
\cO_0^\G(\w) &= \bb \delta_+(\w) \G b
,\\
\cO_1^{\G\mu}(\w) &= \delta_{12}'(\w)
\bb \delta_+(\w_1) \img \cD^\mu \delta_+(\w_2) \G b
,\\
\cO_2^{\G\mu\nu}(\w)
&= \delta_{123}''(\w)
\bb \delta_+(\w_1) \img\cD^\mu \delta_+(\w_2) \img\cD^\nu \delta_+(\w_3) \G b
,\\
\cO_{3,4}^{\G\mu}(\w)
&= \frac{1}{2} \bb \bigl(\img\cD^\mu \delta_+(\w) \pm \delta_+(\w) \img\cD^\mu \bigr) \G b
,\\
\cO_5^{\G\mu\nu}(\w)
&= \delta_{12}'(\w) \bb \delta_+(\w_1) \img\cD^\mu \img\cD^\nu \delta_+(\w_2) \G b
,\\
\cO_{6,7}^{\G\mu\nu}(\w)
&= \frac{1}{2} \delta_{12}'(\w)
\bb \bigl(\img\cD^\mu \delta_+(\w_1) \img\cD^\nu \delta_+(\w_2)
 \pm \delta_+(\w_1) \img\cD^\nu \delta_+(\w_2) \img\cD^\mu \bigr)\G b
,\\
\cO_8^{\G\mu\nu}(\w)
&= \bb \img\cD^\mu \delta_+(\w) \img\cD^\nu \G b
,\\
\cO_{9,10}^{\G\mu\nu}(\w)
&= \frac{1}{2} \bb \bigl(\img\cD^\mu \img\cD^\nu \delta_+(\w) \pm \delta_+(\w) \img\cD^\nu \img\cD^\mu \G \bigl) b
,\end{split}
\end{equation}
where the upper and lower sign belongs to the first and second label, respectively. Note the particular assignment of the Lorentz indices for $\cO^{\G\mu\nu}_{6,7}(\w)$ and $\cO^{\G\mu\nu}_{9,10}(\w)$, which turns out to be useful for parametrizing them. For later convenience we define $\cO_i^\G(\w) = \bar{b} \cO_i(\w) \G b$. That is, we drop the label $\G$ when referring to the derivative structure only, e.g., $\cO_1^\mu(\w) = \delta'_{12} \delta_+(\w_1)\img\cD^\mu \delta_+(\w_2)$.

The $\delta$-function factors $\delta_{12}'(\w)$ and $\delta_{123}''(\w)$ are defined as
\begin{equation}
\label{de_def}
\begin{split}
\delta_{12}'(\w)
&= \int\df\w_1\df\w_2 \frac{\delta(\w - \w_1) - \delta(\w - \w_2)}{\w_1 - \w_2}
= \int\df\w_1\df\w_2 \biggl(\frac{\delta(\w_{01})}{\w_{02}} + \frac{\delta(\w_{02})}{\w_{01}} \biggr)
\\
\delta_{123}''(\w)
&= \int \frac{\df\w_3\delta_{12}'(\w) - \df\w_1 \delta_{23}'(\w)}{\w_1 - \w_3}
= \int \df\w_1\df\w_2\df\w_3 \biggl(\frac{\delta(\w_{01})}{\w_{02}\w_{03}} + \frac{\delta(\w_{02})}{\w_{01}\w_{03}} + \frac{\delta(\w_{03})}{\w_{01}\w_{02}} \biggr)
,\end{split}
\end{equation}
with $\w_{ij} = \w_i - \w_j$ and $\w_0 \equiv \w$. They are completely symmetric in the $\w_i$ and include implicit integrations over $\w_1,\w_2$ and $\w_1,\w_2,\w_3$, respectively. They satisfy
\begin{equation}
\label{de12_prop}
\delta_{12}'(\w) \delta(\w_1) \delta(\w_2) = - \delta'(\w)
\q
\delta_{123}''(\w) \delta(\w_1) \delta(\w_2) \delta(\w_3) = \frac{1}{2} \delta''(\w)
.\end{equation}

The factors in brackets on the right-hand sides of Eqs.~\eqref{de_def} arise as the imaginary parts
\begin{subequations}
\label{de_img}
\begin{align}
\label{de12_img}
- \frac{1}{\pi} \Img \frac{1}{(\w_{01} + \img \eps)(\w_{02} + \img \eps)}
&= \frac{\delta(\w_{01})}{\w_{02}} + \frac{\delta(\w_{02})}{\w_{01}}
,\\
\label{de123_img}
- \frac{1}{\pi} \Img \frac{1}{(\w_{01} + \img \eps)(\w_{02} + \img \eps)(\w_{03} + \img\eps)}
&= \frac{\delta(\w_{01})}{\w_{02}\w_{03}} + \frac{\delta(\w_{02})}{\w_{01}\w_{03}} +
 \frac{\delta(\w_{03})}{\w_{01}\w_{02}}
.\end{align}
\end{subequations}
In Refs.~\cite{Lee:2004ja,Beneke:2004in} the right-hand side of Eq.~\eqref{de123_img} contains an additional piece $-\pi^2 \delta(\w_{01}) \delta(\w_{02}) \delta(\w_{03})$, which we think should not be there. Eqs.~\eqref{de_img} are defined upon integration over $\w$. Taking the imaginary part together with the $\img\eps$ prescription picks out the poles at $\w = \w_1,\w_2,\w_3$, and the replacements in Eqs.~\eqref{de_img} are a formal way of achieving the same. Taking the limit $\w_2,\w_3\to \w_1$ on both sides, Eqs.~\eqref{de_img} reduce to the $n = 1,2$ cases of the standard formula
\begin{equation*}
- \frac{1}{\pi} \Img \frac{1}{(\w_{01} + \img\eps)^{n+1}} = \frac{(-1)^n}{n!} \delta^{(n)}(\w_{01})
.\end{equation*}

The operators in Eq.~\eqref{lc-ops_basis} are not completely independent with respect to their Lorentz structure. An operator of $\ord{\eps^n}$ reduces to one of $\ord{\eps^{n-1}}$ when any derivative next to a Wilson line is contracted with $n_\mu$. For example, using Eq.~\eqref{de12_prop}, we have
\begin{equation}
\label{nO-rel}
n_\mu \cO_1^{\G\mu}(\w) = \bigl(\w \cO_0^\G(\w) \bigr)'
\q
n_\mu \cO_3^{\G\mu}(\w) = -\w \cO_0^\G(\w)
\q
n_\mu \cO_4^{\G\mu}(\w) = 0
,\end{equation}
where the prime denotes the derivative with respect to $\w$. This simply means that only $\cD_-$ and $\cD_\perp$ (or equivalently $v\cdot \cD$ and $\cD_\perp$) in the operators are independent structures, which reduces the number of shape functions needed to parametrize the operators. The full set of such relations is given in the appendix.

Another comment concerns the twist order of the operators in Eq.~\eqref{lc-ops_basis}. Formally, the factors of $\delta'_{12}(\w)$ and $\delta''_{123}(\w)$ reduce the twist order of an operators. For instance, $\cO^\G_{1,2}(\w)$ are formally of leading twist. Nevertheless, the discussion in Sec.~\ref{subsec:powercounting} shows that they do describe sub- or subsubleading twist corrections, because they contain explicit derivatives. Therefore, they must have coefficients of higher twist order. We will see an example of this in Sec.~\ref{subsubsec:zerogluon}. This involves the standard twist power counting, for one has to consider the operators and their coefficients. In this respect, our power counting is more transparent.

\subsection{Shape Functions}
\label{subsec:shapefunctions}

In this subsection we take $\img\cD = \img D - M_B v$. The parametrization of the operators also depends on the specific Dirac structure $\G$. We need $\G = \g^\alpha$ and $\G = \g^\alpha \g_5$, and define
\begin{equation}
\label{OPalpha_def}
\cO_i^\alpha(\w) = \bar{b} \cO_i(\w) \g^\alpha b
\q
\cP_i^\alpha(\w) = \bar{b} \cO_i(\w) \g^\alpha \g_5 b
.\end{equation}
We follow the notation of Ref.~\cite{Mannel:2004as} where possible.\footnote{In Ref.~\cite{Mannel:2004as} the operators are defined in terms of $\img \cD_b$, which shifts the shape functions' argument by $\la$.} Schematically,
\begin{equation}
\begin{split}
\vevB{\cO_i^\alpha(\w)} &= (F_i,G_i)(\w) v^\alpha + (K_i,M_i)(\w) (n - v)^\alpha + L_i(\w) \eta_\perp^\alpha
,\\
\vevB{\cP_i^\alpha(\w)} &= H_i(\w) (n - v)^\alpha + N_i(\w) v^\alpha + R_i(\w) \img \eps_\perp^\alpha
,\end{split}
\end{equation}
where $(F_i,G_i)(\w)$ stands for $F_i(\w)$ or $G_i(\w)$. In the heavy-quark limit, the $\g^\alpha$ in $\cO_i^\alpha(\w)$ is parallel to $v^\alpha$. Therefore, $F_i(\w)$ and $G_i(\w)$ contain the leading contribution, while the functions $K_i(\w)$, $L_i(\w)$, and $M_i(\w)$ are suppressed by $1/m_b$ because $v\cdot(n - v) = v\cdot \eta_\perp = 0$. They contain all higher order corrections in $1/m_b$ perpendicular to $v^\alpha$ that would arise from expanding the $b$-quark field. Similarly, the axial vector $\g^\alpha \g_5$ in $\cP_i^\alpha(\w)$ is perpendicular to $v^\alpha$ at leading order in $1/m_b$, and all contributions parallel to $v^\alpha$ are suppressed by $1/m_b$. Similarly, there will be $1/m_b$ suppressions from the HQET equations of motion, see below.

\subsubsection{The Leading Operator}

The $B$ expectation value of the leading operator is
\begin{equation}
\label{FK0_def}
\vevB{\cO_0^\alpha(\w)} = F_0(\w) v^\alpha + K_0(\w) (n-v)^\alpha
\q
\vevB{\cP_0^\alpha(\w)} = 0
,\end{equation}
which defines the QCD shape functions $F_0(\w)$ and $K_0(\w)$. This is exact, i.e., there are no higher order corrections on the right-hand side. The support of the shape functions is $0\leq \w\leq M_B$. Strictly speaking, the upper limit is $M_B$ rather then $\infty$, because Eq.~\eqref{FK0_def} contains no reference to the heavy-quark limit. The matrix element of $\cP^\alpha_0(\w)$ vanishes by parity invariance.

Using $\img\cD = \img\cD_b - \la v$ instead of $\img\cD_b$ in the leading operator only shifts the argument of the shape functions by $\la$, such that $F_0(\w + \la)$, $K_0(\w + \la)$ correspond to the functions defined in Ref.~\cite{Mannel:2004as}. Their expansion into the usual HQET shape functions \cite{Bauer:2001mh} is
\begin{equation}
\label{FK0_HQETexp}
F_0(\w) = f(\w - \la) + \frac{1}{2m_b} t(\w - \la) + \dotsb
\q
K_0(\w) = \frac{(\w - \la)}{m_b} f(\w - \la) + \frac{1}{m_b} h_1(\w - \la) + \dotsb
,\end{equation}
which explicitly shows the $1/m_b$ suppression of $K_0(\w)$. The QCD shape functions automatically contain the appropriate combinations of HQET shape functions that arise from expanding the QCD fields and states. This is in fact very similar to the local expansion, where the parameters $\mu_{\pi,G}^2$ are defined using the full QCD states, and thus differ from the HQET parameters $\lambda_{1,2}$ by $1/m_b$ corrections.

To constrain the form of $F_0(\w)$ and $K_0(\w)$ we can also directly parametrize their moments in HQET. Using the abbreviations
\begin{equation}
\label{la0_def}
\begin{split}
\tau_1 = \mathcal{T}_1 + 3\mathcal{T}_2
\q
\tau_2 = \mathcal{T}_3/3 + \mathcal{T}_4
\q
\la_0 = \la_1 + \tau_1 + 3(\la_2 + \tau_2)
\q
\rho_0 = \rho_1 + 3 \rho_2
,\end{split}
\end{equation}
where $\la_{1,2}$, $\rho_{1,2}$, and $\mathcal{T}_{1-4}$ are the usual HQET parameters, we find
\begin{equation}
\label{FK0_exp}
\begin{split}
F_0(\w)
&= \delta(\w - \la) - \frac{\la_0}{2m_b} \delta'(\w - \la) - \frac{\la_1 + \tau_1/m_b}{6} \delta''(\w - \la)
- \frac{\rho_1}{18} \delta'''(\w - \la) + \dotsb
,\\
K_0(\w)
&= \frac{2\la_0 - \rho_0/m_b}{6m_b} \delta'(\w - \la) + \frac{\rho_0}{6m_b} \delta''(\w - \la) + \dotsb
.\end{split}
\end{equation}
It is convenient to expand with respect to $\w - \la$, otherwise $\la$ explicitly appears in the moments. This is where the $b$-quark mass reappears. The normalizations of $F_0(\w)$ and $K_0(\w)$ are fixed by $b$-quark number conservation, while all other moments in Eq.~\eqref{FK0_exp} receive higher order corrections starting at order $\LQCD^4$ divided by an appropriate power of $m_b$.

\subsubsection{Subleading Operators}

For the $\ord{\eps}$ operators, parity invariance implies that
\begin{subequations}
\label{PT_inv}
\begin{equation}
\vevB{\cO_i^{\alpha\mu}(n,v)} \stackrel{P}{=} \vevB{\cO_{i\alpha\mu}(n_P,v_P)}
\q
\vevB{\cP_i^{\alpha\mu}(n,v)} \stackrel{P}{=} -\vevB{\cP_{i\alpha\mu}(n_P,v_P)}
,\end{equation}
where we temporarily suppressed the $\w$ dependence, but explicitly showed the dependence on the vectors $n$ and $v$. The transformed vectors satisfy $n_P^\mu = n_\mu$, $v_P^\mu = v_\mu$. Hence, the $\cP_i^{\alpha\mu}(\w)$ must be proportional to $\eps_\perp^{\alpha\mu}$, because $\eps_\perp^{\alpha\mu} = -\eps_{\perp\alpha\mu}$, while the $\cO_i^{\alpha\mu}(\w)$ must not contain $\eps_\perp^{\alpha\mu}$. Time-reversal invariance requires
\begin{equation}
\begin{split}
\vevB{\cO_{1,3}^{\alpha\mu}(n,v)} &\stackrel{T}{=} \vevB{\cO_{1,3\alpha\mu}(n_P,v_P)}^*
= \vevB{\cO_{1,3\alpha\mu}(n_P,v_P)}
\\
\vevB{\cO_{4}^{\alpha\mu}(n,v)} &\stackrel{T}{=} \vevB{\cO_{4\alpha\mu}(n_P,v_P)}^*
= -\vevB{\cO_{4\alpha\mu}(n_P,v_P)}
,\end{split}
\end{equation}
\end{subequations}
and identical relations hold for the $\cP_i^{\alpha\mu}(\w)$. The complex conjugation reverses the order of all derivatives in the operators, yielding the additional minus sign for $\cO_4^{\alpha\mu}(\w)$ and $\cP_4^{\alpha\mu}(\w)$. Eqs.~\eqref{PT_inv} show that the matrix elements of $\cP_{1,3}^{\alpha\mu}(\w)$ and $\cO_4^{\alpha\mu}(\w)$ have to vanish.

The remaining nonvanishing matrix elements are parametrized as\footnote{The definition of $R_4(\w)$ is slightly different in Ref.~\cite{Mannel:2004as}, $R_4'(\w)$ there corresponds to $2 R_4(\w)$ here.}
\begin{equation}
\label{O1_param}
\begin{split}
\vevB{\cO_1^{\alpha\mu}(\w)}
&= - [(\w F_0(\w) v^\alpha + \w K_0(\w) (n - v)^\alpha]' (n - v)^\mu
\\ & \quad
 - [(F_1-\la F_0)(\w) v^\alpha + (K_1 - \la K_0)(\w) (n - v)^\alpha]' n^\mu
 - \frac{1}{2} L_1'(\w) \eta^{\perp\alpha\mu}
,\\
\vevB{\cO_3^{\alpha\mu}(\w)}
&= [(\w F_0(\w) v^\alpha + \w K_0(\w) (n - v)^\alpha] (n - v)^\mu
\\ & \quad
 + [(F_3 - \la F_0)(\w) v^\alpha + (K_3 - \la K_0)(\w) (n - v)^\alpha] n^\mu
 + \frac{1}{2} L_3(\w) \eta^{\perp\alpha\mu}
,\\
\vevB{\cP_4^{\alpha\mu}(\w)} &= - \frac{\img}{2} R_4(\w) \eps_\perp^{\alpha\mu}
,\end{split}
\end{equation}
and we already took into account the constraints from Eq.~\eqref{nO-rel}. The $\delta_{12}'(\w)$ inside $\cO_1^{\alpha\mu}(\w)$ makes it formally twist $\ord{1}$. Mainly for cosmetical reasons, we want its shape functions to be of the twist order at which they actually appear, which is why we use derivatives of shape functions to parametrize the operator. Eqs.~\eqref{O1_param} are chosen such that
\begin{equation}
\label{FK1_def}
\begin{aligned}
\vevB{v_\mu\cO_1^{\alpha\mu}(\w)} &= - (F_1 - \la F_0)'(\w) v^\alpha - (K_1 - \la K_0)'(\w) (n - v)^\alpha
,\\
\vevB{\eta_{\perp\alpha\mu}\cO_1^{\alpha\mu}(\w)} &= - L_1'(\w)
,\\
\vevB{v_\mu\cO_3^{\alpha\mu}(\w)} &= (F_3 - \la F_0)(\w) v^\alpha + (K_3 - \la K_0)(\w) (n - v)^\alpha
,\\
\vevB{\eta_{\perp\alpha\mu} \cO_3^{\alpha\mu}(\w)} &=L_3(\w)
,\\
\vevB{\img\eps_{\perp\alpha\mu}\cP_4^{\alpha\mu}(\w)} &= R_4(\w)
.\end{aligned}
\end{equation}
To leading order in $1/m_b$, $R_4(\w)$ equals $-h_1(\w-\la)$ of Ref.~\cite{Bauer:2001mh}. The $K_{1,3}(\w)$ and $L_{1,3}(\w)$ are suppressed by $1/m_b$, as argued before. In the heavy-quark limit the HQET equations of motion imply $F_3(\w) = 0$, and therefore $F_3(\w)$ is also suppressed by $1/m_b$, which is why we choose $(n - v)^\mu$ and $n^\mu$ as independent vectors in Eqs.~\eqref{O1_param}.

Considering the $\ord{\eps^2}$ operators, by the same arguments as in Eqs.~\eqref{PT_inv}, the only nonvanishing matrix elements are
\begin{equation}
\label{OP_e2_nonzero}
\vevB{\cO_{2,5,8}^{\alpha(\mu\nu)}(\w)}
\q
\vevB{\cP_{2,5,8}^{\alpha[\mu\nu]}(\w)}
\q
\vevB{\cO_{6,9}^{\alpha\mu\nu}(\w)}
\q
\vevB{\cP_{7,10}^{\alpha\mu\nu}(\w)}
.\end{equation}
Due to the three indices the decompositions become rather lengthy, so we will not write them out explicitly.\footnote{A complete decomposition for $\cO_2^{\alpha(\mu\nu)}(\w)$ is given in Ref.~\cite{Mannel:2004as}.} Instead, we define the shape functions as in Eqs.~\eqref{FK1_def} by projecting out the independent Lorentz structures, which is done in the appendix. Here, we only list those needed in the following,
\begin{equation}
\begin{aligned}
\vevB{v_\alpha\eta_{\perp\mu\nu}\cO_2^{\alpha(\mu\nu)}(\w)}
&= - \frac{1}{2} G_2'(\w)
\q &
\vevB{\eta_{\perp\alpha(\mu} v_{\nu)} \cO_2^{\alpha(\mu\nu)}(\w)}
&= \frac{1}{2} (L_2 - \la L_1)''(\w)
,\\
\vevB{v_\alpha\eta_{\perp\mu\nu}\cO_5^{\alpha(\mu\nu)}(\w)}
&= G_5(\w)
\q &
\vevB{(n - v)_\alpha\img \eps_{\perp\mu\nu}\cP_5^{\alpha[\mu\nu]}(\w)}
&= H_5(\w)
,\\
\vevB{\bn_\mu \img\eps_{\perp\alpha\nu} \cP_{10}^{\alpha\mu\nu}(\w)}
&= (R_{10} - \la R_4)(\w)
.\end{aligned}
\end{equation}
The functions $G_5(\w)$, $H_5(\w)$ equal $G_3(\w-\la)$, $H_4(\w-\la)$ of Ref.~\cite{Mannel:2004as}\footnote{The numbering of the operators is changed to account for $\cO_{3,4}(\w)$. The operators $\cO_3(\w)$, $\cP_4(\w)$ of Ref.~\cite{Mannel:2004as} correspond to $\cO_5(\w)$, $\cP_5(\w)$ here.}, and to leading order in $1/m_b$, $G_2(\w-\la)$, $H_2(\w-\la)$ of Ref.~\cite{Bauer:2001mh}, respectively.

\subsection{Reparametrization Invariance}
\label{subsec:RPI}

The operators $\cO_{0,3,4,5}(\w)$ correspond to the operator basis originally introduced in Ref.~\cite{Bauer:2001mh}, and appear in the OPE for the triple differential rate, see Eq.~\eqref{lc-ops_btous} below. The operators $\cO_{0,1,2,5}(\w)$ are the complete set of operators needed for the direct computation of the lepton energy spectrum in $B\to X_u\ell\bar{\nu}_\ell$ \cite{Mannel:2004as}. There, the momentum of the charged lepton is used to define the light-cone direction, i.e., one chooses $p_\ell = E_\ell \bn$. For $B\to X_s \g$ with $q = p_\g$ this choice is equivalent to ours in Eq.~\eqref{lc_fix}. For the triple differential rate with $q = p_\ell + p_\nu$ the two choices of light-cone directions are rotated with respect to each other by an angle depending on the three-momenta of charged lepton and neutrino.

The lepton energy spectrum is independent of the choice of the light-cone direction. Therefore, integrating the triple differential rate should give the same result as the direct computation. Since the two approaches require different subsets of operators, not all operators in Eq.~\eqref{lc-ops_basis} can contain independent nonperturbative information, and therefore, some shape functions appearing in their parametrizations should be related, beyond simple relations like Eq.~\eqref{nO-rel}. The shape functions basically describe the momentum distribution of the $B$ meson, and because it has zero spin, they cannot contain any spatial directional information. However, the light-cone coordinates separate the spatial directions into $\eta_\perp$ and $n - \bn$. Hence, shape functions referring to $\eta_\perp$ and $n - \bn$ should be somehow related.

The independence of physical quantities on the choice of the light-cone direction is described by reparametrization invariance (RPI). More generally, there are two types of ambiguities related to RPI in our setting. First, the ambiguity in the decomposition of the heavy-quark momentum leads to the well-known reparametrization invariance of HQET \cite{Luke:1992cs}. Its implications for the twist expansion have been studied in HQET in Ref.~\cite{Campanario:2002fy}. The authors there consider the case of a single outgoing particle with $q_- = q_\perp = 0$. Their results are thus not applicable to the triple differential rate, and it is not surprising that they do not hold in this case.

The second type of ambiguity arises from the arbitrariness in the definition of the light-cone vectors. This has been studied in some detail in SCET \cite{Chay:2002vy,Manohar:2002fd,Pirjol:2002km}, where it places many constraints on the form of allowed operators. Following Ref.~\cite{Manohar:2002fd}, there are three classes of transformations that preserve the fundamental properties $n^2 = \bn^2 = 0$ and $n\cdot \bn = 2$,
\begin{equation*}
\text{(I)} \quad \bigg\{\begin{aligned} n &\to n + \delta_\perp, \\ \bn &\to \bn, \end{aligned}
\qquad
\text{(II)} \quad \bigg\{\begin{aligned} n &\to n, \\ \bn &\to \bn + \bar{\delta}_\perp, \end{aligned}
\qquad
\text{(III)} \quad \bigg\{\begin{aligned} n &\to (1 + \alpha) n, \\ \bn &\to (1 - \alpha) \bn, \end{aligned}
\end{equation*}
which are generated by the five infinitesimal parameters $\{\delta_\perp,\bar{\delta}_\perp,\alpha\}$. The RPI transformation studied in Ref.~\cite{Campanario:2002fy} corresponds to a combined action of (I) and (III).

We want to study the effect of rotating the light-cone direction, while keeping $2v = n + \bn$ fixed. Thus, we set $\alpha = 0$ and $\bar{\delta}_\perp = -\delta_\perp$ and consider the infinitesimal rotation $\delta_R$
\begin{subequations}
\label{deR}
\begin{equation}
n \to n + \delta_R n
\q
\bn \to \bn + \delta_R \bn \quad\text{with}\quad
\delta_R n = - \delta_R \bn = \delta_\perp
,\end{equation}
under which
\begin{equation}
\delta_R v = 0 \q \delta_R (n - \bn) = 2 \delta_\perp \q
\delta_R \eta_\perp^{\mu\nu} = (n - \bn)^{(\mu} \delta_\perp^{\nu)}
\q
\delta_R \eps_\perp^{\mu\nu} = (n - \bn)^{[\mu} \eps_\perp^{\nu]\la} \delta_{\perp\la}
.\end{equation}
\end{subequations}
The last two transformations can be found by requiring that metric and Levi-Civita tensor stay invariant. Similarly, any four-vector is invariant under $\delta_R$, only its light-cone components change according to Eqs.~\eqref{deR}.

We also need the transformation of the Wilson line $\delta_+ = \delta(\img\cD_+ + \w)$,
\begin{equation}
\label{deR_d+}
\delta_R \delta_+(\w) = -\delta'_{12}(\w) \delta_+(\w_2) (\delta_\perp\cdot\img\cD_\perp) \delta_+(\w_1)
,\end{equation}
which can be found using $\delta_+(\w) = (-1/\pi)\Img (\img\cD_+ + \w + \img\eps)^{-1}$. Note, that $\delta_\perp$ need not be formally $\ord{\La}$, but can be $\ord{1}$, because $\delta_R$ only rotates the light-cone components of $k$ into each other, but leaves $v$ invariant. In particular, the transformation \eqref{deR_d+} does not change the twist order of an operator. However, it connects different orders in $\eps$, which yields constraints on the shape functions arising from operators of different order in $\eps$.

The Dirac structure $\G$ and the $b$-quark fields are unaffected by $\delta_R$, so Eq.~\eqref{deR_d+} yields
\begin{subequations}
\label{deR_Os}
\begin{equation}
\label{deR_O0}
\delta_R \cO_0^\G(\w) = - \delta_{\perp\mu} \cO_1^{\G\mu}(\w)
,\end{equation}
and for the $\ord{\eps}$ operators
\begin{equation}
\delta_R \cO_1^{\G\mu}(\w) = - 2 \delta_{\perp\nu} \cO_2^{\G(\mu\nu)}
\q
\delta_R \cO_{3,4}^{\G\mu}(\w) = - \delta_{\perp\nu} \cO_{6,7}^{\G\mu\nu}
.\end{equation}
\end{subequations}
Similarly, the operators of $\ord{\eps^2}$ are transformed into $\ord{\eps^3}$ operators. Taking the $B$ expectation values of Eqs.~\eqref{deR_Os}, we can pull $\delta_R$ out of the matrix elements, because it has no effect on the $B$-meson states, as well. Thus, the same relations also hold for the $B$ expectation values, which reduces the number of independent shape functions.

Taking $\G=\g^\alpha$ and employing Eqs.~\eqref{FK0_def} and \eqref{O1_param}, the matrix element of Eq.~\ref{deR_O0} yields
\begin{equation}
\delta_R \bigl(F_0(\w) v^\alpha + K_0(\w) (n - v)^\alpha \bigr) = K_0(\w) \delta_\perp^\alpha
= \frac{1}{2} L_1'(\w) \delta_\perp^\alpha
,\end{equation}
from which it follows that
\begin{subequations}
\begin{equation}
L_1'(\w) = 2K_0(\w)
.\end{equation}
This relation has the expected form, since $K_0(\w)$ and $L_1'(\w)$ are proportional to $(n - v) = (n - \bn)/2$ and $\eta_\perp$, respectively. Similarly, the relation between $\cO_1^{\alpha\mu}(\w)$ and $\cO_2^{\alpha(\mu\nu)}(\w)$ yields%
\begin{equation}
L_2'(\w) = 2K_1(\w)
\quad\text{and}\quad
G_2(\w) = -2 (\w - \la) F_0(\w) - 2F_1(\w)
.\end{equation}
\end{subequations}
The remaining relations following from Eqs.~\eqref{deR_Os} are given in the appendix.

Writing the second relation as $G_2(\w) + 2F_1(\w) = -2 (\w - \la) F_0(\w)$ one can easily see that the $B\to X_u \ell \bar{\nu}_\ell$ lepton energy spectrum in Ref.~\cite{Mannel:2004as} expanded to subleading twist agrees with the results obtained in Refs.~\cite{Lee:2004ja,Bosch:2004cb,Beneke:2004in}. The appearance of the different operator structure in Ref.~\cite{Mannel:2004as} is not related to the use of QCD vs HQET fields as presumed in Ref.~\cite{Lee:2004ja}, but arises from choosing the light-cone direction to be parallel to the lepton momentum. We disagree with the statement in Ref.~\cite{Beneke:2004in} that this choice can lead (by itself) to incorrect results. Part of the reason why Ref.~\cite{Bauer:2002yu} obtained an incorrect result is that it tried to match on an (for this choice) incomplete operator basis.

\section{The Light-Cone OPE}
\label{sec:light-coneOPE}

In this section, which is mainly technical, we compute the light-cone OPE of the current correlator $T_{\alpha\beta}$ in Eq.~\eqref{Tf_def} to $\ord{\eps^2}$ in the power counting \eqref{pc_mod}. We start with considering generic currents $J^\alpha = \bar{b} \G^\alpha f$ and $J^\beta = \bar{f} \G^\beta b$ with two arbitrary Dirac structures $\Gamma^\alpha$ and $\Gamma^\beta$. For completeness, Sec.~\ref{subsec:matching} contains the actual matching calculation. In Secs.~\ref{subsec:generalcurrents} and \ref{subsec:slandradcurrents} we give the results for general currents, Eq.~\eqref{lc-OPE}, and semileptonic and radiative currents, Eqs.~\eqref{lc-OPE_f}. The latter are used in Sec.\ref{subsec:hadronictensor} to obtain the hadronic tensor.

\subsection{Matching Calculation}
\label{subsec:matching}

\subsubsection{Zero-Gluon Matrix Element}
\label{subsubsec:zerogluon}

\begin{figure}%
\includegraphics[width=0.45\columnwidth]{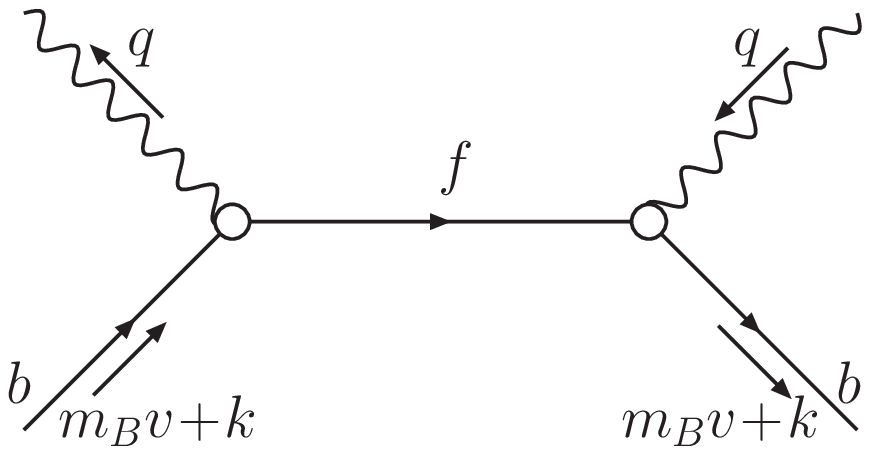}%
\hfill%
\includegraphics[width=0.45\columnwidth]{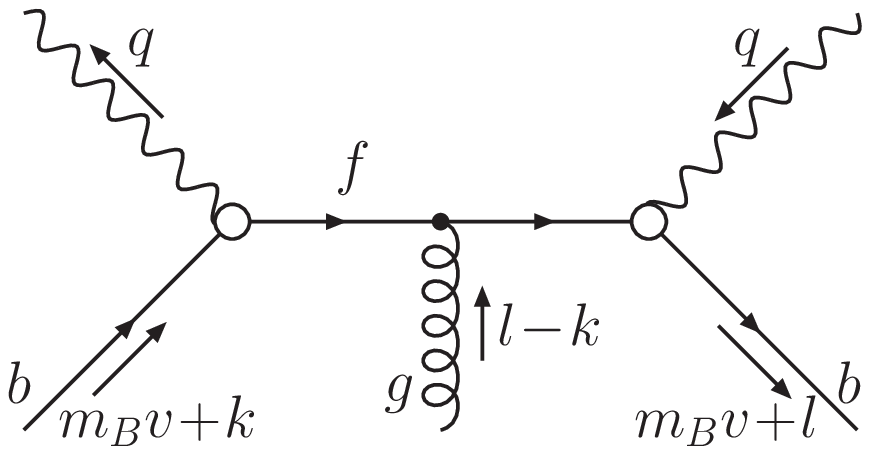}%
\caption{\label{fig:mes}Tree-level Feynman diagrams for the zero- and one-gluon matrix element.}
\end{figure}

The matrix element of $T^{\alpha\beta}$ between $b$-quark states with momentum $p_b = M_B v + k$ is shown on the left in Fig.~\ref{fig:mes}. With $P = M_B v - q$ it is
\begin{equation}
\label{A0_def}
\mae{b}{T^{\alpha\beta}}{b}
= -\frac{1}{\pi} \Img \biggl(
\bu_b \G^\alpha \frac{1}{\slP + \gslash{k} + \img\eps} \G^\beta u_b \biggr)
= \bu_b A_{0\mu} \G^\alpha \g^\mu \G^\beta u_b
.\end{equation}
Using the shorthand $\delta_k \equiv \delta(P_+ + k_+)$, the expansion of $A_0^\mu$ to $\ord{\eps^2}$ is
\begin{equation}
\label{A0_exp}
\begin{split}
2A_0^\mu
=  n^\mu \delta_k + \frac{2k_\perp^\mu}{P_-} \delta_k
+ \frac{k_\perp^2}{P_-} \biggl(n^\mu\delta'_k - \frac{1}{P_-} \bn^\mu \delta_k \biggr)
- \frac{2k_\perp^\mu k_-}{P_-^2} \delta_k + \ord{\eps^3}
.\end{split}
\end{equation}
The derivative is with respect to the argument of the $\delta$ function.
There are no terms proportional to $k_-$ and $k_-^2$. By first expanding the $\delta$ function in $k_\perp^2$ only, one can see that there are no contributions proportional to $k_-^n$.

Since $\cO_0(\w)$ is the only operator at order $\eps^0$, we can extract its coefficient from Eq.~\eqref{A0_exp}
\begin{equation}
\label{lc-OPE_O0}
A_0^\mu = \int\df\w \delta(P_+ - \w) \frac{n^\mu}{2} \mae{b}{\cO_0(\w)}{b} + \ord{\eps}
,\end{equation}
where $\mae{b}{\cO_0(\w)}{b} = \delta(k_+ + \w)$. At higher orders, this extraction becomes ambiguous. For example, by partial integration we can rewrite $k_\perp^2 \delta'_k$ as
\begin{equation}
\begin{split}
k_\perp^2 \delta'(P_+ + k_+)
&= \int\df\w \delta'(P_+ - \w) k_\perp^2 \delta(k_+ + \w)
= \int\df\w \delta(P_+ - \w) k_\perp^2 \delta'(k_+ + \w)
\\
&= \int\df\w \theta(P_+ - \w) k_\perp^2 \delta''(k_+ + \w)
.\end{split}
\end{equation}
The first expression requires $\cO_{8,9}(\w)$, the second $\cO_{5,6}(\w)$, and the last $\cO_2(\w)$. The operators themselves correspond to twist $\ord{\La^2}$ through $\ord{1}$. The difference in their order is canceled by their coefficients in the convolution, which are of relative $\ord{\La^{-1}}$ through $\ord{\La}$, so that the total order is the same. In our case the convolution always involves $\delta(P_+ - \w)$.

Because $k$ and $\delta(k_+ + \w)$ commute, while $\img\cD$ and $\delta(\img\cD_+ + \w)$ do not, the zero-gluon matrix element cannot distinguish these operators, and thus only fixes a linear combination of their Wilson coefficients. This is avoided by either directly expanding the propagator $1/(\slP + \img\gslash{\cD})$, or gets resolved by the one-gluon matrix element, which can distinguish the operators. For terms containing $k_-$ this means that one is a priori not allowed to use the HQET equations of motion to replace $k_- = -k_+$, because in some operators the derivatives are separated from the $b$-quark fields by Wilson lines.

\subsubsection{One-Gluon Matrix Element}

The one-gluon matrix element is depicted on the right in Fig.~\ref{fig:mes}. It has an additional soft background gluon, which we take to be in the initial state with momentum $l - k$. Working in light-cone gauge $A_+ = 0$, we have
\begin{equation}
\label{A1_def}
\mae{b}{T^{\alpha\beta}}{bg}
=-\frac{1}{\pi} \Img \biggl(
- g t^a \bu_b \G^\alpha \frac{1}{\slP + \gslash{l} + \img\eps)} \gslash{\e}^a
\frac{1}{\slP + \gslash{k} + \img\eps} \G^\beta u_b \biggr)
= \bu_b A_{1\mu} \G^\alpha \g^\mu \G^\beta u_b
.\end{equation}
where
\begin{equation}
\label{A1_unexp}
A_1^\mu \g_\mu
=- (\slP + \gslash{l}) \gslash{\e} (\slP + \gslash{k})
\frac{\delta[(P + k)^2] - \delta[(P + l)^2]}{(P + l)^2 - (P + k)^2}
.\end{equation}
We absorbed all factors from the gluon vertex into the polarization vector $\e \equiv g t^a \e^a$, such that $\e$ corresponds to a covariant derivative $\img \cD = (\img \partial - M_B v) + g A$.

According to Eq.~\eqref{pc_ops} we expand this in $k_-, l_-, \e_-$, and $k_\perp, l_\perp, \e_\perp$. Employing the identity%
\begin{equation}
\label{dirac}
\g^\alpha \g^\mu \g^\beta = \eta^{\alpha\mu} \g^\beta + \eta^{\mu\beta} \g^\alpha
- \eta^{\alpha\beta} \g^\mu - \img \eps^{\alpha\mu\beta\nu} \g_\nu \g_5
\end{equation}
to reduce the product of three $\g$ matrices, we find to $\ord{\eps^2}$
\begin{equation}
\label{A1_exp}
\begin{split}
2A_1^\mu
&= \frac{1}{P_-} \biggl( (\eta - \img\eps)_\perp^{\mu\nu} \e_\nu (\delta_k \pm \delta_l)
- n^\mu (\eta - \img\eps)_\perp^{\nu\la} (l \pm k)_\nu \e_\la \frac{\delta_k - \delta_l}{l_+ - k_+} \biggr)
\\ & \quad
- \frac{1}{P_-^2} \biggl(\bn^\mu (\eta + \img\eps)_\perp^{\nu\la} (\e_\nu k_\la \delta_k + l_\nu \e_\la \delta_l)
\\ & \quad
+ (\eta - \img\eps)_\perp^{\mu\nu} \bigl( (\e_- k_\nu + l_- \e_\nu) \delta_k  \pm (l_\nu \e_- + \e_\nu k_-) \delta_l \bigr)
\biggr) + \ord{\eps^3}
.\end{split}
\end{equation}
Here, $\delta_k \equiv \delta(P_+ + k_+)$, $\delta_l \equiv \delta(P_+ + l_+)$, and the upper and lower signs belong to $\eta_\perp$ and $\img\eps_\perp$, respectively.

\subsubsection{Four-Quark Contributions}

\begin{figure}%
\includegraphics[width=0.45\columnwidth]{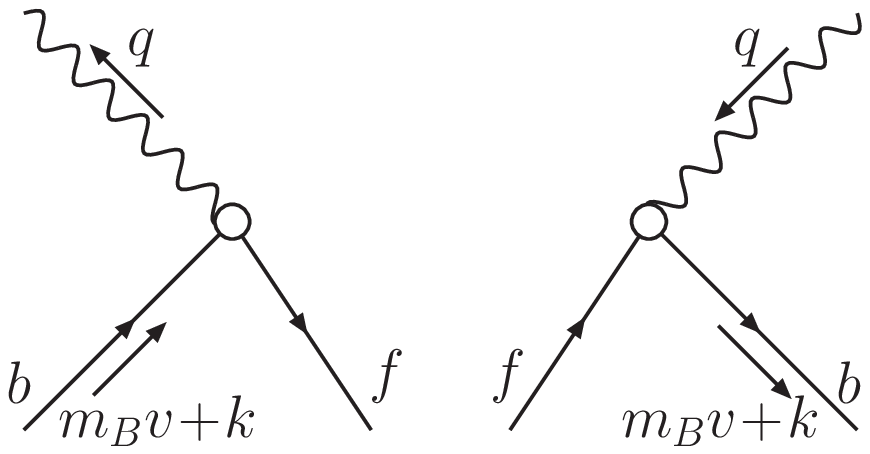}%
\hfill%
\includegraphics[width=0.45\columnwidth]{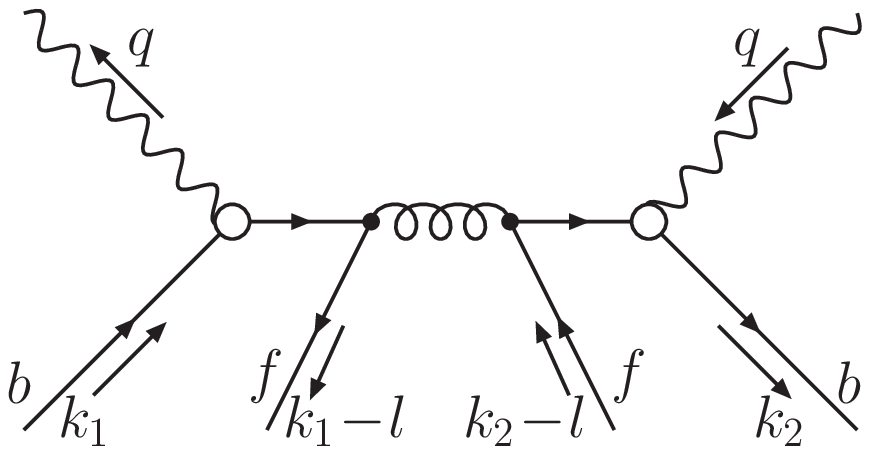}%
\caption{\label{fig:4q}Four-quark diagrams. On the right we only show the routing of residual momenta.}
\end{figure}

We include the four-quark matrix element on the right in Fig~\ref{fig:4q}, which was first considered in Ref.~\cite{Lee:2004ja}. Its size has been subject to some recent discussion \cite{Lee:2004ja,Bosch:2004cb,Neubert:2004cu,Beneke:2004in}. Although the corresponding operator is formally of subleading twist and third local order, it is unclear at present how well, or if at all, this represents its actual size. We thus refrain from assigning it a power in $\eps$. Instead, we treat it as a separate contribution and keep only its leading twist term. The graph on the left in Fig.~\ref{fig:4q} has no imaginary part contributing to the triple differential rate. Its contribution to single differential spectra can be computed, where the lepton or neutrino lines are connected \cite{Leibovich:2002ys}.

We route the momenta such that the quarks carry residual momenta $k_1$ and $k_2$, and the gluon residual momentum $l$. To leading twist the matrix element of $T^{\alpha\beta}$ then becomes \cite{Lee:2004ja}
\begin{equation}
\label{A4q_def}
\begin{split}
\mae{bf}{T^{\alpha\beta}}{fb}
&=-\frac{1}{\pi} \Img \biggl(-g^2
\bu_b \G^\alpha \frac{1}{\slP + \gslash{k}_2 + \img\eps} \g_\mu t^a u_f
\frac{\eta_\perp^{\mu\nu}}{(P + l)^2 + \img \eps} \bu_f
 t^a \g_\nu \frac{1}{\slP + \gslash{k}_1 + \img\eps} \G^\beta u_b \biggr)
\\
&= -g^2 A_{4q}^{\mu\nu} n_{(\la} \bn_{\ka)} \bu_b^i \G^\alpha \g_\mu \g^\la \g_\nu \G^\beta u_b^j (\bu_f t^a)^j \g^\ka P_L (t^a u_f)^i
,\end{split}
\raisetag{3ex}
\end{equation}
where $i,j$ are color indices. The second form is obtained by noting that for the currents in Eq.~\eqref{currents} $u^L_4 = (\g^\nu\G^\beta u_b)^L$ and $\bar{u}^L_1 = (\bar{u}_b \G^\alpha \g^\mu)^L$ are left-handed spinors and employing the Fierz identity
\begin{equation*}
(\bar{u}^L_1 \g_{\perp\la} u^L_2)(\bar{u}^L_3 \g_\perp^\la u^L_4)
= n_{(\la} \bn_{\ka)} (\bar{u}^L_1 \g^\la u^L_4)(\bar{u}^L_3 \g^\ka u^L_2)
.\end{equation*}
The latter follows from contracting the general Fierz identity for left-handed vector currents
\begin{equation*}
(\bar{u}^L_1 \g^\mu u^L_2)(\bar{u}^L_3 \g^\nu u^L_4)
=\frac{1}{2}(\eta^{\mu\nu} \eta^{\la\ka} - \eta^{\mu\la} \eta^{\nu\ka} - \eta^{\mu\ka} \eta^{\nu\la}
 - \img \eps^{\mu\nu\la\ka}) (\bar{u}^L_1 \g_\la u^L_4)(\bar{u}^L_3 \g_\ka u^L_2)
.\end{equation*}

The expansion of $A_{4q}^{\mu\nu}$ to leading twist is
\begin{equation}
\begin{split}
4A_{4q}^{\mu\nu}
&=\frac{1}{P_-} n^\mu n^\nu  \biggl(\frac{\delta_{k_1}}{(P + k_2)_+(P + l)_+}
 + \frac{\delta_{k_2}}{(P + l)_+(P + k_1)_+} + \frac{\delta_l}{(P + k_2)_+(P + k_1)_+} \biggr)
.\end{split}
\end{equation}
Plugged into Eq.~\eqref{A4q_def} this is matched onto
\begin{subequations}
\label{T4q}
\begin{equation}
\begin{split}
T_{4q}^{\alpha\beta}
&= -\frac{1}{2}\int\df\w \delta(P_+ - \w) \frac{1}{P_-} \bar{b} n_\mu \cQ^f(\w) \G^\alpha \g^\mu \G^\beta b
,\end{split}
\end{equation}
with the four-quark operator
\begin{equation}
\label{lc-ops_4q}
\begin{split}
\cQ^{fij}(\w)
&= g^2\delta_{123}''(\w) \delta_+(\w_1) \delta^f_+(\w_2) (\bar{f} t^a)^j \gslash{n}P_L (t^af)^i \delta_+(\w_3)
.\end{split}
\end{equation}
\end{subequations}
Here, $\delta_+^f(\w)$ acts on everything on its right except $f$. We include the $g^2$ in the operators, and do not think of it as $4\pi\alpha_s$, but treat it like the $g^2$ inside $(\img\cD)^2$, as suggested in Ref.~\cite{Beneke:2004in}.

\subsection{The Result for General Currents}
\label{subsec:generalcurrents}

From the general basis of light-cone operators in Eq.~\eqref{lc-ops_basis} we define the combinations
\begin{equation}
\label{lc-ops_btous}
\begin{split}
\cO_{5,8\perp}^\G(\w) &= \eta_{\perp\mu\nu} \cO_{5,8}^{\G\mu\nu}(\w)
,\\
\cP_{5,8\perp}^\G(\w) &= \img \eps_{\perp\mu\nu} \cO_{5,8}^{\G\mu\nu}(\w)
,\\
\cR_{4\perp}^{\G\mu}(\w)
&= \eta_{\perp\la}^\mu \cO_3^{\G\la}(\w) - \img\eps_{\perp\la}^\mu \cO_4^{\G\la}(\w)
,\\
\cR_{10\perp}^{\G\mu}(\w)
&= \bn_\la(\eta_{\perp\ka}^\mu \cO_9^{\G\la\ka}(\w) - \img \eps_{\perp\ka}^\mu \cO_{10}^{\G\la\ka}(\w))
.\end{split}
\end{equation}
The operators $\cR_{4,10\perp}(\w)$ are only needed for $B\to X_u \ell \bar{\nu}_\ell$, but not for $B\to X_s \g$.
Their major contributions arise from the ``gluonic'' parts $\cO_{4,10}(\w)$. Employing Eqs.~\eqref{de12_prop} the zero- and one-gluon matrix elements of the operators are straightforward to calculate, and comparing with Eqs.~\eqref{A0_exp} and \eqref{A1_exp} we can read off their OPE coefficients.

To write down the light-cone OPE of $T^{\alpha\beta}$, we write it as
\begin{subequations}
\label{lc-OPE}
\begin{equation}
\label{Amu_def}
T^{\alpha\beta} = \bb A_\mu \G^\alpha \g^\mu \G^\beta b + T^{\alpha\beta}_{4q}
.\end{equation}
The four-quark contribution $T^{\alpha\beta}_{4q}$ is given in Eqs.~\eqref{T4q}. The expansion of $A^\mu$ to $\ord{\eps^2}$ reads
\begin{equation}
\label{lc-OPE_Amu}
\begin{split}
A^\mu &= \frac{1}{2} \int\df\w \delta(P_+ - \w) \biggl\{
n^\mu \cO_0(\w) + \frac{2}{P_-} \cR_{4\perp}^\mu(\w)
\\ & \quad
- \frac{1}{P_-} \biggl[ n^\mu (\cO - \cP)_{5\perp}(\w)
+ \frac{1}{P_-} \bn^\mu (\cO + \cP)_{8\perp}(\w) \biggr]
- \frac{2}{P_-^2} \cR_{10\perp}^\mu(\w)
\biggr\} + \ord{\eps^3}
.\end{split}
\end{equation}
\end{subequations}
For comparison, we wrote the terms in the same order as the corresponding ones in Eq.~\eqref{A0_exp}. Eqs.~\eqref{lc-OPE} represent the light-cone OPE of $T^{\alpha\beta}$ to $\ord{\eps^2}$ and are the starting point for the further analysis.

\subsection{Semileptonic and Radiative Currents}
\label{subsec:slandradcurrents}

To continue we consider the currents $J_\alpha^f$ in Eq.~\eqref{currents}. For $f = u$, corresponding to $B\to X_u \ell\bar{\nu}_\ell$, we have $\G^\alpha = \g^\alpha P_L$, and the Dirac structure in Eq.~\eqref{Amu_def} becomes $\G = \G^\alpha \g^\mu \G^\beta = \g^\alpha \g^\mu \g^\beta P_L$. Therefore, with the help of Eq.~\eqref{dirac},
\begin{equation}
\label{dirac_btoq}
\begin{split}
n_\mu \G^\alpha \g^\mu \G^\beta
&= \bigl(n^\alpha n^\beta \bn_\nu - (\eta + \img\eps)_\perp^{\alpha\beta} n_\nu \bigr) \g^\nu P_L
,\\
\bn_\mu \G^\alpha \g^\mu \G^\beta
&= \bigl( \bn^\alpha \bn^\beta n_\nu - (\eta - \img\eps)_\perp^{\alpha\beta} \bn_\nu \bigr) \g^\nu P_L
,\\
\eta_{\perp\mu\nu} \G^\alpha \g^\mu \G^\beta
&= \bigl( - n^{(\alpha}\bn^{\beta)} \eta_{\perp\mu\nu}
 - \img n^{[\alpha} \bn^{\beta]} \eps_{\perp\mu\nu} \bigr) \g^\nu P_L
.\end{split}
\end{equation}
Similarly, for $B\to X_s \g$, Eq.~\eqref{currents} gives $\G^\alpha = \g_\perp^\alpha \gslash{\bn} P_L$. Using $\gslash{\bn}^2 = 0$ and Eq.~\eqref{dirac} yields
\begin{equation}
\label{dirac_btos}
\G^\alpha \g^\mu \G^\beta
= \g_\perp^\alpha \gslash{\bn} \g^\mu \gslash{\bn} \g_\perp^\beta P_R
= -2 \bn^\mu (\eta + \img\eps)_\perp^{\alpha\beta} \bn_\nu \g^\nu P_R
.\end{equation}
In this case only the terms proportional to $n^\mu$ in Eq.~\eqref{lc-OPE_Amu} contribute.

The appearing Dirac structures are $\G = \g^\alpha P_L$ and $\G = \g^\alpha P_R$. The discussion in Sec.~\ref{subsec:shapefunctions} shows that the $\g^\alpha \g_5$ part of $\cO^\G_{5,8\perp}$ and the $\g^\alpha$ part of $\cP^\G_{5,8\perp}(\w)$ vanish in the $B$ expectation value, and can thus be dropped. The same is true for the parity even and odd parts of the $\cR^\G_{4,10\perp}(\w)$, which only come with $\G = \g^\alpha P_L$. From Eqs.~\eqref{lc-ops_btous} we define
\begin{equation}
\label{lc-ops_alpha}
\begin{split}
\cO_{5,8\perp}^\alpha(\w) &= \eta_{\perp\mu\nu} \cO_{5,8}^{\alpha\mu\nu}(\w)
,\\
\cP_{5,8\perp}^\alpha(\w) &= \img\eps_{\perp\mu\nu} \cP_{5,8}^{\alpha\mu\nu}(\w)
,\\
\cR_{4\perp}^{\alpha\mu}(\w) &= \eta_{\perp\la}^\mu \cO_3^{\alpha\la}(\w) + \img\eps_{\perp\la}^\mu \cP_4^{\alpha\la}(\w)
,\\
\cR_{10\perp}^{\alpha\mu}(\w) &= \bn_\la(\eta_{\perp\ka}^\mu \cO_9^{\alpha\la\ka}(\w) + \img \eps_{\perp\ka}^\mu \cP_{10}^{\alpha\la\ka}(\w))
.\end{split}
\end{equation}
The sign of the parity-odd parts in $\cR_{4,10\perp}^\G(\w)$ has changed due the minus sign from $P_L$.

We parametrize $T^{\alpha\beta}$ in analogy to the hadronic tensor in Eq.~\eqref{W_param} as
\begin{subequations}
\label{lc-OPE_f}
\begin{equation}
T^{\alpha\beta}
= \int\df\w \delta(P_+ - \w)\biggl(- \frac{1}{2}(\eta + \img\eps)_\perp^{\alpha\beta} t_1
- \frac{1}{2} (\eta - \img \eps)_\perp^{\alpha\beta} t_2
- n^{(\alpha} \bn^{\beta)} t_3 + n^\alpha n^\beta t_4 + \bn^\alpha \bn^\beta t_5 \biggr)
,\end{equation}
where the structure functions $t_i$ are scalar functions of $\w$ and $P_-$. They can be read off from Eq.~\eqref{lc-OPE_Amu} using Eqs.~\eqref{dirac_btoq} and \eqref{dirac_btos}. For $B\to X_u \ell \bar{\nu}_\ell$ we obtain
\begin{equation}
\label{ti_u}
\begin{aligned}
t^u_1 &= \frac{1}{2} n_\alpha \biggl( \cO_0^\alpha(\w)
- \frac{1}{P_-} (\cO + \cP)_{5\perp}^\alpha(\w) \biggr)
\q &
t^u_2 &= - \frac{1}{2P_-^2} \bn_\alpha (\cO - \cP)_{8\perp}^\alpha(\w)
,\\
t^u_4 &= \frac{1}{4} \bn_\alpha \biggl( \cO_0^\alpha(\w)
- \frac{1}{P_-} (\cO + \cP)_{5\perp}^\alpha(\w) \biggr)
\q &
t^u_5 &= - \frac{1}{4P_-^2} n_\alpha (\cO - \cP)_{8\perp}^\alpha(\w)
,\\
t^u_3 &= \frac{1}{2P_-} \eta_{\perp\alpha\mu} \biggl(\cR_{4\perp}^{\alpha\mu}(\w)
- \frac{1}{P_-} \cR_{10\perp}^{\alpha\mu}(\w) \biggr)
.\end{aligned}
\end{equation}
The $P_L$ gives an additional factor $1/2$ for the $\cO^\alpha_i(\w)$ and $\cR^\alpha_{4,10}(\w)$, and $-1/2$ for $\cP^\alpha_{5,8}(\w)$ compared to Eq.~\eqref{lc-OPE_Amu}. For $B\to X_s \g$ only $t^s_1$ is nonzero, while the $t^s_{2-5}$ vanish identically,%
\begin{equation}
\label{ti_s}
t^s_1 = 2 \bn_\alpha \biggl( \cO_0^\alpha(\w) - \frac{1}{P_-} (\cO - \cP)_{5\perp}^\alpha(\w) \biggr)
\q
t^s_{2-5} = 0
.\end{equation}
\end{subequations}

Note that the only operator structures for $f = s$ are $\cO_{0,5}(\w)$. In particular, as in Ref.~\cite{Mannel:2004as}, $\cO_{3,4}(\w)$ do not appear in the QCD light-cone OPE. In both cases they arise only if the QCD light-cone operators are expanded into HQET ones.

Finally, we consider the four-quark contribution $T_{4q}^{\alpha\beta}$. It contains the same Dirac structure $\G = \G^\alpha \g^\mu \G^\beta$. Therefore, from Eq.~\eqref{lc-ops_4q} we define
\begin{equation}
\label{cQi_def}
\begin{split}
\cQ_1^{f\alpha}(\w) &= \bar{b}^i \cQ^{fij}(\w) \g^\alpha b^j
,\\
\cQ_2^{f\alpha}(\w) &= -\bar{b}^i \cQ^{fij}(\w) \g^\alpha \g_5 b^j
.\end{split}
\end{equation}
We included a minus sign in the second definition, because parity only allows the axial part of the left-handed light-quark bilinear to contribute. The four-quark contributions to the $B\to X_u \ell \bar{\nu}_\ell$ structure functions are
\begin{subequations}
\label{t4qi}
\begin{equation}
\begin{aligned}
t^u_{4q,1} &= -\frac{1}{2P_-} n_\alpha (\cQ_1^u + \cQ_2^u)^{\alpha}(\w)
\q &
t^u_{4q,4} &= -\frac{1}{4P_-} \bn_\alpha (\cQ_1^u + \cQ_2^u)^{\alpha}(\w)
,\end{aligned}
\end{equation}
and for $B\to X_s \g$
\begin{equation}
\begin{aligned}
t^s_{4q,1} &= - \frac{2}{P_-} \bn_\alpha (\cQ_1^s - \cQ_2^s)^{\alpha}(\w)
.\end{aligned}
\end{equation}
\end{subequations}
Eqs.~\eqref{lc-OPE_f} and \eqref{t4qi} provide the light-cone OPE of $T^{\alpha\beta}$ to $\ord{\eps^2}$ for $B\to X_u \ell \bar{\nu}_\ell$ and $B \to X_s \g$.

As a nontrivial check we tested our results against the known local expressions by plugging the shape-function parametrizations of the operators into Eqs.~\eqref{lc-OPE_f} and employing their moment expansions. This reproduces the full local result to $\ord{\La^2}$ \cite{manoharwise,Falk:1994dh}. We also checked all local $\ord{\La^3}$ contributions that should be fully contained in our results with the expressions in Ref.~\cite{Gremm:1997df}. We find agreement for the form factors $T^{(3)}_{1-4}$ in the notation of Ref.~\cite{Gremm:1997df}. Concerning $T^{(3)}_5$, we actually disagree with Ref.~\cite{Gremm:1997df}. We believe the contribution proportional to $1/\D_0^3$, corresponding to a subleading twist term, should read
\begin{equation*}
-\frac{2(\rho_1 + 3\rho_2)}{3m_b \D_0^3} (2m_b + q\cdot v)(m_b - q\cdot v) + \frac{2\rho_2 m_b}{\D_0^3}
.\end{equation*}
We explicitly verified this by directly computing this term in the local OPE. Ref.~\cite{Gremm:1997df} misses the $2m_b$ in the first term. This might have been overlooked so far because $T_5^{(3)}$ is not needed in the decay rates for massless leptons, only for $\tau$ leptons.

\section{Differential Decay Distributions}
\label{sec:decayrates}

We will now use our results from the previous section to derive expressions for various differential decay distributions. The decay rates in this section can be used over the entire phase space to study arbitrary cuts on kinematic variables. They are valid to $\ord{\La}$ for small hadronic masses $s_H\sim\ord{\LQCD M_B}$ and to $\ord{\La^2}$ for large hadronic masses $s_H \sim \ord{M_B^2}$. For practical purposes this holds provided all shape functions are modelled with correct moments up to $\ord{\La^2}$. In the resonance region, the expansion necessarily breaks down, and one has to integrate the rates over a sufficiently large region to trust the results.

\subsection{Hadronic Tensor}
\label{subsec:hadronictensor}

It is straightforward to take the $B$ expectation value of Eqs.~\eqref{lc-OPE_f} and \eqref{t4qi} and use the parametrizations in the appendix to express the hadronic tensor in terms of the full set of shape functions appearing to $\ord{\eps^2}$ at the operator level. However, this includes many higher order corrections beyond subleading twist and second local order, as we saw in Sec.~\ref{subsec:shapefunctions}. For phenomenological purposes it is more desirable to reduce the number of shape functions as much as possible.

Since our expansion is accurate to subleading twist and second local order, we can neglect all shape functions of twist $\ord{\La^3}$. In addition, it suffices to keep only those subsubleading shape functions which have moments at $\ord{\La^2}$. The final parametrizations of the operators appearing in the OPE including these simplifications are given in Eqs.~\eqref{Oi_param}.

Writing the structure functions $W_i$ in Eq.~\eqref{W_param} as
\begin{subequations}
\begin{equation}
W_i(P_+,P_-) = \int \df \w \delta(P_+ - \w ) w_i(\w,P_-)
,\end{equation}
and using Eqs.~\eqref{Oi_param} to take the $B$ expectation value of the $t_i^u$ in Eqs.~\eqref{ti_u}, the hadronic tensor for $B\to X_u \ell \bar{\nu}_\ell$ becomes
\begin{align}
\label{wi_u}
\nonumber
w_1^u &= \frac{1}{2} \biggl(F_0 - K_0 - \frac{1}{P_-} (G_5 - H_5) \biggr)(\w)
\q
w_4^u = \frac{1}{4} \biggl(F_0 + K_0 - \frac{1}{P_-} (G_5 + H_5) \biggr)(\w)
,\\
\nonumber
w_2^u &= \frac{1}{2P_-^2} (\w - \la)\bigl[2 (\w - \la) F_0 + R_4\bigr](\w)
\q
w_5^u = \frac{1}{4P_-^2} (\w - \la) \bigl[2 (\w - \la) F_0 - R_4 \bigr](\w)
,\\
w_3^u &= \frac{1}{2P_-} \biggl(R_4 - 2(\w - \la)K_0
- \frac{1}{P_-} \bigl[R_{10} - \la R_4 + 2\la(\w - \la) K_0\bigr] \biggr)(\w)
.\end{align}
The $w_i^u$ are regarded as functions of $\w$ and $P_-$.
For $B \to X_s \g$ we set $P_- = M_B$, and Eq.~\eqref{ti_s} yields
\begin{equation}
\label{wi_s}
w_1^s(\w)
= 2\biggl( F_0 + K_0 - \frac{1}{M_B} (G_5 - H_5) \biggr)(\w)
\q w_{2-5}^s(\w) = 0
.\end{equation}
\end{subequations}
Expanding all shape functions into HQET shape functions to subleading twist, the $w_{2,5}^u$ vanish and the $w_{1,3,5}^u$ and $w_1^s$ reproduce the expressions for the hadronic tensor in Refs.~\cite{Bosch:2004cb,Beneke:2004in}. The additional $\ord{\La^2}$ contributions have not been computed before.

Considering the four-quark operators, we define [see Eqs.~\eqref{GH4q_def}]
\begin{equation}
\begin{split}
\vevB{\cQ_1^{f\alpha}(\w)} = G^f_1(\w) v^\alpha + \dotsb
\q
\vevB{\cQ_2^{f\alpha}(\w)} = H^f_2(\w) (n - v)^\alpha + \dotsb
\end{split}
\end{equation}
The four-quark shape functions are different for charged and neutral $B$-mesons and $f = u,s$. To avoid having to distinguish between these cases we will not consider them explicitly in the following. If desired, they are incorporated by replacing $G_5(\w) \to (G_5 + G^f_1)(\w)$, $H_5(\w) \to (H_5 + H^f_2)(\w)$ in all expressions.

\subsection{Shape-Function Models}
\label{subsec:sf-models}

To illustrate our results in the next subsection we employ three models for the shape functions, based on the two model functions
\begin{equation}
\begin{split}
F_\mathrm{mod1}(\w) &=
c a \frac{\pi}{2 b^2} \w \exp\biggl[-\frac{\pi}{4} \Bigr(\frac{\w}{b}\Bigl)^2 \biggr] \theta(\w)
+ c(1 - a)\frac{32}{\pi^2 b^3} \w^2 \exp\biggl[-\frac{4}{\pi} \Bigr(\frac{\w}{b}\Bigl)^2 \biggr]\theta(\w)
,\\
F_\mathrm{mod2}(\w) &=
c \frac{a^{ab}}{\Gamma(ab)} \w^{ab - 1} e^{- a \w} \theta(\w)
.\end{split}
\end{equation}
The first function is an extension of the one given in Ref.~\cite{Mannel:1994pm} and is used in the first model. The second one is taken from Ref.~\cite{Leibovich:2002ys} and is used for the second and third model. The moment expansions of the shape functions are given in Eqs.~\eqref{sf0_exp}, \eqref{sf1_exp}, and \eqref{sf2_exp}. Note that the moments are taken with respect to $\w - \la$.

\begin{figure}%
\centering
\includegraphics[width=0.49\columnwidth]{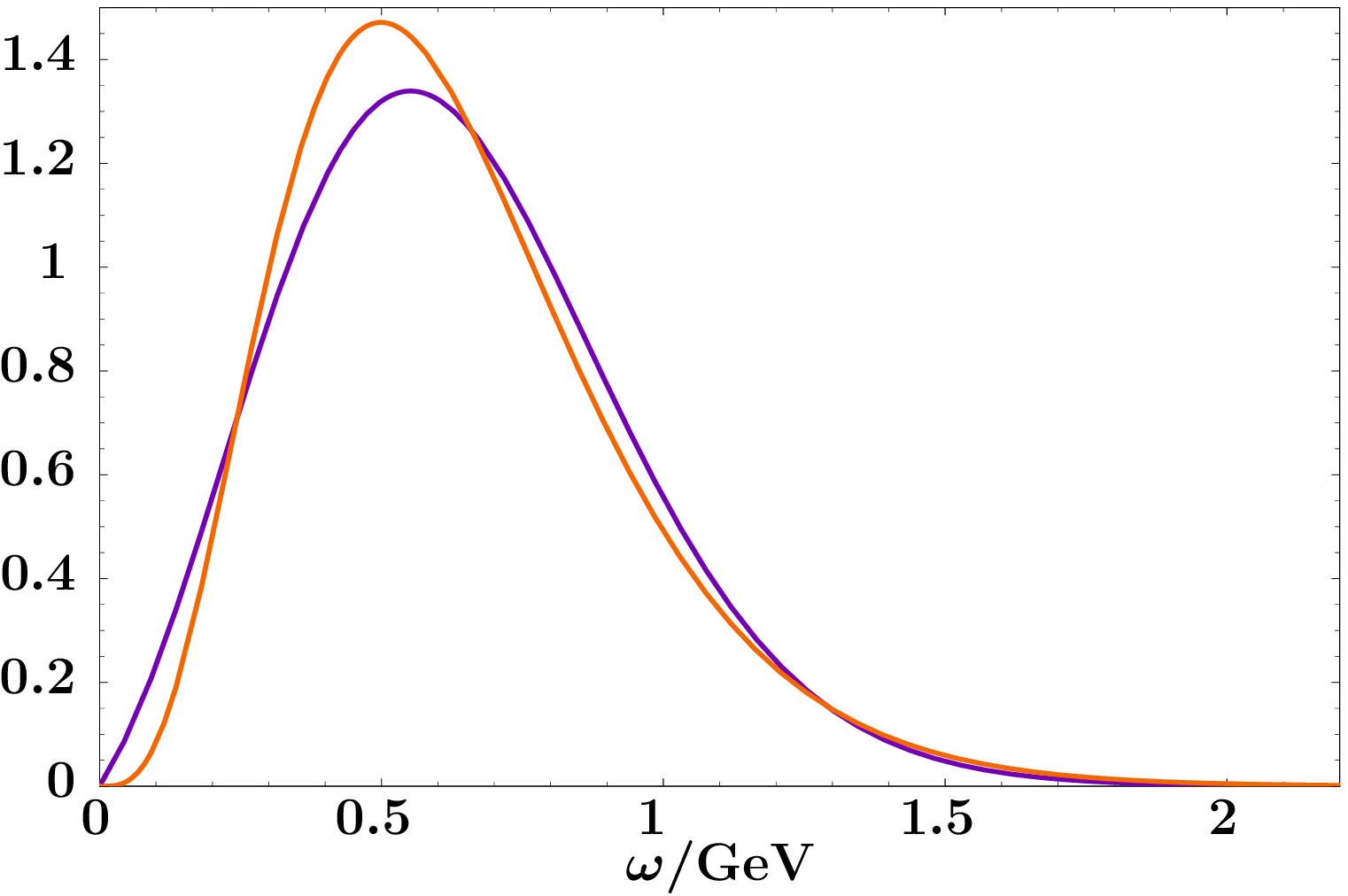}\hfill
\includegraphics[width=0.495\columnwidth]{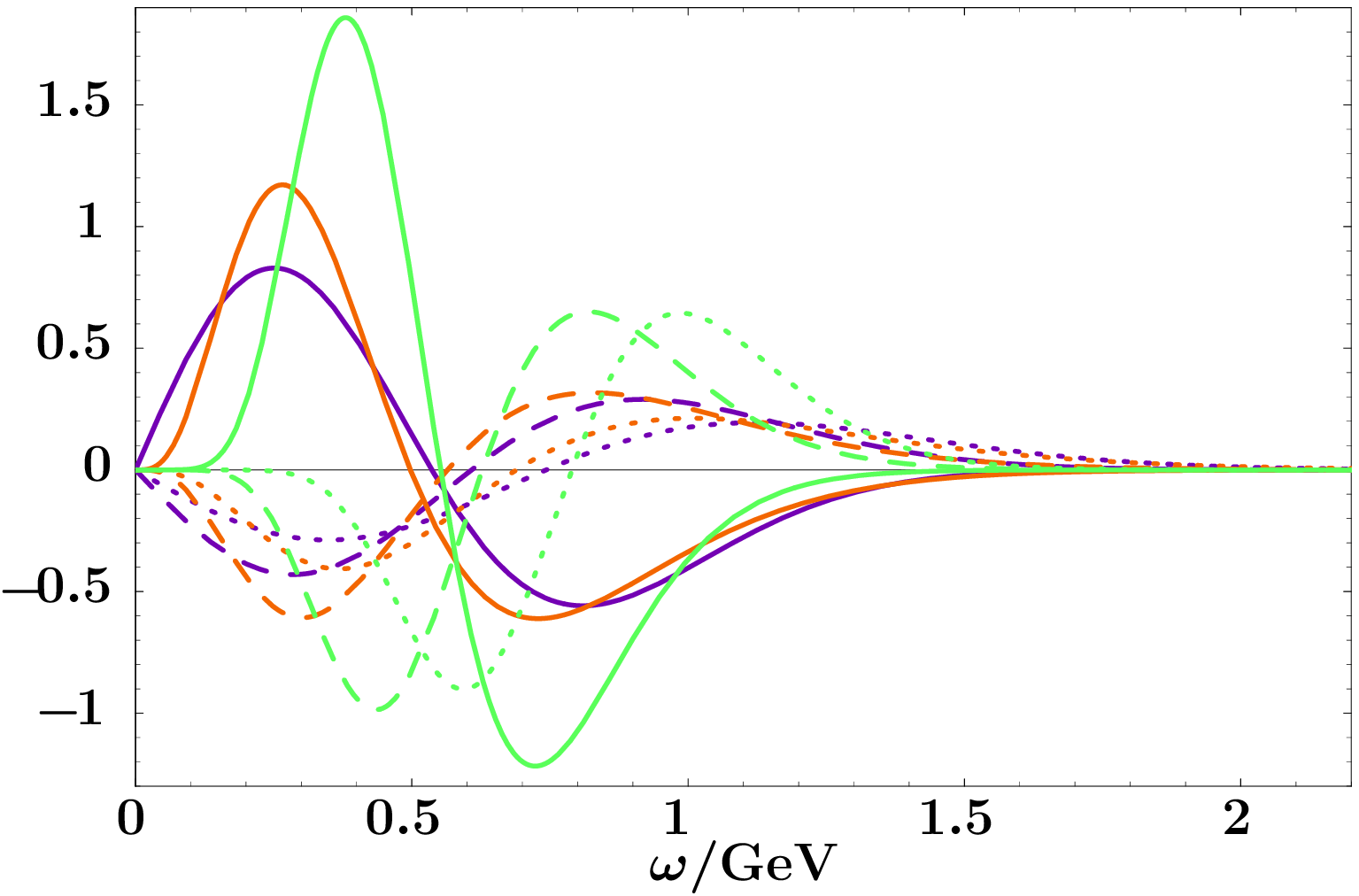}%
\caption{\label{fig:sfs}(color online) Shape functions. Model 1, 2, and 3 are dark (violet), medium (orange), and light (green). The left plot shows $F_0(\w)$. The right one shows $G_5(\w)$ (solid), $H_5(\w)$ (dashed), and $R_4(\w)$ (dotted).}
\end{figure}

The leading shape function $F_0(\w)$ is modelled from $F_\mathrm{mod1,2}(\w)$ by adjusting the parameters $a$, $b$, $c$ to produce the correct zeroth, first, and second moment. For this purpose we set $\tau_1 = \tau_2 = 0$ and use $\la_1 = -0.27 \GeV^2$, $\la_2 = 0.12 \GeV^2$, and $m_b = 4.65 \GeV$, corresponding to $\la = M_B - m_b = 0.63 \GeV$, as our default values. They are inspired by the values obtained in Refs.~\cite{Aubert:2004aw} and \cite{Bauer:2004ve}. The third moment of $F_0(\w)$ predicts $\rho_1 = 0.055\GeV^3$ in model 1 and $\rho_1 = 0.084 \GeV^3$ in model 2. The left plot in Fig.~\ref{fig:sfs} shows $F_0(\w)$ in model 1 and 2. $F_0(\w)$ is the same in model 2 and 3.

The zeroth moments of the subleading shape functions $G_5(\w)$, $H_5(\w)$, and $R_4(\w)$ vanish to all orders in $1/m_b$, because the functions arise from operators containing the derivative $\delta_{12}'(\w)$\footnote{For $R_4(\w)$ this can be seen, for instance, from Eq.~\eqref{O67_RPI}.}. Therefore, it seems natural to model them by the derivatives $F'_\mathrm{mod1,2}(\w)$. In the first model we set $a = 0$, to ensure that the functions vanish at $\w = 0$, and adjust $b$ and $c$ to reproduce the correct first and second moments. For the second moment of $R_4(\w)$ we use $\rho_2 = -0.05 \GeV^3$. The second moments of $G_5(\w)$ and $H_5(\w)$ vanish at $\ord{\LQCD^3}$. We set them to $(2/3)(0.5 \GeV)^4/m_b$ and $(0.5 \GeV)^4/m_b$, respectively.

In the second model we adjust $a$, $b$, $c$ such that the functions have the same first, second, and third moment as in model 1. Their shape is actually quite sensitive to the value of the third moment, which is of $\ord{\LQCD^4}$. In our third model we adjust the third moments of $G_5(\w)$, $H_5(\w)$, and $R_4(\w)$ to one half their values in the second model. Therefore, model 2 and 3 differ only in the subleading shape functions' third and higher moments, which are $\ord{\LQCD^4}$ and higher.
The three functions are shown for each model in Fig.~\ref{fig:sfs}. Throughout the paper we plot model 1, 2, and 3 in dark (violet), medium (orange), and light (green), respectively. Notice that $G_5(\w)$ and $H_5(\w)$ behave roughly oppositely, which means the combination $(G_5 - H_5)(\w)$ is rather large, while $(G_5 + H_5)(\w)$ is small, as one would expect from their first moments.

For modelling purposes we set [see Eqs.~\eqref{FK0_HQETexp} and \eqref{R10_rel}] $K_0(\w) = [(\w - \la)F_0 - R_4](\w)/m_b$, where in this case the first moment of $F_0(\w)$ is set to zero, and $R_{10}(\w) = -(\w - \la) H_5(\w)$.

Note that in our approach we regard the local parameters, e.g., $\la_1$ and $m_b$, as known input parameters. To a first approximation, the error due to the uncertainty in their values should be treated separately from the error due to the unknown form of the shape functions, i.e., their unknown higher order moments. For example, the total rate is very sensitive to $m_b$, but basically shape-function independent. Therefore, we do not vary $\la_1$ and $m_b$ over a large range to produce different shapes for the functions, because this exaggerates the uncertainty for any quantity which is dominated by the local expansion or the first few moments of the shape functions. Instead, to assess the true sensitivity of a given quantity to the specific form of the shape functions, we look at its variation between different shape-function models, while keeping the zeroth, first, and second moments of all shape functions fixed. To do so, we use the three models described above. In a more extensive treatment one should include more model functions and also scan over generic values for the higher order moments. Of course, it is not possible to completely disentangle the two uncertainties, for example, the shape-function dependence itself might be different to some extent for different values of $\la_1$ or $m_b$.

Our main interest is in if and by how much a given quantity is influenced by shape functions effects and how sensitive it is to the specific form or higher moments of the shape functions. We emphasize that the variations we will see in the plots do not represent total uncertainties. They give a measure of the shape-function dependence alone, and are only one, in some cases small, part in the total uncertainty. To estimate the latter, one has to vary the local parameters as well. To illustrate this, we will vary $m_b$ in the range $\pm 50\MeV$ in a few cases.

\subsection{Decay Spectra}

We are now ready to assemble the expressions for various decay spectra. Since there are no phase space restrictions on our results, we can easily switch to any desired set of kinematic variables. We use the notation
\begin{equation*}
\D = M_B - 2E_\ell \q M_\w = M_B - \w \q \D_\w = \D - \w = M_\w - 2E_\ell
.\end{equation*}
The spectra in all plots are normalized to the partonic rate $\G_p^u = \G_0^u m_b^5$ or $\G_p^s = \G_0^s m_b^3$, respectively.

\subsubsection{Photon Energy Spectrum}

\begin{figure}%
\centering
\includegraphics[width=0.495\columnwidth]{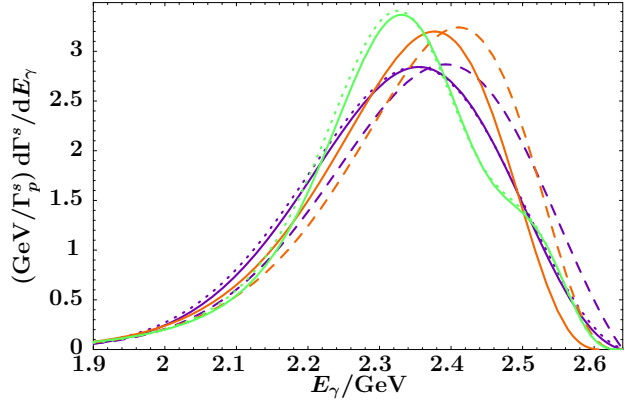}
\caption{\label{fig:dGEg}(color online) $E_\g$ spectrum. The solid lines show the full result, the dashed lines include only the contributions proportional to $F_0(\w)$, and the dotted ones show the subleading twist result.}
\end{figure}

We start by writing down the $B \to X_s \g$ photon energy spectrum. From Eqs.~\eqref{dGs} and \eqref{wi_s} we have
\begin{equation}
\label{dG_Eg}
\frac{\df\G^s}{\df E_\g}
= 16\G^s_0 E_\g^3 \theta(E_\g) \biggl(F_0 + K_0 - \frac{1}{M_B}(G_5 - H_5)\biggr)(M_B - 2E_\g)
.\end{equation}
We do not expand the overall $E_\g^3$, because there is no need to do so. Interestingly, Eq.~\eqref{dG_Eg} does not contain any subsubleading shape functions. Instead, $F_0(\w)$ and $K_0(\w)$ already contain all local $\ord{\La^2}$ pieces that are of subsubleading twist. The photon energy spectrum, normalized to the partonic rate $\G^s_p = \G^s_0 m_b^3$, is shown in Fig.~\ref{fig:dGEg}. Since at tree level, its support lies entirely in the shape-function region, our expansion yields only a small correction of $\ord{\La^2}$ to the subleading twist result.

\subsubsection{Spectra in $P_+$ and $P_-$}

With the $w_i^u = w_i^u(\w,P_-)$ given in Eqs.~\eqref{wi_u}, the triple differential decay rate, Eq.~\eqref{dG3}, becomes
\begin{equation}
\label{dG3_wq}
\begin{split}
&\frac{\df^3\G^u}{\df E_\ell \df P_+ \df P_-}
= 48 \G^u_0 \int\df\w \frac{\delta(P_+ - \w)}{P_- - \w} \theta(\D_\w)\theta(P_- - \D)\theta(M_B - P_-)
\\ &  \quad \times
\biggl\{
M_\w(M_B - P_-) \Bigl[(\D - P_-)^2 w^u_1(\w,P_-) + \D_\w^2 w^u_2(\w,P_-) \Bigr]
- 2\D_\w(\D - P_-)
\\ &  \quad \times
\Bigl[M_\w(M_B - P_-) w^u_3(\w,P_-) + M_\w^2 w^u_4(\w,P_-)
+ (M_B - P_-)^2 w^u_5(\w,P_-) \Bigr] \biggr\}
.\end{split}
\end{equation}
The double differential rate, Eq.~\eqref{dG2}, reads
\begin{equation}
\label{dG2_wq}
\begin{split}
&\frac{\df^2\G^u}{\df P_+ \df P_-}
= 8 \G^u_0 \int\df\w \delta(P_+ - \w) \theta(P_- - \w)\theta(M_B - P_-) (P_- - \w)^2
\\ & \quad \times
\Bigl[ M_\w(M_B - P_-)(w^u_1 + w^u_2 + w^u_3)(\w,P_-) + M_\w^2 w^u_4(\w,P_-)
+ (M_B - P_-)^2 w^u_5(\w,P_-) \Bigr]
.\end{split}
\raisetag{7ex}
\end{equation}
To use $q_\pm$ one has to replace $P_\pm = M_B - q_\pm$.

The spectrum in the variable $P_+$ is interesting, since it can be directly compared to the photon energy spectrum in $B\to X_s\gamma$ to determine the ratio $\abs{V_{ub}}/\abs{V_{ts}}$ \cite{Mannel:1999gs}.
Integrating Eq.~\eqref{dG2_wq} over $P_-$, we obtain
\begin{equation}
\label{dG_Ppu}
\begin{split}
\frac{\df\G^u}{\df P_+}
&= \G_0^u \int\df\w \delta(P_+ - \w) \theta(M_B - \w) \biggl\{
M_\w^5 \biggl( F_0(\w) + \frac{1}{3} K_0(\w) \biggl)
\\ & \quad
+\frac{2M_\w}{3} \Bigl(M_\w (-M_\w^2 + 3\w(M_B + \w)) + 6M_B \w^2 \ln(\w/M_B) \Bigr)
(G_5 - H_5 - R_4 - L_3)(\w)
\\ & \quad
+ M_\w^2 \Bigl(-M_\w(M_\w - 2\w) + 2\w^2 \ln(\w/M_B) \Bigr)(G_5 + H_5)(\w)
\\ & \quad
-2 M_\w \Bigl(M_\w(M_B + 5\w) + 2\w(2M_B + \w)\ln(\w/M_B) \Bigr)[G_8 - H_8 + R_{10} - \la (R_4 + L_3)](\w)
\\ & \quad
-\frac{2}{3} \Bigl(M_\w (M_\w^2 + 12\w M_B) + 6 M_B\w(M_B + \w)\ln(\w/M_B) \Bigr) (G_8 + H_8)(\w)
\biggr\}
,\end{split}
\raisetag{4ex}
\end{equation}
with [see Eqs.~\eqref{eom1}, \eqref{eom2}]
\begin{equation}
\label{GH8}
L_3(\w) = -2(\w - \la) K_0(\w)
\q
G_8(\w) = -2(\w - \la)^2 F_0(\w)
\q
H_8(\w) = (\w - \la) R_4(\w)
.\end{equation}
Notice that the leading term comes indeed with a power $M_\w^5$, as suggested by the subleading result, confirming the leading order result obtained in Ref.~\cite{Mannel:1999gs}. The integration over $P_+$ to obtain the total rate amounts to dropping the $\delta(P_+ - \w)$ under the integral in Eq.~\eqref{dG_Ppu}.
Expanding Eq.~\eqref{dG_Ppu} to subleading twist reproduces the results in Refs.~\cite{Lee:2004ja,Bosch:2004cb}\footnote{In comparing our result with Ref.~\cite{Lee:2004ja} we set the additional $P_-$ cut employed there to zero.}.

\begin{figure}%
\centering
\includegraphics[width=0.495\columnwidth]{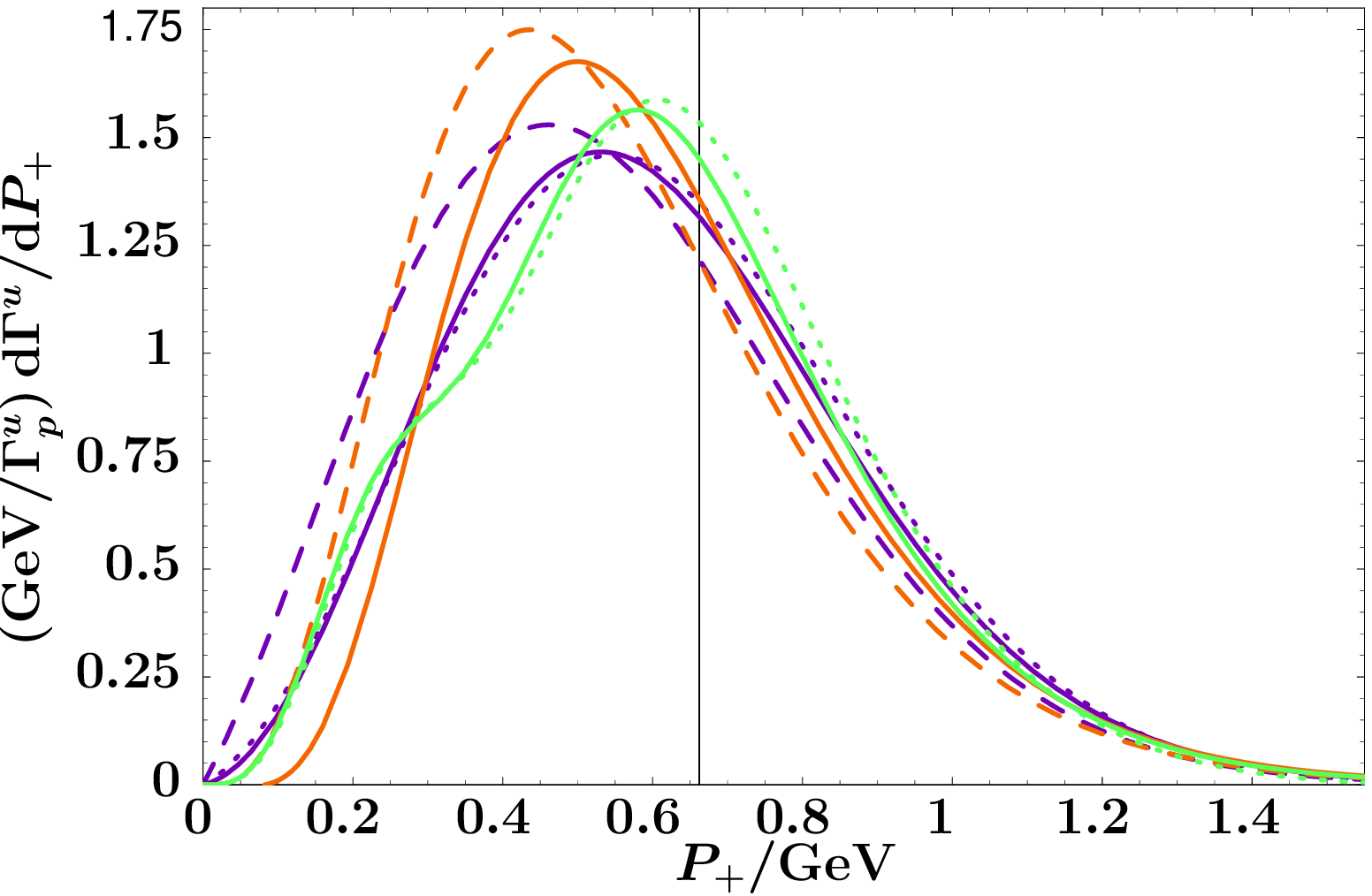}\hfill
\includegraphics[width=0.495\columnwidth]{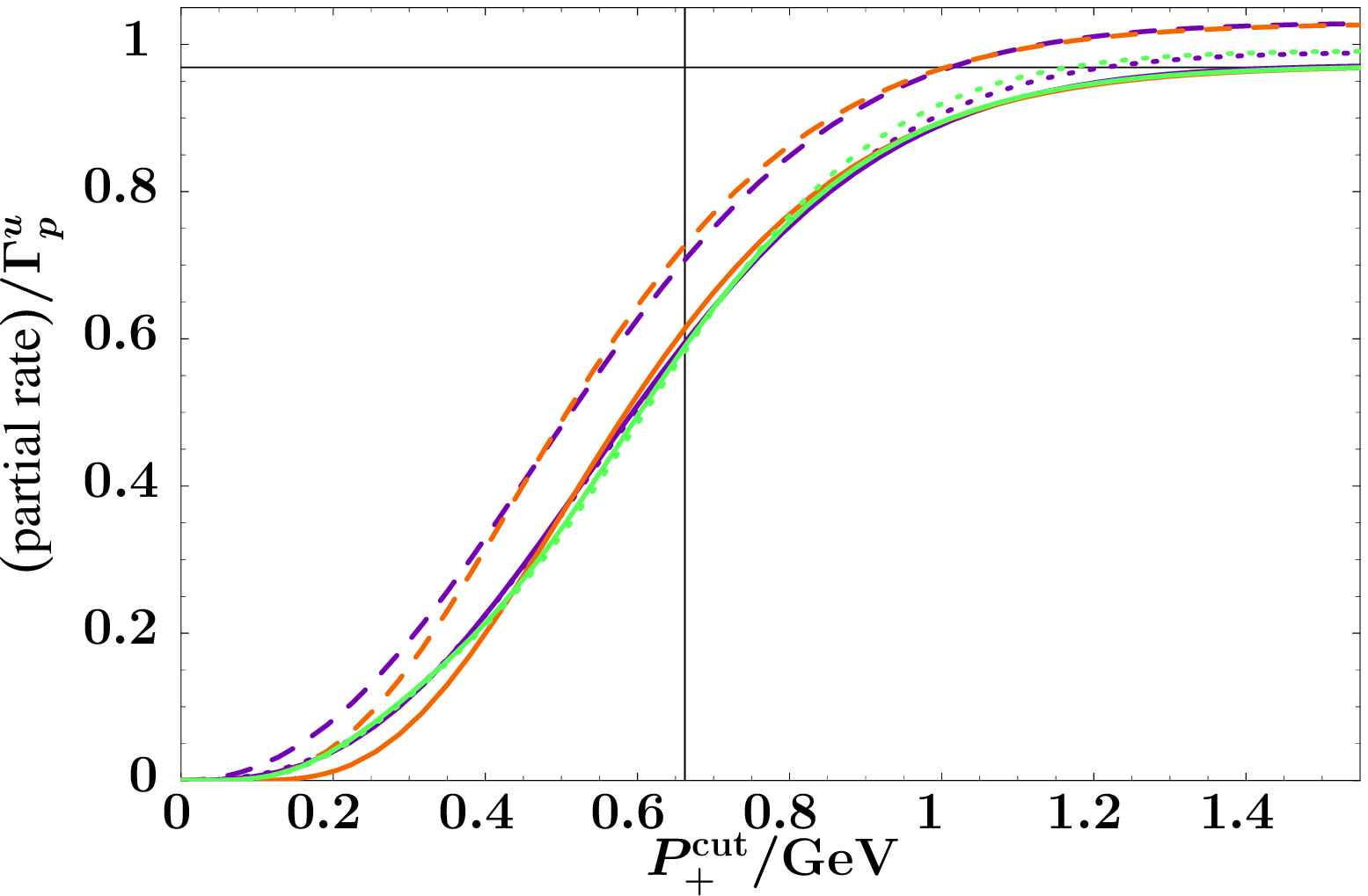}%
\caption{\label{fig:dGP}(color online) $P_+$ spectrum (left) and its partial rate (right). The meaning of the curves is the same as in Fig.~\ref{fig:dGEg}. The vertical line denotes $P_+ = m_D^2/M_B$, the kinematic limit for $b\to c$ transitions. The horizontal line on the right denotes the value of the total rate to $\ord{\La^2}$.}
\end{figure}

The $P_+$ spectrum \eqref{dG_Ppu} normalized to $\G_p^u$ is depicted on the left in Fig.~\ref{fig:dGP}. As for the photon energy spectrum, at tree level it has only support in the shape-function region. The corrections of the full result (solid) to the subleading twist result (dotted) are thus small, although larger then in case of $E_\g$ in Fig.~\ref{fig:dGEg}. The right plot shows the corresponding partial rate, i.e., the spectrum integrated up to $P_+ \leq P_+^\mathrm{cut}$ (still normalized to $\G_p^u)$. Beyond $P_+^\mathrm{cut} = 1.5 \GeV$ the curves stay constant. One can see that independently of the used model our result indeed approaches the value for the total rate including $\ord{\La^2}$ corrections.

\subsubsection{Spectra Containing the Hadronic Invariant Mass}

The hadronic invariant mass $s_H$ is useful for the extraction of $\abs{V_{ub}}$ \cite{Barger:1990tz,Falk:1997gj,Bigi:1997dn}. Using
\begin{equation*}
s_H = P_+ P_- \q q^2 = (M_B - P_+)(M_B - P_-)
\q \df s_H \df q^2 = M_B(P_- - P_+) \df P_+ \df P_-
,\end{equation*}
and defining $s_\w = s_H/\w$ the triple and double differential rates in terms of $s_H$ and $q^2$ are%
\begin{equation}
\label{dG3_sHq2}
\begin{split}
\frac{\df^3\G^u}{\df E_\ell \df s_H \df q^2}
&= 48 \G^u_0 \int\df\w \frac{\delta[q^2 - M_\w(M_B - s_\w)]}{\w(s_\w - \w)}
\theta(\D_\w)\theta(s_\w - \D)\theta(M_B - s_\w)
\\ & \quad \times
\biggl\{ M_\w (M_B - s_\w) \Bigl[(\D - s_\w)^2 w^u_1(\w, s_\w) + \D_\w^2 w^u_2(\w, s_\w) \Bigr]
- 2\D_\w(\D - s_\w)
\\ & \quad \times
 \Bigl[M_\w(M_B - s_\w) w^u_3(\w, s_\w)
+ M_\w^2 w^u_4(\w, s_\w) + (M_B - s_\w)^2 w^u_5(\w, s_\w) \Bigr] \biggr\}
,\end{split}
\end{equation}
and
\begin{equation}
\label{dG2_sHq2}
\begin{split}
&\frac{\df^2\G^u}{\df s_H \df q^2}
= 8 \G^u_0 \int\df\w \delta[q^2 - M_\w(M_B - s_\w)] \theta(s_\w - \w)\theta(M_B - s_\w)
\frac{(s_\w - \w)^2}{\w}
\\ & \quad \times
\Bigl[ M_\w(M_B - s_\w)(w^u_1 + w^u_2 + w^u_3)(\w,s_\w)
+ M_\w^2 w^u_4(\w,s_\w) + (M_B - s_\w)^2 w^u_5(\w,s_\w) \Bigr]
.\end{split}
\end{equation}

Eqs.~\eqref{dG3_sHq2} and \eqref{dG2_sHq2} can easily be integrated to give $\df^2\G^u/\df E_\ell \df s_H$ and $\df\G^u/\df s_H$ by dropping the $\delta$ function. In this case, the phase space limits yield the limits on the $\w$ integration
\begin{equation}
\begin{split}
0 &\leq \frac{s_H}{M_B} \leq \w \leq
\left\{ \begin{aligned}
&s_H/\D &&\text{for} & \sqrt{s_H} &\leq \D \leq M_B
,\\
&\D s_H/s_H \quad &&\text{for} & s_H/M_B &\leq \D \leq \sqrt{s_H} \leq M_B
,\end{aligned} \right.
\\
0 &\leq \frac{s_H}{M_B} \leq \w \leq \sqrt{s_H}
,\end{split}
\end{equation}
for $\df^2\G^u/\df E_\ell \df s_H$ and $\df\G^u/\df s_H$, respectively. For the latter, upon integration over $\w$, the limits on $s_H$ are $0 \leq s_H \leq M_B^2$.

\begin{figure}%
\centering
\includegraphics[width=0.495\columnwidth]{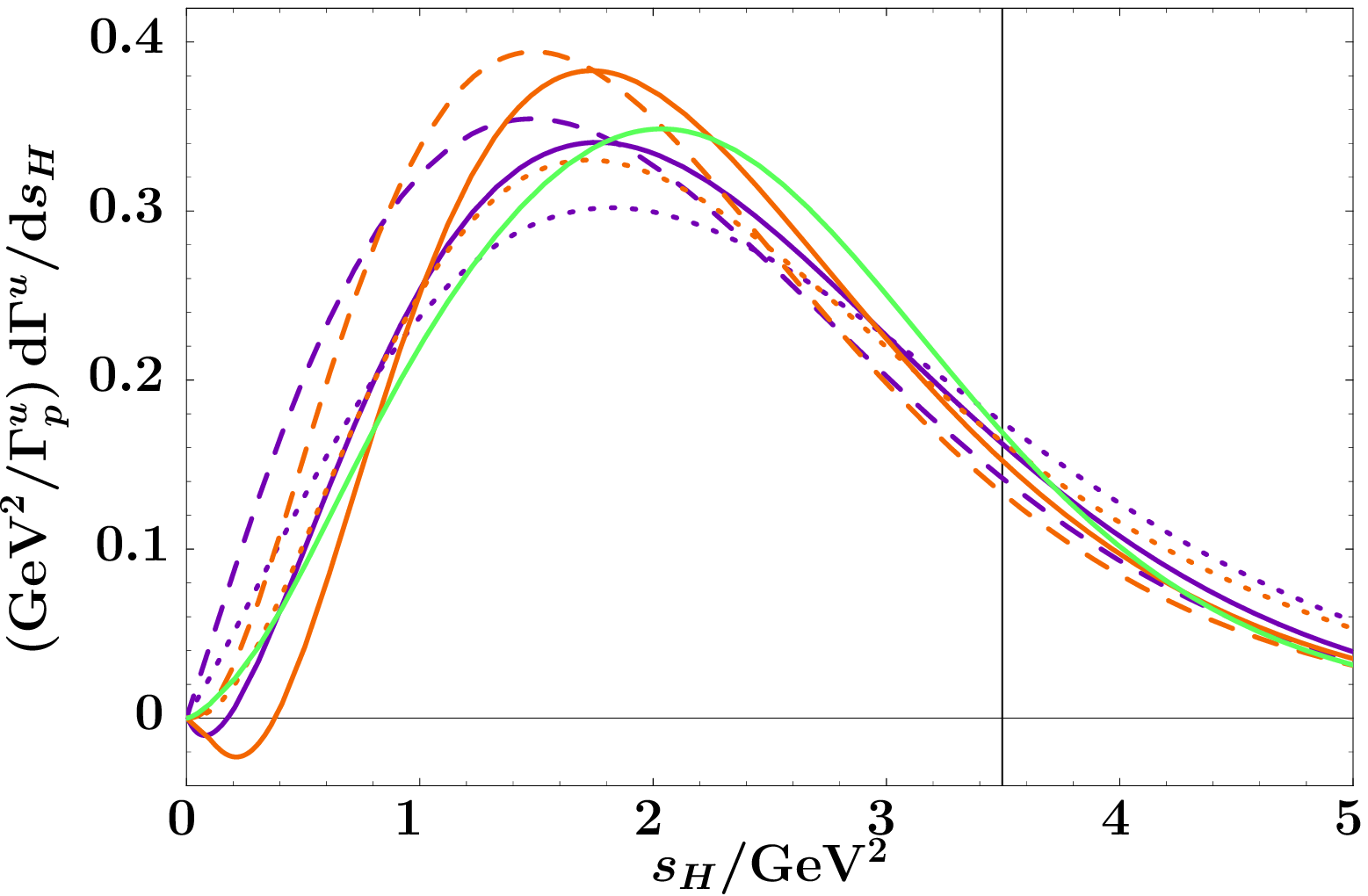}\hfill
\includegraphics[width=0.495\columnwidth]{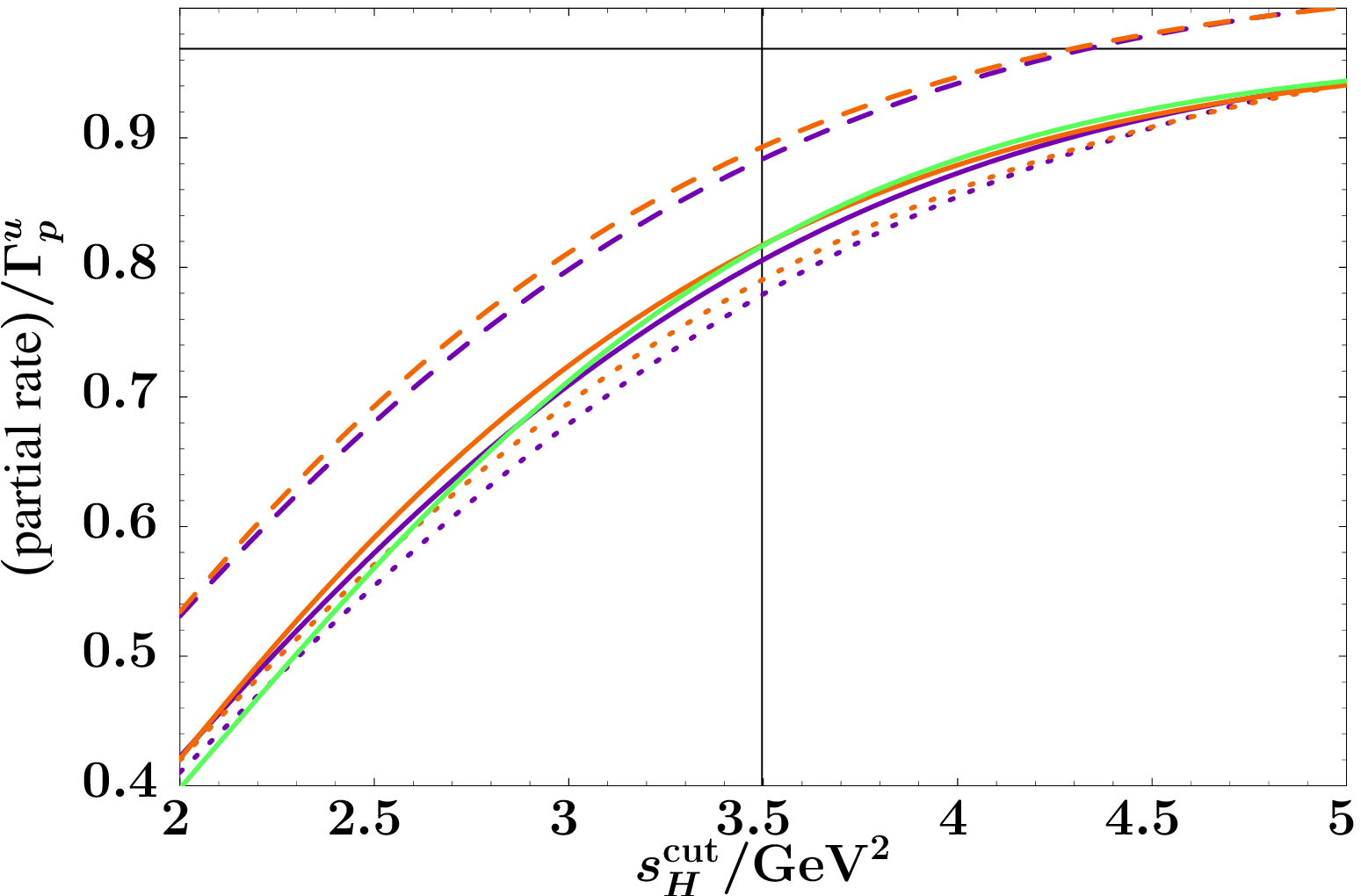}%
\caption{\label{fig:dGsH1}(color online) $s_H$ spectrum (left) and partial rate (right). The solid lines show the full result, the dashed lines only include the contributions from $F_0(\w)$, and the dotted lines correspond to the prescription of Ref.~\cite{DeFazio:1999sv}. The vertical line denotes $s_H = m_D^2$. The horizontal line on the right denotes the value of the total rate to $\ord{\La^2}$ in units of $\G^u_p$.}
\end{figure}
\begin{figure}%
\centering
\includegraphics[width=0.495\columnwidth]{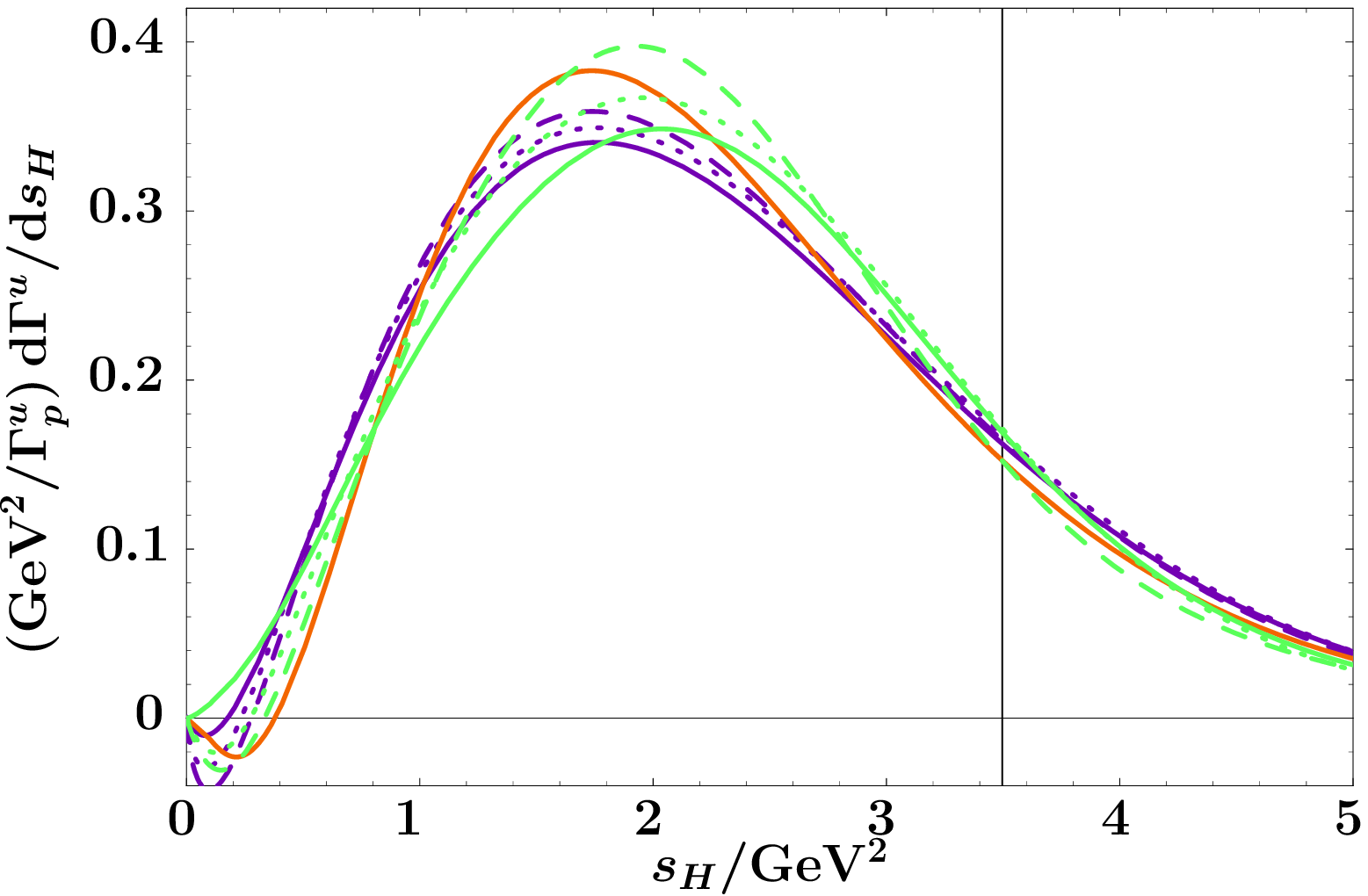}\hfill
\includegraphics[width=0.495\columnwidth]{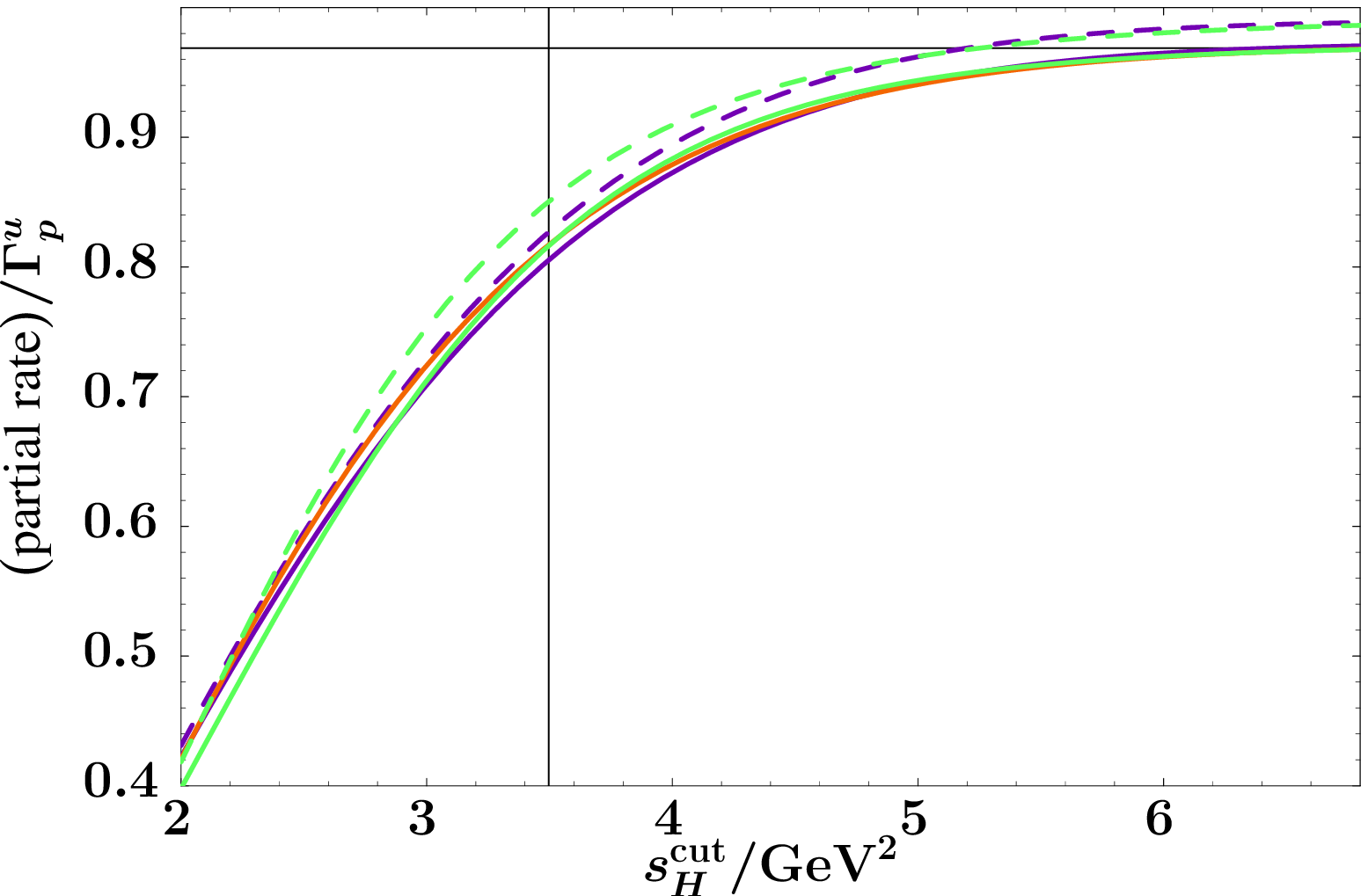}%
\caption{\label{fig:dGsH2}(color online) Comparison of $s_H$ spectrum (left) and partial rate (right) with subleading twist results. The solid lines show our result, the dashed lines the subleading twist result of Ref.~\cite{Burrell:2003cf}, and the dotted lines the result of Ref.~\cite{Bosch:2004cb}. The vertical line denotes $s_H = m_D^2$. The horizontal line on the right denotes the value of the total rate to $\ord{\La^2}$ in units of $\G^u_p$.}
\end{figure}
\begin{figure}%
\centering
\includegraphics[width=0.495\columnwidth]{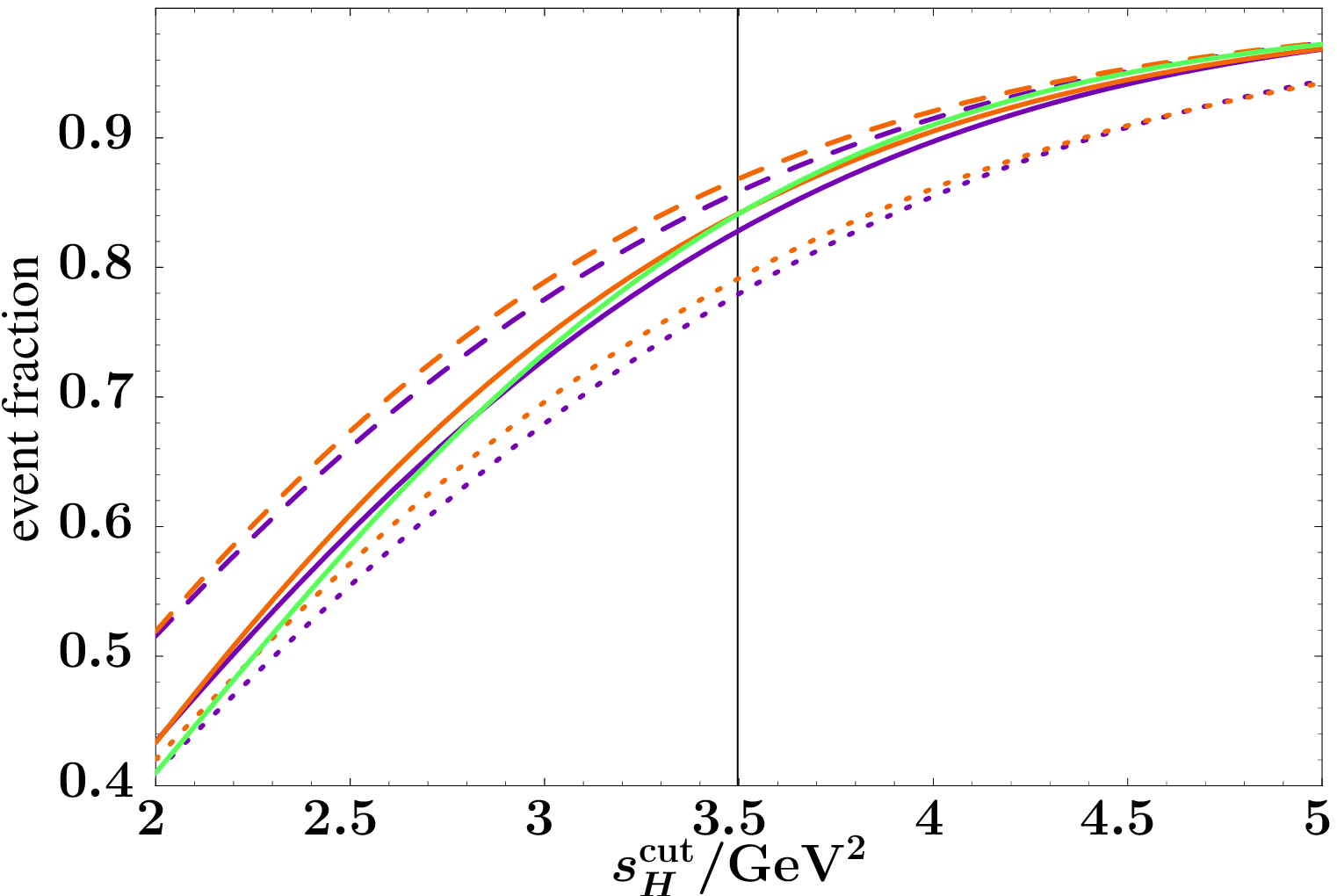}\hfill
\includegraphics[width=0.495\columnwidth]{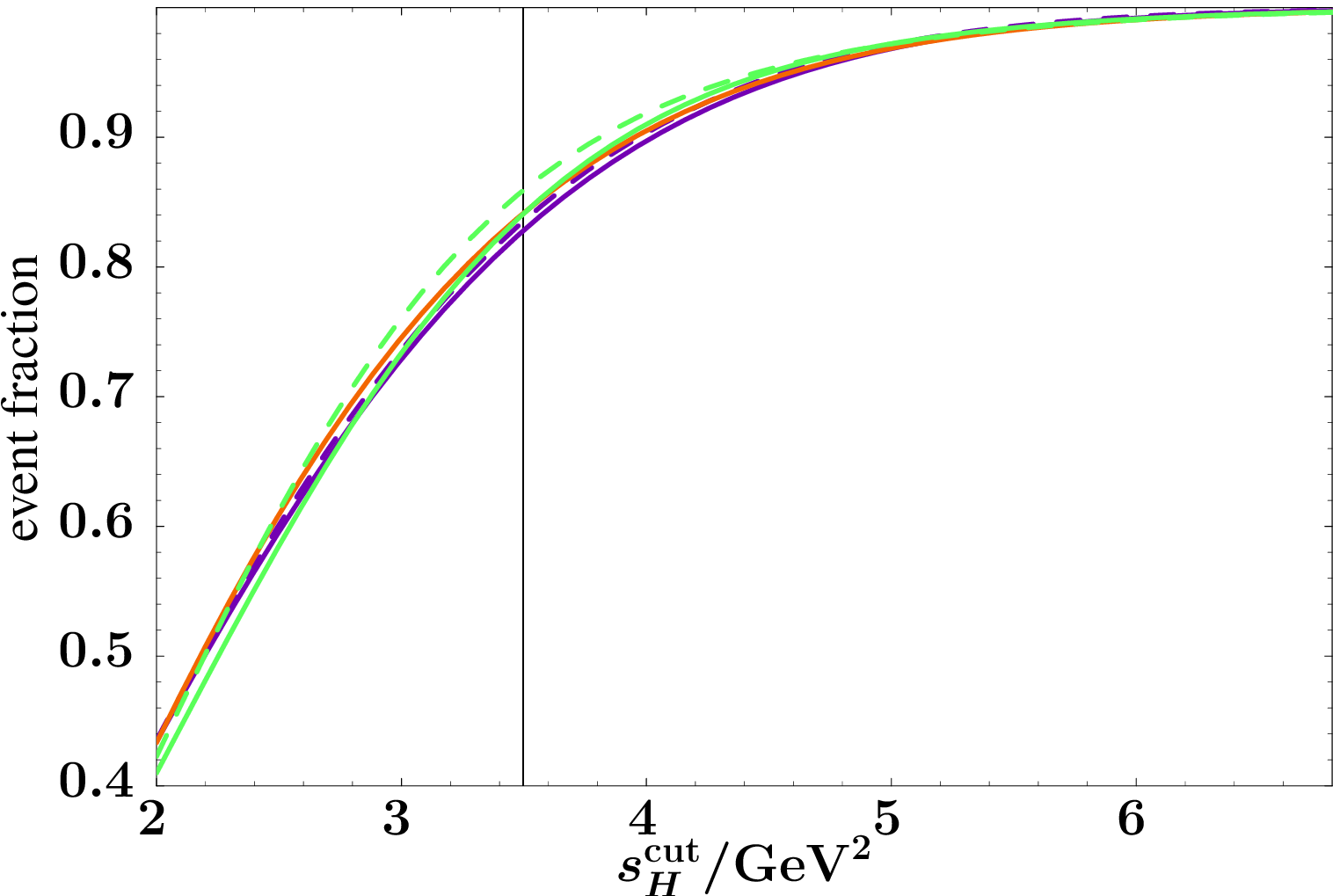}%
\caption{\label{fig:nIsH}(color online) Event fractions for $s_H$ spectrum. The left plot corresponds to the right one of Fig.~\ref{fig:dGsH1} and the right plot to the right one of Fig.~\ref{fig:dGsH2}.}
\end{figure}

The hadronic invariant mass spectrum $\df\G^u/\df s_H$ obtained from Eq.~\eqref{dG2_sHq2} is shown on the left of Fig.~\ref{fig:dGsH1}. The right plot contains the partial rate for an upper cut $s_H \leq s_H^\mathrm{cut}$. The solid lines correspond to the full result. The dashed ones show the result keeping only the contributions from $F_0(\w)$, in which the complete $m_b$ dependence of the partonic spectrum is convoluted with $F_0(\w)$. For comparison, the dotted curves give the result from using $F_0(\w)$ in the prescription of Ref.~\cite{DeFazio:1999sv}\footnote{For consistency this includes only the tree-level results of Ref.~\cite{DeFazio:1999sv}.}, where an overall $m_b^5$ is excluded from the convolution. The full result lies in between the two and neither gives a better approximation than the other. One should also note that the solid medium and light (orange and green) curves only differ in the third and higher moments, which are $\ord{\LQCD^4}$ and higher, of the subleading shape functions. In particular, they share the same medium (orange) dashed and dotted curves.

Expanding the $s_H$ spectrum obtained from Eq.~\eqref{dG2_sHq2} to subleading twist reproduces the result of Ref.~\cite{Burrell:2003cf}. In Fig.~\ref{fig:dGsH2} we compare our result (solid) for the spectrum and the corresponding partial rate with the subleading twist result (dashed). In addition, the dotted lines show the result of Ref.~\cite{Bosch:2004cb}, which keeps certain factors of $M_B$, and hence contains some higher order terms compared to the result of Ref.~\cite{Burrell:2003cf}. The corrections to the subleading twist result from our result are more significant then in the $E_\g$ or $P_+$ spectrum. For the partial rate, in the third model (light, green) they are bigger then the difference between the individual models. Our result also has less sensitivity to the form of the shape functions then the result to subleading twist.

The solid lines in the right plot in Fig.~\ref{fig:dGsH2} approach the horizontal line, which is again a manifestation of the fact that our result contains the total rate to $\ord{\La^2}$. For practical purposes, the partial rates are usually translated into event fractions by normalizing them to the respective predicted total rate. This introduces an additional error, if the total rate is not reproduced correctly. Fig.~\ref{fig:nIsH} shows the event fractions corresponding to the partial rates on the right of Figs.~\ref{fig:dGsH1} and \ref{fig:dGsH2}.

\begin{figure}%
\centering
\includegraphics[width=0.495\columnwidth]{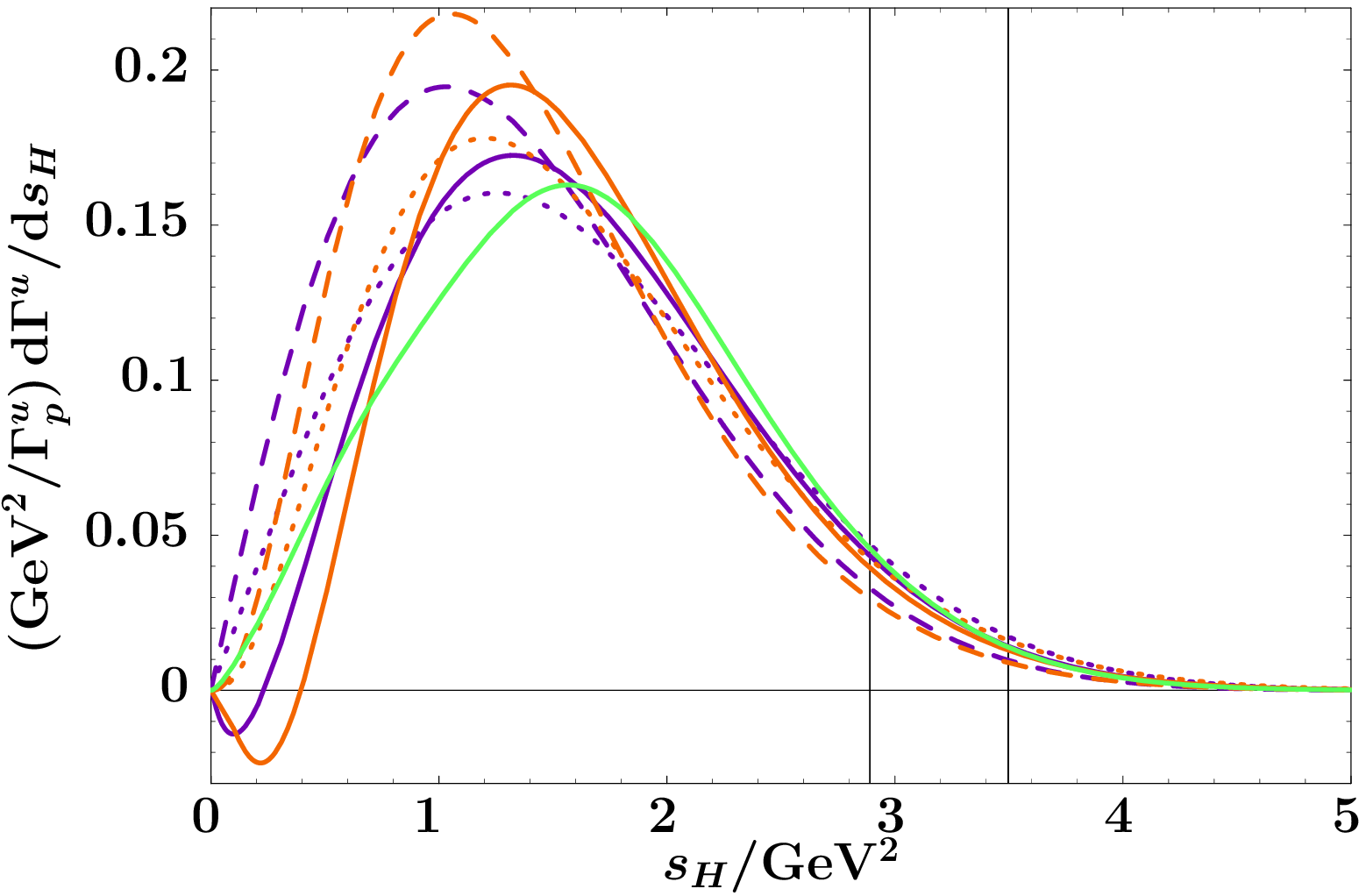}\hfill
\includegraphics[width=0.495\columnwidth]{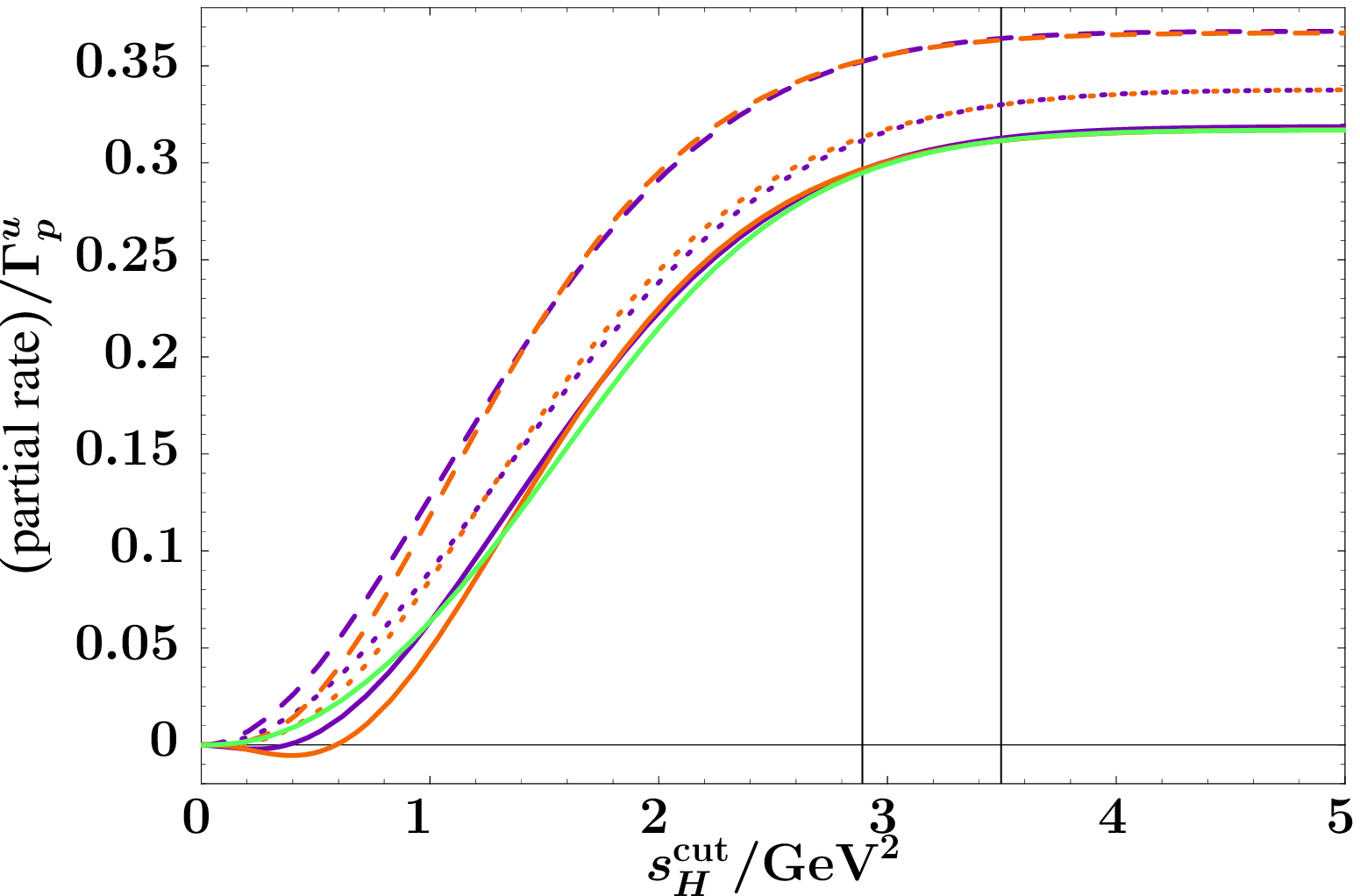}%
\caption{\label{fig:dGsH_q28}(color online) $s_H$ spectrum (left) and partial rate (right) with a cut $q^2 > 8\GeV^2$. The meaning of the curves is the same as in Fig.~\ref{fig:dGsH1}. The vertical lines are $s_H = m_D^2$ and $s_H = (1.7\GeV)^2$.}
\end{figure}

Eq.~\eqref{dG2_sHq2} also allows us to obtain the $s_H$ spectrum with an additional lower cut on $q^2$, as proposed in Ref.~\cite{Bauer:2001rc}. Fig.~\ref{fig:dGsH_q28} shows the spectrum and its partial rate for the cut $q^2 > 8 \GeV^2$ employed by BaBar and Belle \cite{Aubert:2004bq,Abe:2004sc}. As expected, the cut on high $q^2$ significantly reduces the shape-function dependence compared to Fig.~\ref{fig:dGsH1}. This comes at the price of having a much smaller number of contained $b\to u$ events. In Ref.~\cite{Bauer:2001rc} the correction from smearing the local result with the leading shape function is translated to 100\% into an uncertainty on the partial rate. Our results can be used to improve on that, and additionally allow one to include $\ord{\La}$ corrections.

Another possibility is to replace the $q^2$ cut by a cut on the hadronic energy $E_H = (P_+ + P_-)/2$. The rates in terms of $E_H$ are obtained by replacing $\delta[q^2 - M_\w(M_B - s_\w)] \to \delta[E_H - (s_\w + \w)/2]$ in Eqs.~\eqref{dG3_sHq2} and \eqref{dG2_sHq2}. A cut on $E_H < M_B - \sqrt{(q^2)^\mathrm{min}}$ produces the same upper limit on $s_H$ as $q^2 > (q^2)^\mathrm{min}$.

\begin{figure}%
\centering
\includegraphics[width=0.495\columnwidth]{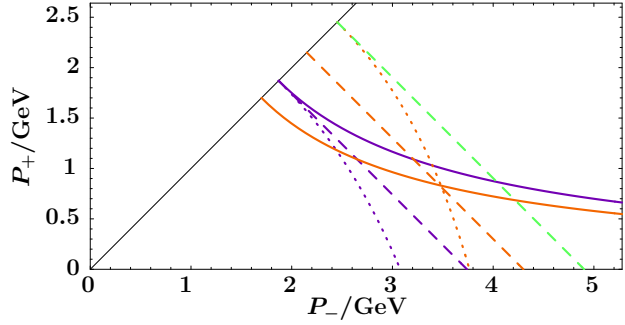}
\caption{\label{fig:phsp_cuts}(color online) Phase space cuts. Solid, dashed, and dotted lines are constant $s_H$, $E_H$, and $q^2$. The dark (violet) ones are $s_H = m_D^2$, $E_H = m_D$, and $q^2 = (M_B - m_D)^2$. The medium (orange) ones are $s_H = (1.7\GeV)^2$, $E_H = 2.15 \GeV$, and $q^2 = 8\GeV^2$, and the light (green) dashed line is $E_H = 2.45\GeV$.}
\end{figure}
\begin{figure}%
\centering
\includegraphics[width=0.495\columnwidth]{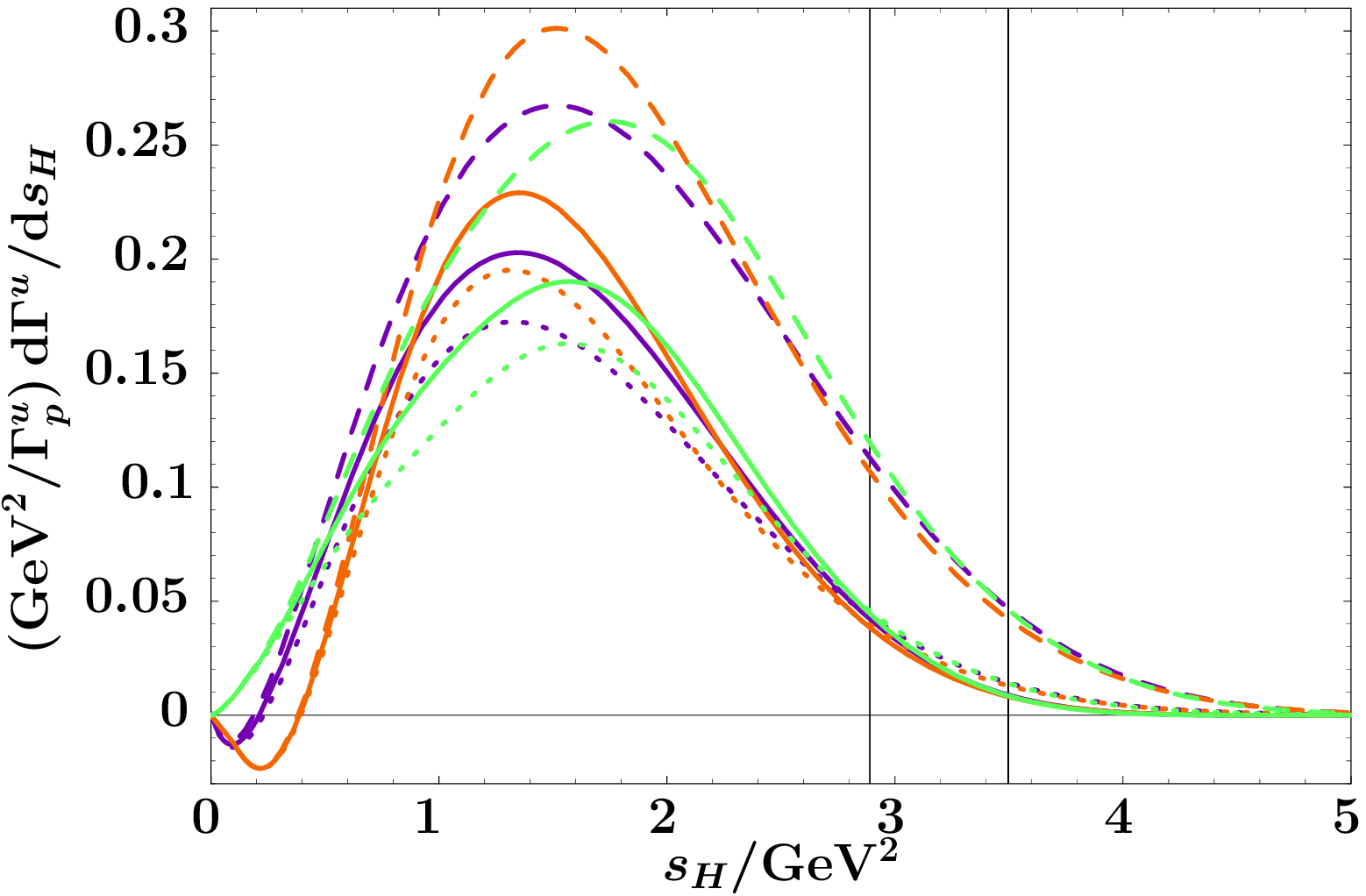}\hfill
\includegraphics[width=0.495\columnwidth]{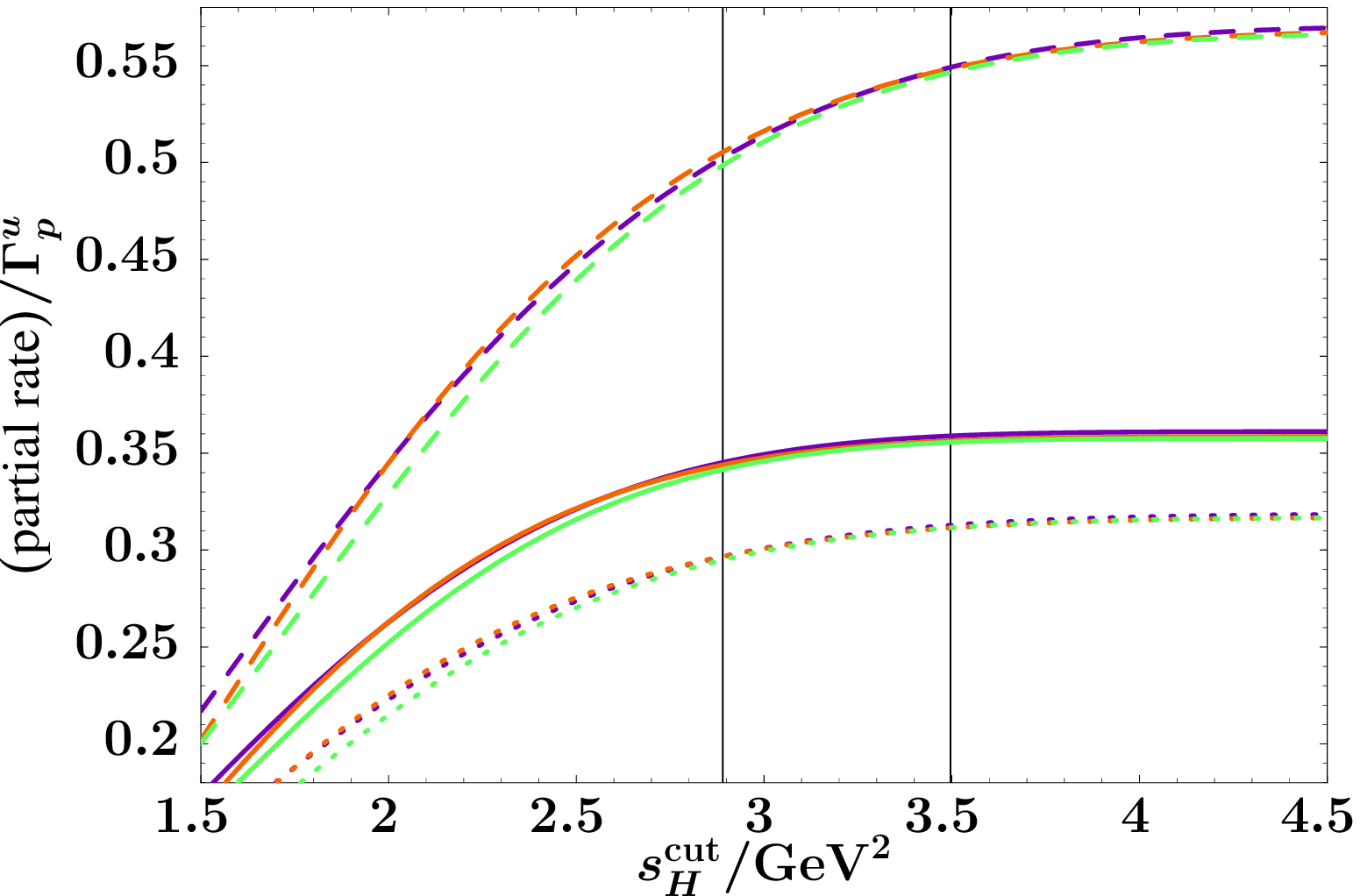}%
\caption{\label{fig:dGsH_EH}(color online) $s_H$ spectrum (left) and partial rate (right) with cuts $E_H < 2.15\GeV$ (solid), $E_H < 2.45\GeV$ (dashed), and for comparison $q^2 > 8\GeV^2$ (dotted).}
\end{figure}

The various phase space cuts are depicted in Fig.~\ref{fig:phsp_cuts}. The solid dark (violet) and medium (orange) lines are $s_H = m_D^2$ and $s_H = (1.7\GeV)^2$. For $(q^2)^\mathrm{min} > 8\GeV^2$ (medium or orange dotted) this corresponds to the cut $E_H < 2.45 \GeV$ (light or green dashed). The spectrum and partial rate for this cut are given by the dashed lines in Fig.~\ref{fig:dGsH_EH}. For $s_H < (1.7 \GeV)^2$ the $E_H$ cut significantly increases the fraction of $b\to u$ events from about 30\% to about 50\% (at tree level), while achieving almost the same shape-function independence. On the other hand this cut has less separation power then the $q^2$ cut to reject contamination from $b\to c$ transitions. Lowering it to the intersection of $q^2 = 8 \GeV^2$ and $s_H = (1.7\GeV)^2$ yields $E_H < 2.15 \GeV$ (medium or orange dashed). This still retains a larger fraction of the signal, and at the same time cuts out a somewhat larger portion of the $b\to c$ phase space, which should in principle provide an equal or better suppression of contamination from $b\to c$.

The $s_H$ spectrum and partial rate for this cut are shown by the solid lines of Fig.~\ref{fig:dGsH_EH}. For comparison, the dotted lines show the result for the $q^2$ cut, i.e., they are identical to the solid lines of Fig.~\ref{fig:dGsH_q28}. The $E_H$ cut has basically the same shape-function independence, but retains additional 5\% of signal events and should provide an equal or better $b\to c$ separation. Of course, eventually this depends on the experimental resolution. We conclude that a combined analysis of $s_H$ and $E_H$ provides a viable alternative for measuring $\abs{V_{ub}}$, with a potentially higher accuracy than the analogous measurement of $s_H$ and $q^2$.

\begin{figure}%
\centering
\includegraphics[width=0.495\columnwidth]{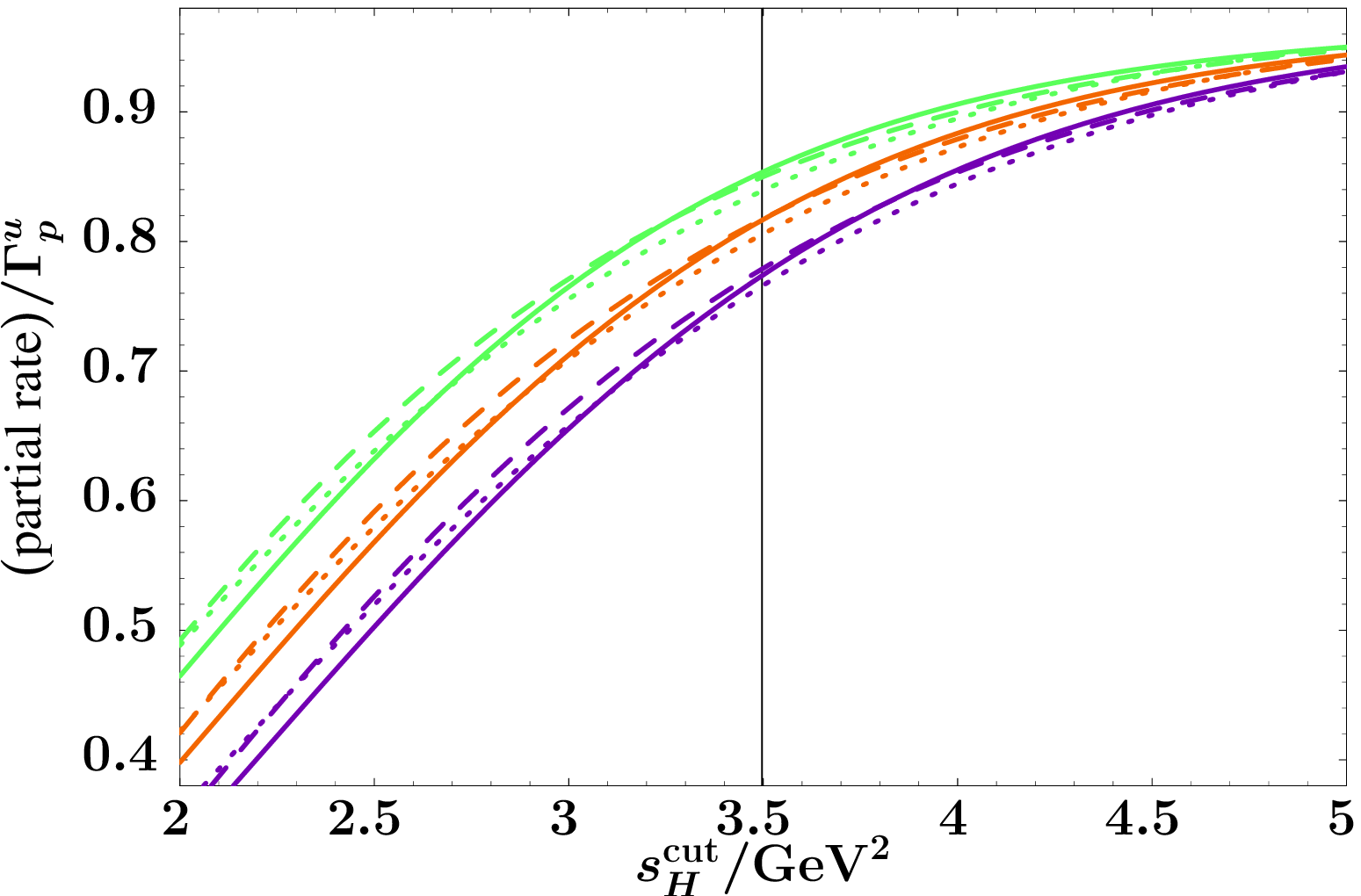}\hfill
\includegraphics[width=0.495\columnwidth]{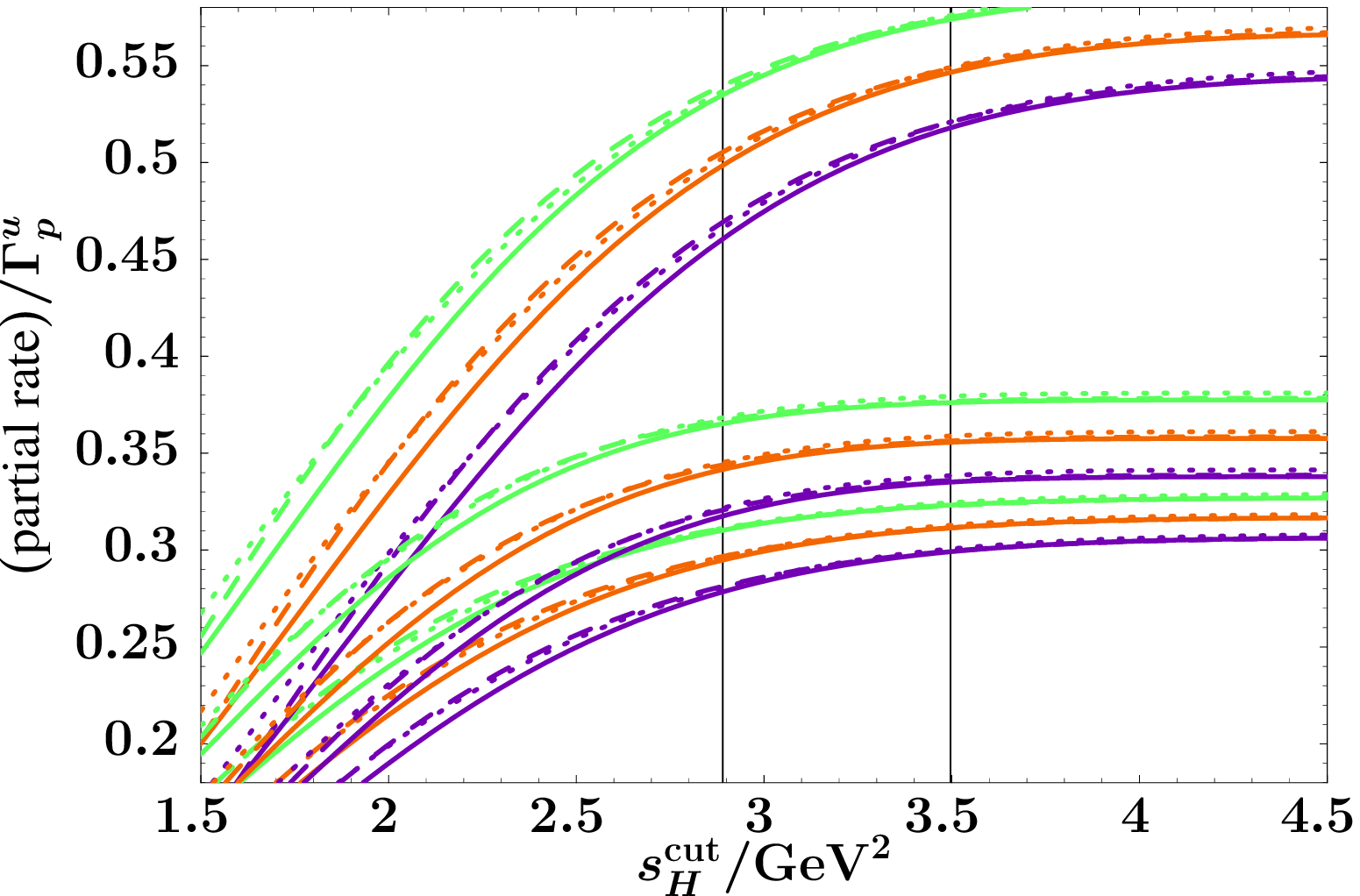}%
\caption{\label{fig:IsHv}(color online) $m_b$ dependence of partial rate for $s_H \leq s_H^\mathrm{cut}$. Here, dark (violet), medium (orange), and light (green) lines correspond to $m_b = 4.6$, $4.65$, $4.7\GeV$, respectively, while the three different models are dotted, dashed, and solid. The left plot contains no additional cut, corresponding to Fig.~\ref{fig:dGsH1}. The right plot corresponds to Fig.~\ref{fig:dGsH_EH}, where the upper set of curves is for $E_H < 2.45\GeV$, the middle set is for $E_H < 2.15\GeV$, and the lower set is for $q^2 > 8\GeV^2$.}
\end{figure}

Fig.~\ref{fig:IsHv} shows the variation of the partial rate for $s_H \leq s_H^\mathrm{max}$ for the various cuts when changing $m_b$ by $\pm 50\MeV$, where curves of the same color correspond to the same value of $m_b$. It shows, that the uncertainty in $m_b$ is clearly a separate, in these cases much bigger, effect then the sensitivity to the specific form of the shape functions. The two effects should therefore be distinguished and treated as separate uncertainties, as argued at the end of Sec.~\ref{subsec:sf-models}.

\subsubsection{Hadronic Energy Spectrum}

A fixed hadronic energy corresponds to a line with slope $-1$ in the $P_\pm$ plane, see Fig.~\ref{fig:phsp_cuts}. The hadronic energy spectrum thus receives contributions from both local and twist phase space regions, and is therefore interesting to study in its own right. In addition, considering the single differential spectra, a cut on $E_H < m_D$ (dark or violet dashed in Fig.~\ref{fig:phsp_cuts}) retains substantially more signal events than the equivalent cut on $q^2 > (m_B - m_D)^2$ (dark or violet dotted). Both methods have been suggested to determine $\abs{V_{ub}}$ \cite{Bouzas:1994cr,Greub:1996ed,Bauer:2000xf}.

Changing variables from $q^2$ to $E_H$ in Eq.~\eqref{dG2_sHq2} and integrating over $s_H$, we obtain
\begin{equation}
\label{dG_EH}
\begin{split}
\frac{\df\G^u}{\df E_H}
&= 64 \G^u_0 \int\df\w \theta(E_H - \w)\theta(M_B - 2E_H + \w)(E_H - \w)^2
\\ & \quad \times
\Bigl[M_\w(M_B - 2E_H + \w)(w^u_1 + w^u_2 + w^u_3)(\w,2E_H - \w)
+ M_\w^2 w^u_4(\w,2E_H - \w)
\\ & \quad
+ (M_B - 2E_H + \w)^2 w^u_5(\w,2E_H - \w) \Bigr]
.\end{split}
\end{equation}
Using the fact that $0 \leq \w$, the phase space limits yield the integration limits
\begin{equation}
\begin{aligned}
0 &\leq \w \leq E_H &&\text{for} & 0 &\leq E_H \leq M_B/2
,\\
2E_H - M &\leq \w \leq E_H \quad &&\text{for} & M_B/2 &\leq E_H \leq M_B
.\end{aligned}
\end{equation}

\begin{figure}%
\centering
\includegraphics[width=0.495\columnwidth]{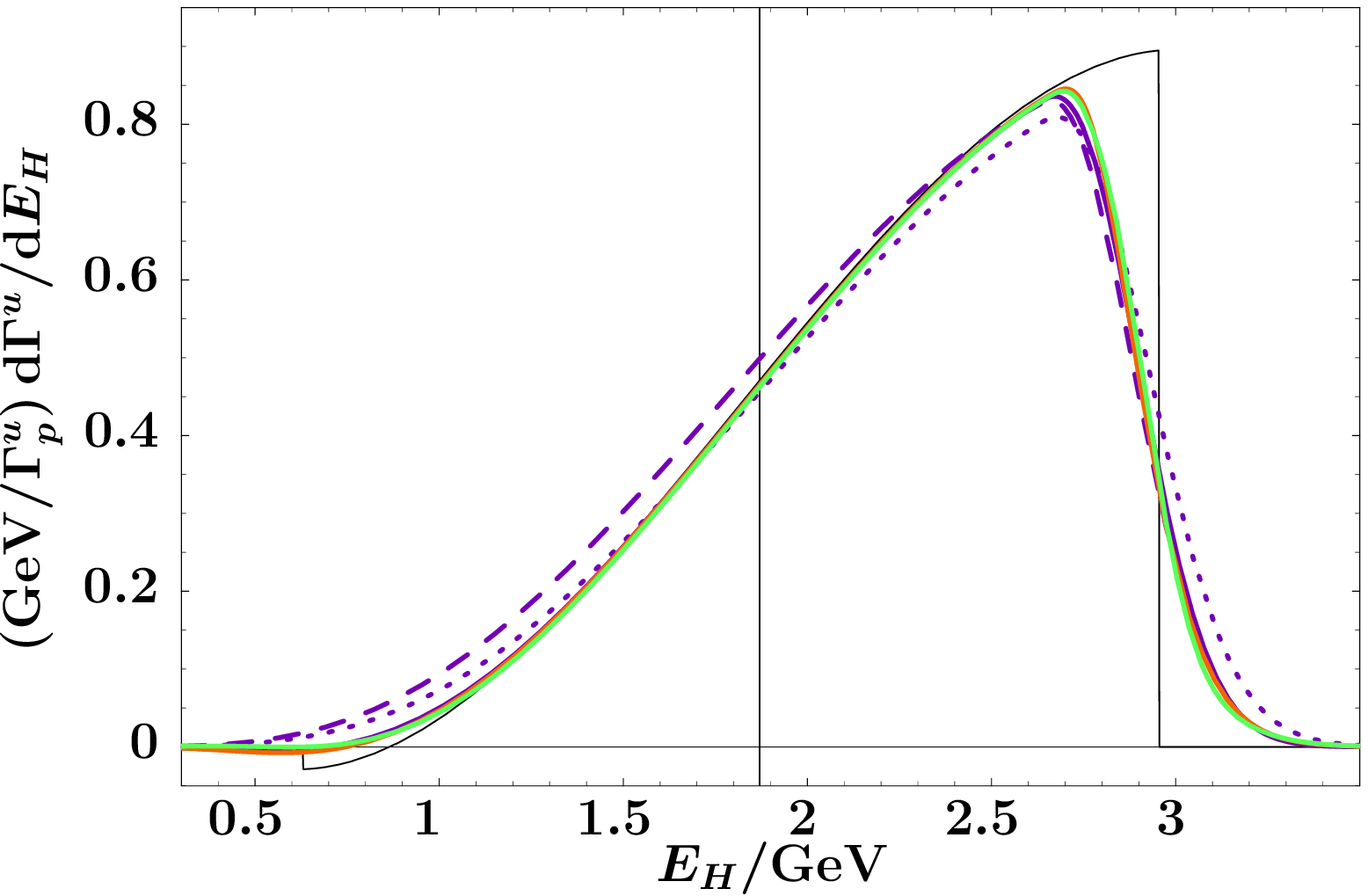}\hfill
\includegraphics[width=0.495\columnwidth]{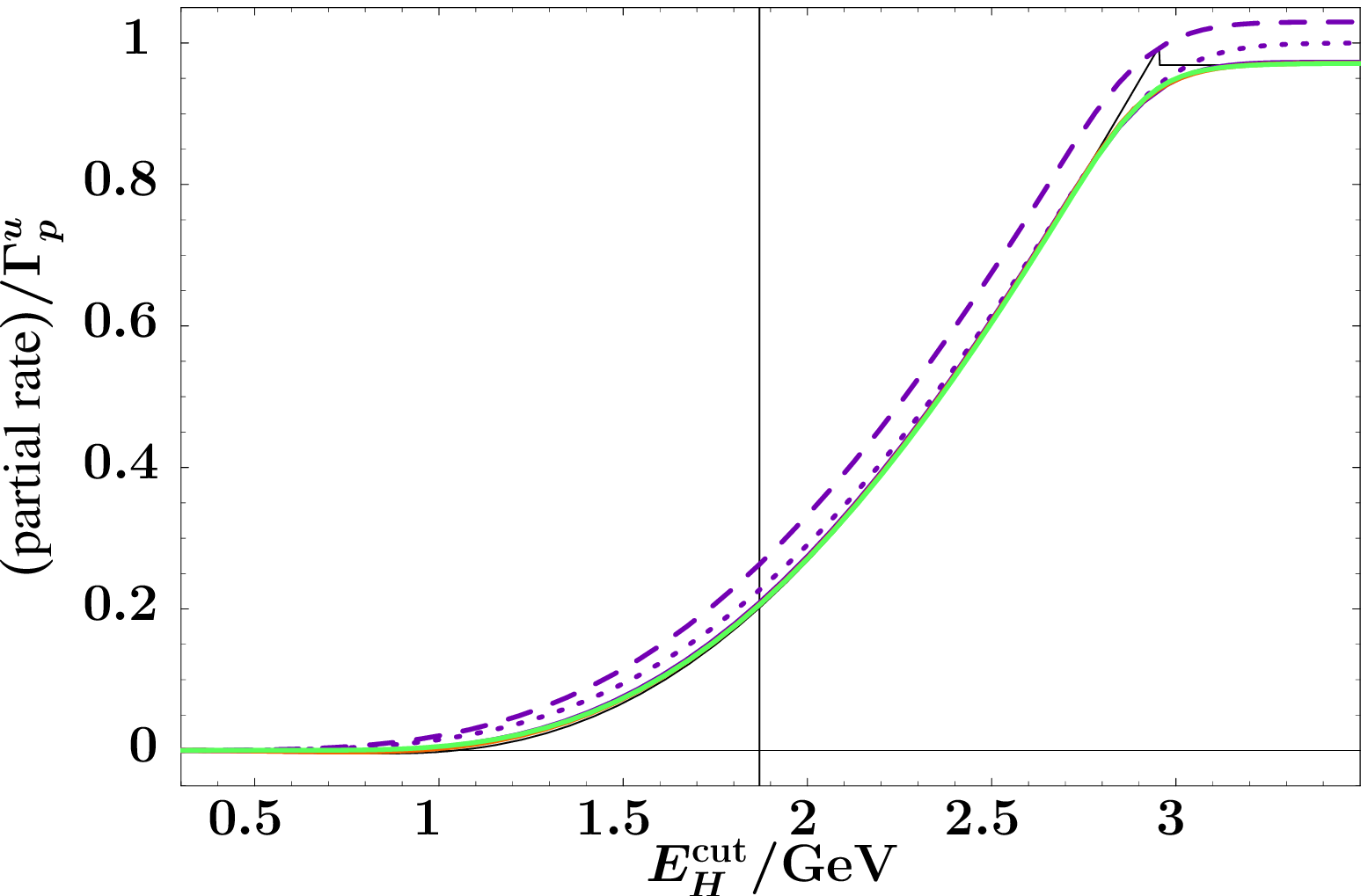}%
\caption{\label{fig:dGEH1}(color online) $E_H$ spectrum (left) and partial rate (right). The solid lines show the full result, the dashed lines only include the contributions from $F_0(\w)$, and the dotted lines correspond to the prescription of Ref.~\cite{DeFazio:1999sv}. The thin black curve is the local $\ord{\La^2}$ result. The vertical line denotes $E_H = m_D$.}
\end{figure}

The $E_H$ spectrum and the partial rate obtained from integrating it up to $E_H \leq E_H^\mathrm{cut}$ are shown in Fig.~\ref{fig:dGEH1}. Our result (solid) matches the local result (thin black) over a wide range of energies and smooths it out near the partonic phase space boundaries. The partonic boundary however does not lie in the shape-function region, and hence, the shape-function dependence in the dropoff at $E_H = m_b/2 + \la$, i.e., the differences between the three models, are very mild, for example compared to the $s_H$ spectrum or the lepton energy spectrum (see below). In the $E_H$ spectrum the prescription of Ref.~\cite{DeFazio:1999sv} seems to give a better approximation than convoluting the full $m_b$ dependence of the partonic spectrum. A cut on $E_H < m_D$ alone keeps 21\% of the $b\to u$ signal (normalized to the partonic rate), which is 50\% more than the cut on $q^2 > (m_B - m_D)^2$, which keeps 14\%. This relative increase should not be changed much by radiative corrections. At the same time the spectrum and partial rate in this region are completely shape-function independent. Hence, measuring the hadronic energy spectrum alone to extract $\abs{V_{ub}}$ seems worth pursuing, too.

\begin{figure}%
\centering
\includegraphics[width=0.495\columnwidth]{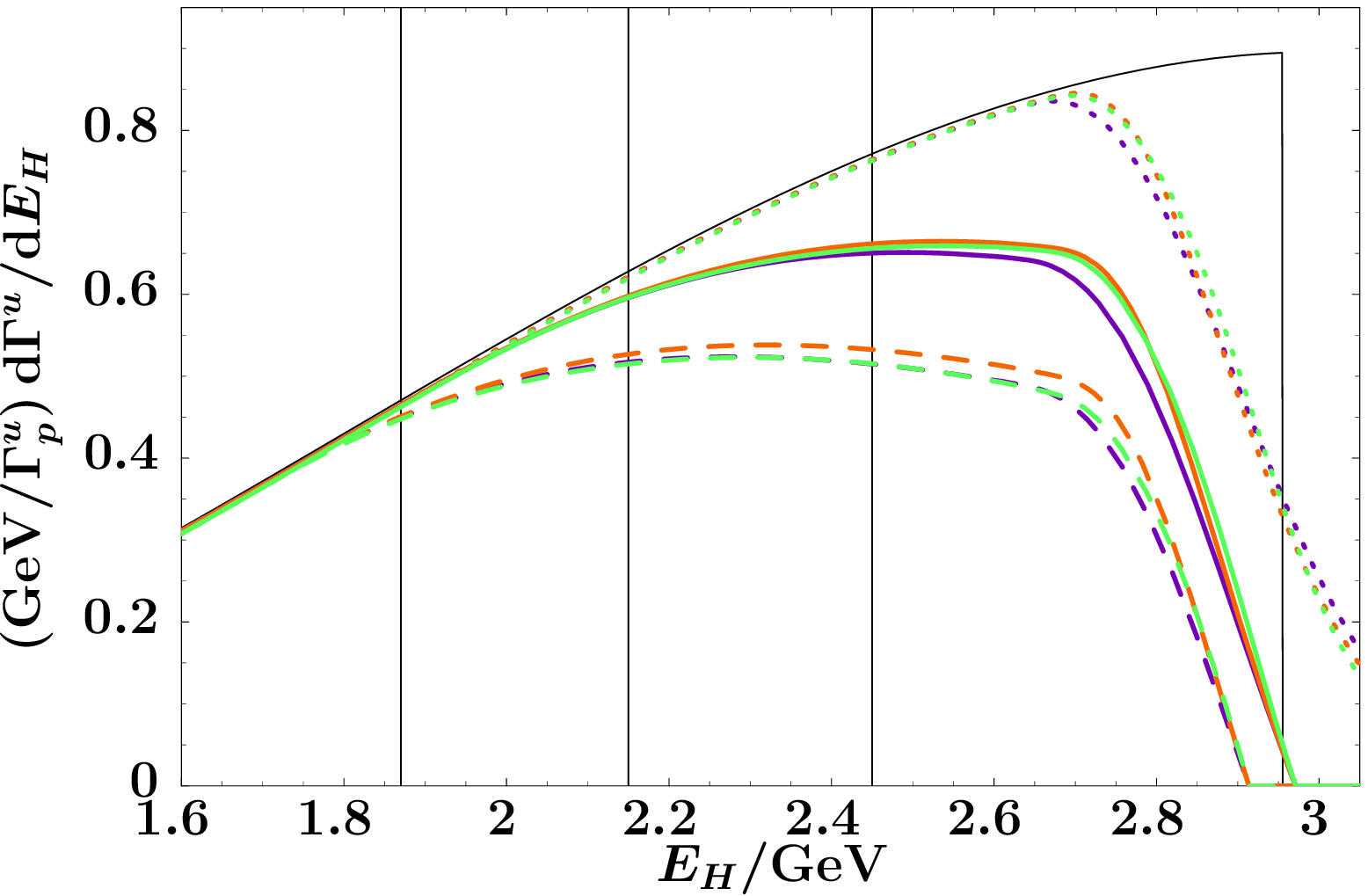}\hfill
\includegraphics[width=0.495\columnwidth]{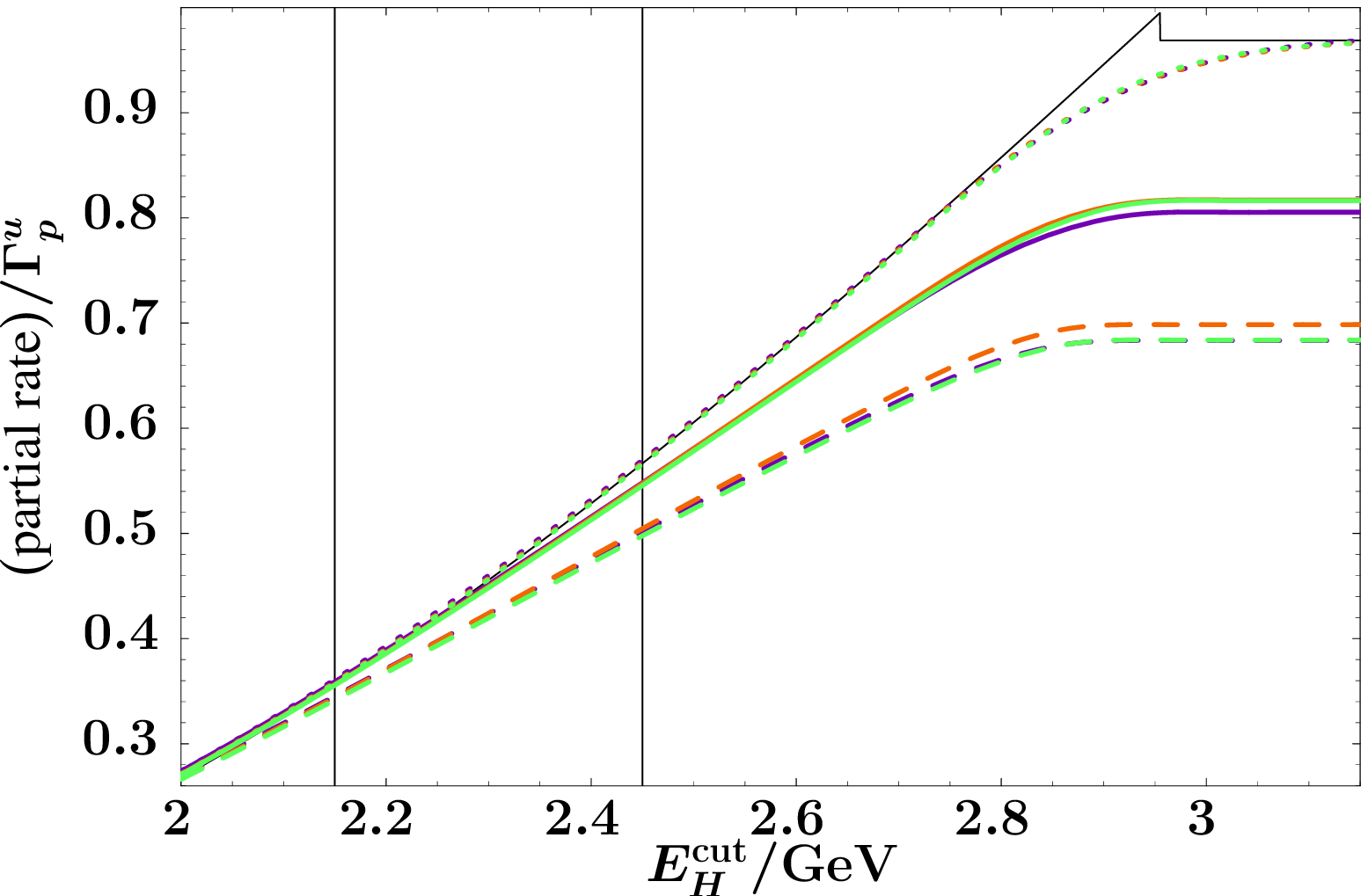}%
\caption{\label{fig:dGEH2}(color online) Hadronic energy spectrum (left) and partial rate (right) with a cut $s_H < m_D^2$ (solid), $s_H < (1.7\GeV)^2$ (dashed), and no cut (dotted). The thin black curve is the local $\ord{\La^2}$ result, and the vertical lines denote $E_H = m_D$, $E_H = 2.15\GeV$, and $E_H = 2.45\GeV$.}
\end{figure}
\begin{figure}%
\centering
\includegraphics[width=0.495\columnwidth]{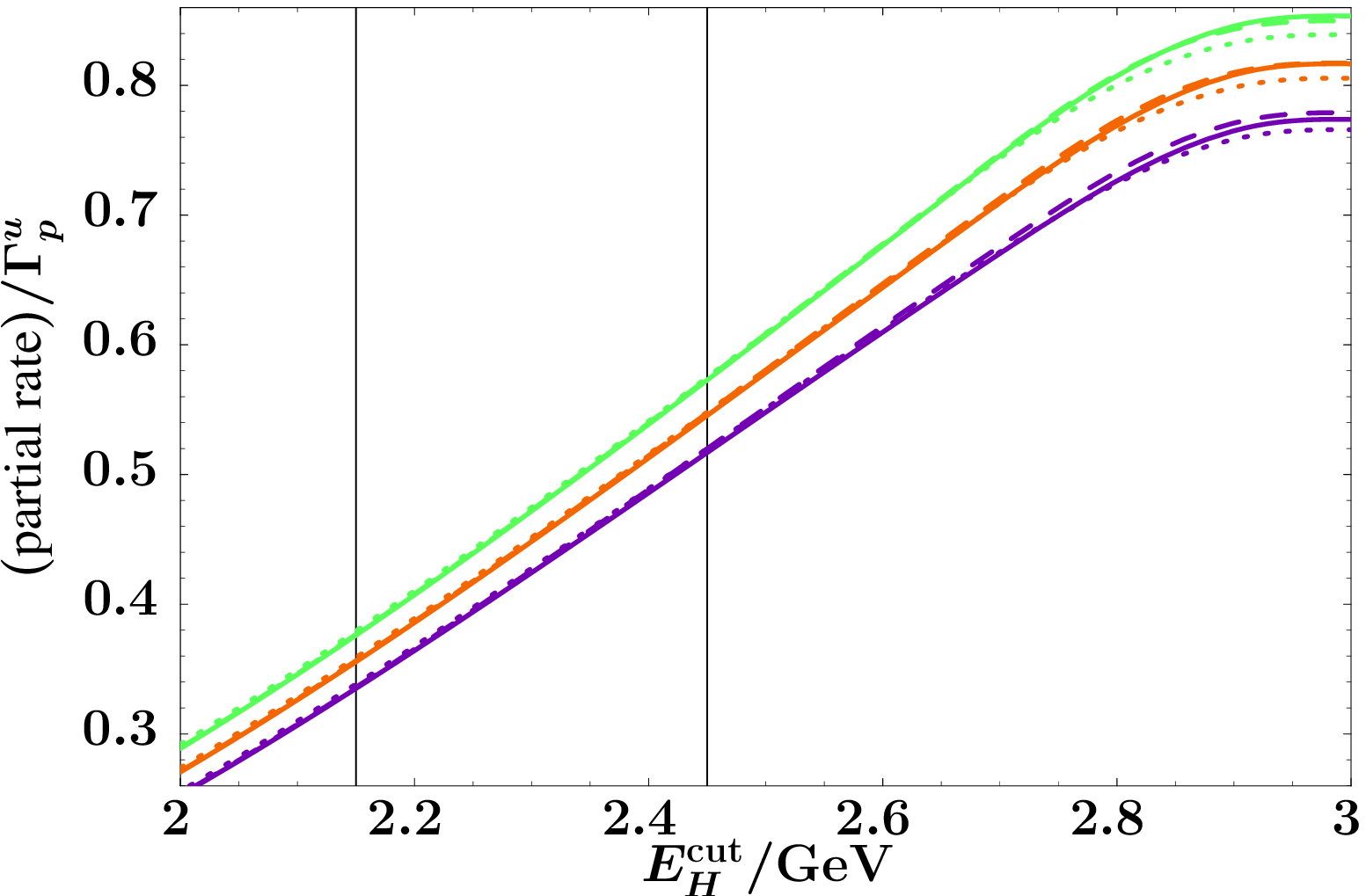}\hfill
\includegraphics[width=0.495\columnwidth]{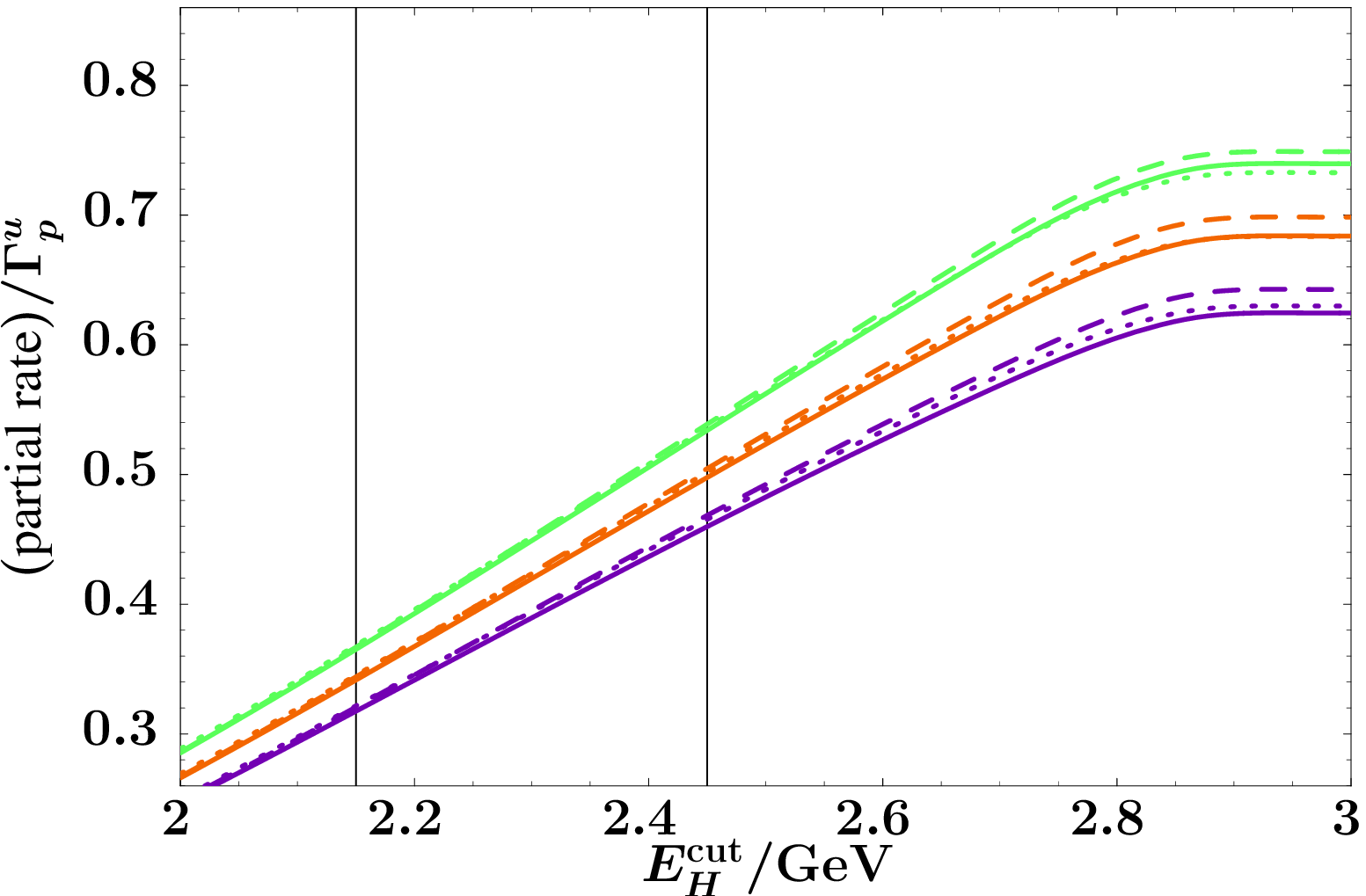}%
\caption{\label{fig:IEHv}(color online) $m_b$ dependence of partial rate for $E_H \leq E_H^\mathrm{cut}$ with a cut $s_H < m_D^2$ (left) and $s_H < (1.7\GeV)^2$ (right). The meaning of the curves is the same as in Fig.~\ref{fig:IsHv}.}
\end{figure}

The spectrum and partial rate with an additional cut $s_H < m_D^2$ (solid) and $s_H < (1.7 \GeV)^2$ (dashed) are given in Fig.~\ref{fig:dGEH2}. A slight increase in the $s_H$ cut allows one to substantially raise the $E_H$ cut while still keeping the partial rate practically shape-function independent. Ideally, if the cut $s_H < m_D^2$ would remove all charm background, the $E_H$ cut could be raised up to $E_H < 2.7\GeV$, which would yield a partial rate around 70\%. The $m_b$ dependence of the partial rate with these cuts is shown in Fig.~\ref{fig:IEHv}. The uncertainty in $m_b$ is again important, as one would expect, but distinct from the shape-function uncertainty.

\subsubsection{$q^2$-$E_\ell$ Spectrum}

\begin{figure}%
\centering
\includegraphics[width=0.495\columnwidth]{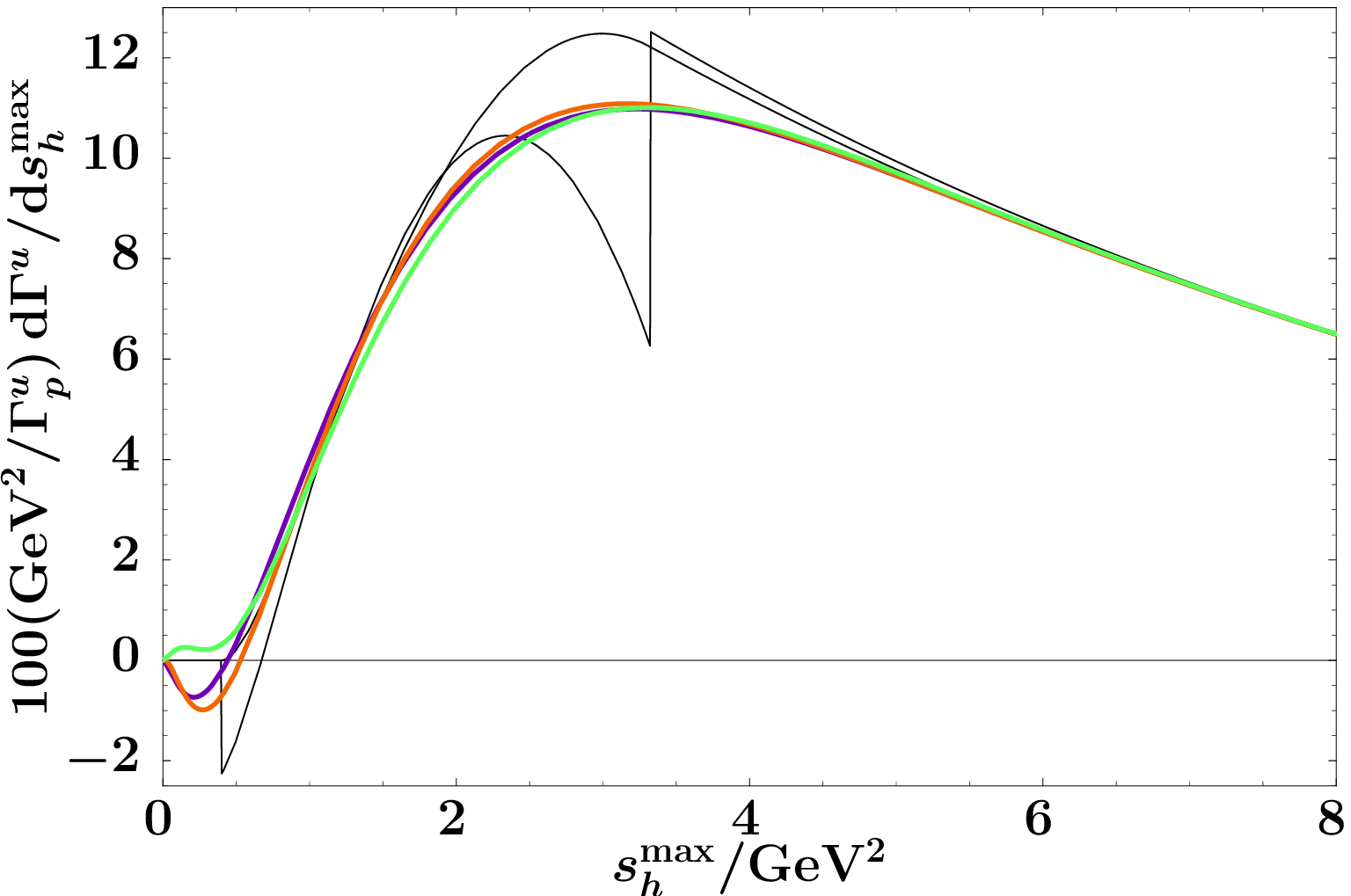}\hfill
\includegraphics[width=0.495\columnwidth]{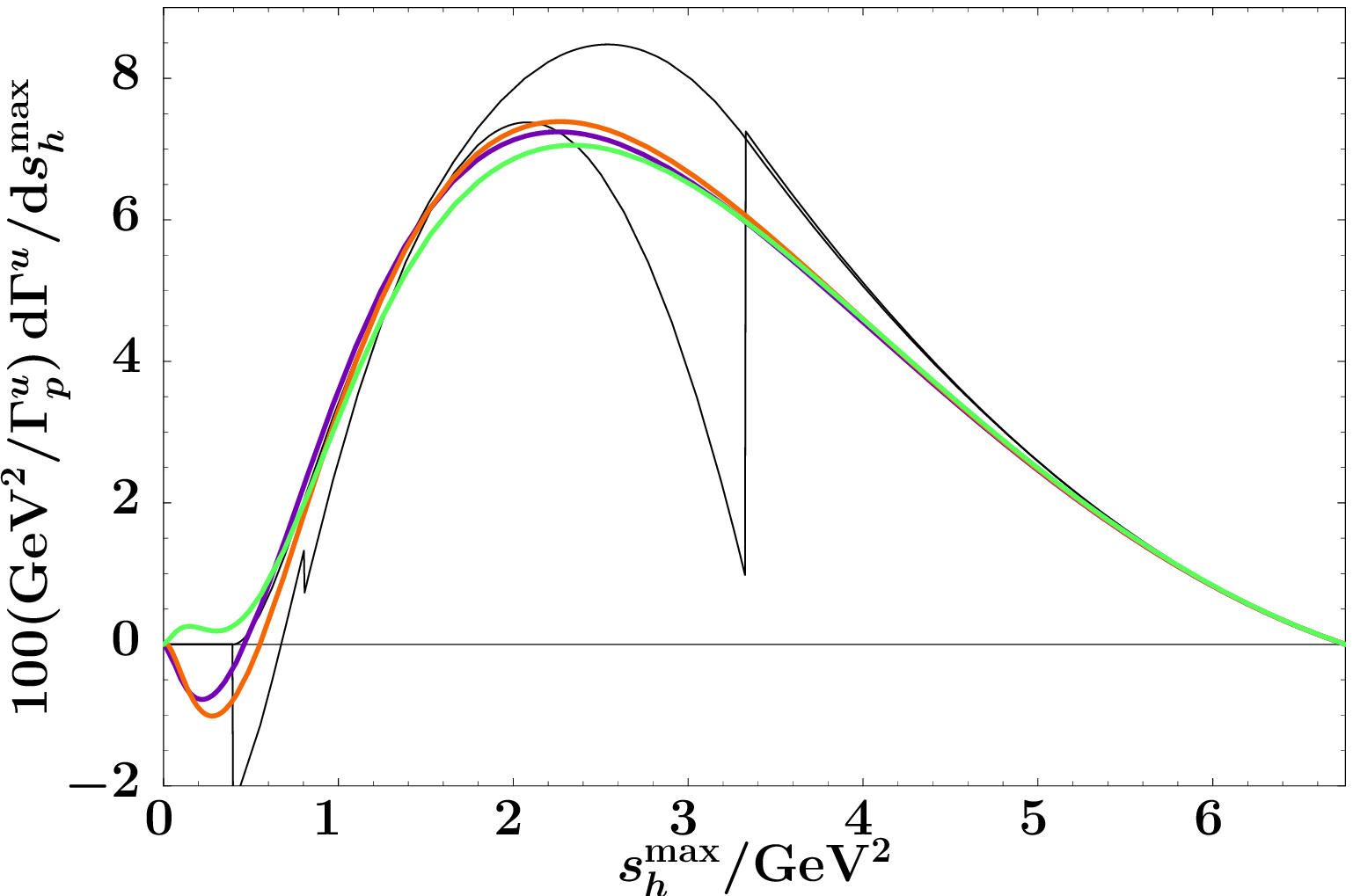}%
\caption{\label{fig:dGs1}(color online) $s_h^\mathrm{max}$ spectrum without cut on $E_\ell$ (left) and with $E_\ell > 2\GeV$ (right). In each case, the smooth black curve is the partonic result and the one with the edge is the local $\ord{\La^2}$ result.}
\end{figure}
\begin{figure}%
\centering
\includegraphics[width=0.495\columnwidth]{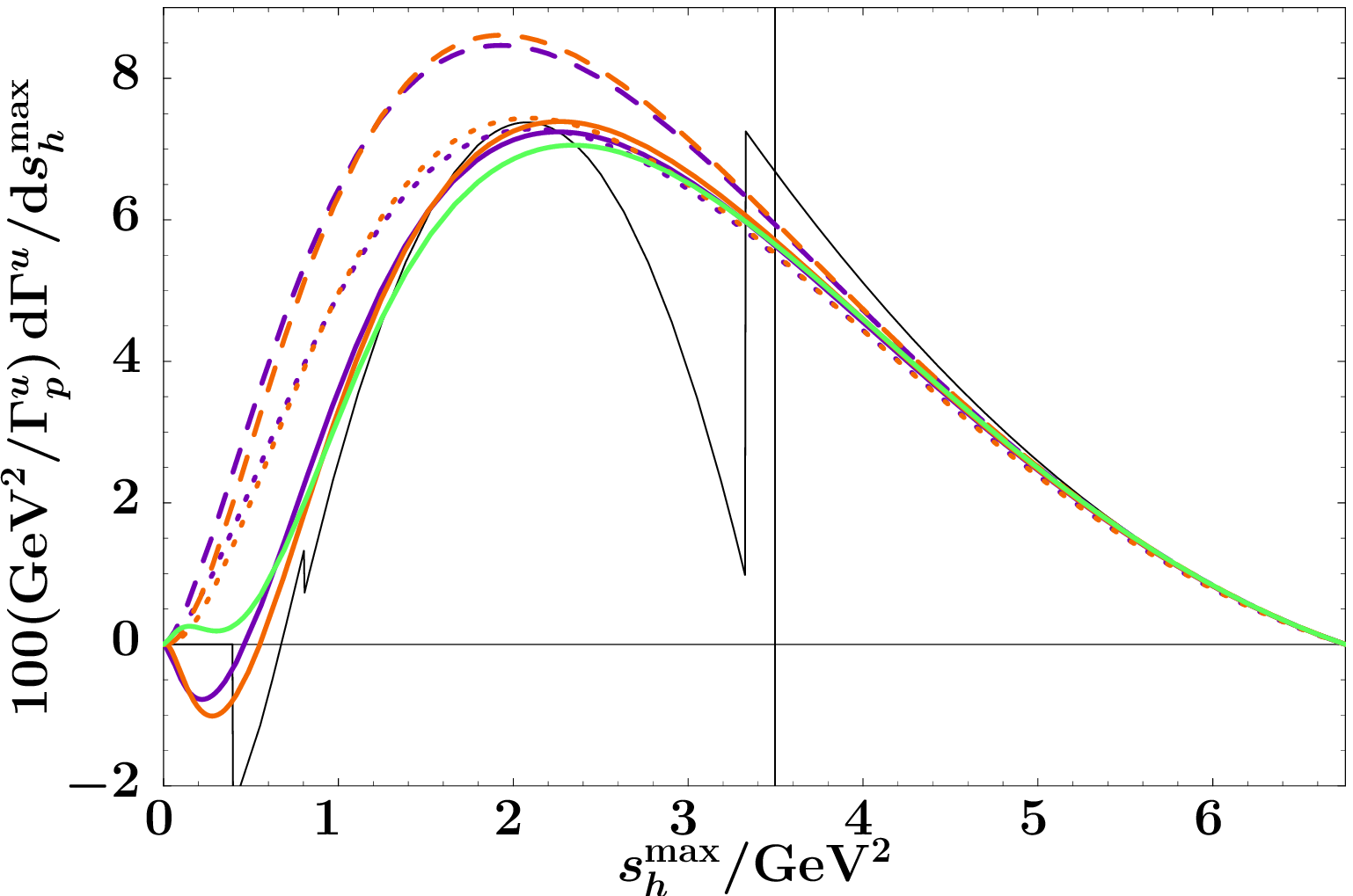}\hfill
\includegraphics[width=0.495\columnwidth]{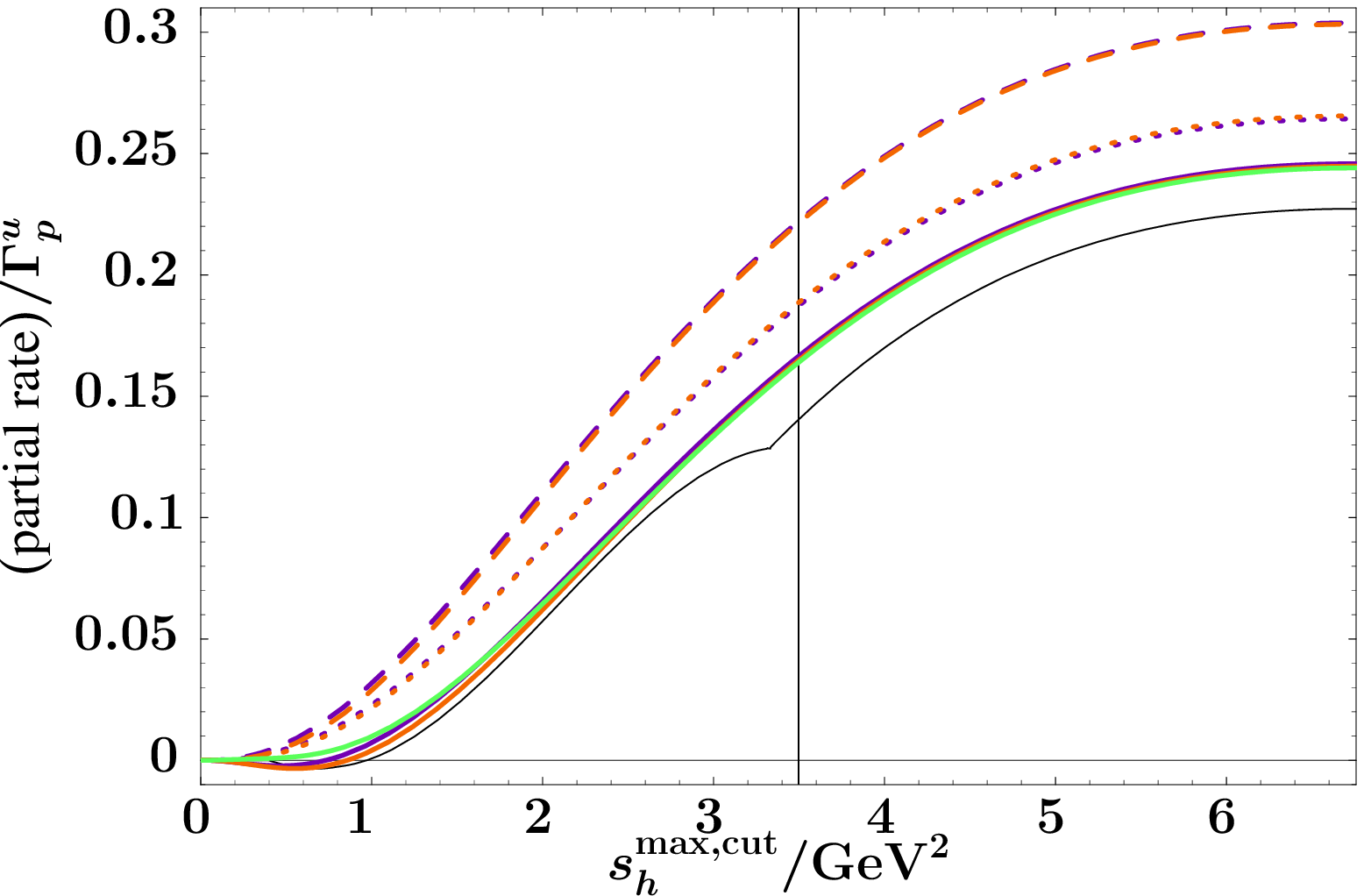}%
\caption{\label{fig:dGs2}(color online) $s_h^\mathrm{max}$ spectrum (left) and partial rate (right) with $E_\ell > 2\GeV$.
The solid lines show the full result, the dashed lines only include the contributions from $F_0(\w)$, and the dotted lines correspond to the prescription of Ref.~\cite{DeFazio:1999sv}. The thin black curve is the local result to $\ord{\La^2}$, and the vertical line denotes $s_h^\mathrm{max} = m_D^2$.}
\end{figure}

The $q^2$-$E_\ell$ spectrum is also of great interest for measuring $\abs{V_{ub}}$.
Integrating Eq.~\eqref{dG3_sHq2} over $s_H$, and defining $q_\w = q^2/M_\w$, we obtain
\begin{equation}
\label{dG2_Elq2}
\begin{split}
\frac{\df^2\G^u}{\df E_\ell \df q^2}
&= 48 \G^u_0 \int \frac{\df\w}{M_\w (M_\w - q_\w)} \theta(\D_\w)\theta(2E_\ell - q_\w)\theta(q_\w)
\biggl\{ q^2 \Bigl[(q_\w - 2E_\ell)^2 w^u_1(\w, M_B - q_\w)
\\ & \quad
+ \D_\w^2 w^u_2(\w, M_B - q_\w) \Bigr]
- 2\D_\w(q_\w - 2E_\ell) \Bigl[q^2 w^u_3(\w, M_B - q_\w)
\\ & \quad
+ M_\w^2 w^u_4(\w, M_B - q_\w) + (q_\w)^2 w^u_5(\w, M_B - q_\w) \Bigr] \biggr\}
.\end{split}
\raisetag{7ex}
\end{equation}
The resulting integration limits on $\w$ are
\begin{equation}
0 \leq \w \leq
\left\{ \begin{aligned}
&M_B - 2E_\ell &&\text{for} & 0 &\leq 2E_\ell \leq \sqrt{q^2} \leq M_B
,\\
&M_B - q^2/(2E_\ell) \quad &&\text{for} & 0 &\leq \sqrt{q^2} \leq 2E_\ell \leq M_B
.\end{aligned} \right.
\end{equation}

The maximally allowed hadronic invariant mass for given $q^2$ and $E_\ell$ defines the variable
\begin{equation*}
s_h^\mathrm{max} = \Bigl(M_B - \frac{q^2}{2E_\ell}\Bigr)(M_B - 2E_\ell)
.\end{equation*}
Requiring $s_h^\mathrm{max} < m_D^2$ is an efficient way to discriminate the $b\to c$ background \cite{Kowalewski:2002qs} and has been implemented by BaBar \cite{Aubert:2004tw}. The distribution in $s_h^\mathrm{max}$ provides a quite nontrivial application for our expansion, because fixed $s_h^\mathrm{max}$ contains contributions from large $q^2$, which should be shape-function independent \cite{Bauer:2000xf}, as well as large lepton energies, which are shape-function sensitive. The $s_h^\mathrm{max}$ spectrum is obtained from Eq.~\eqref{dG2_Elq2} by changing variables from $q^2$ to $s_h^\mathrm{max}$ and integrating over the lepton energy. It is depicted in Fig.~\ref{fig:dGs1} with and without an additional cut on $E_\ell > 2\GeV$. Above $s_h^\mathrm{max}\sim 5\GeV^2$ the spectrum is determined by the local result, which extends to $s_h^\mathrm{max}=M_B^2$, where it goes to zero. Below that our result smooths out the local spectrum. The shape-function sensitivity in the lower part of the spectrum is somewhat larger than for $E_H$, but still much smaller then for $s_H$ or $E_\ell$. With a cut $E_\ell > 2\GeV$ the maximum value of $s_h^\mathrm{max}$ is $M_B(M_B - 4\GeV) = 6.76\GeV^2$. Although this cut removes a large fraction of the local OPE part of the included phase space, it is still low enough that the increase in the shape-function sensitivity is insignificant.

For comparison, Fig.~\ref{fig:dGs2} shows the spectrum and partial rate with a cut $E_\ell > 2\GeV$, where the dashed lines only include the contributions from $F_0(\w)$, and the dotted lines implement the prescription of Ref.~\cite{DeFazio:1999sv}. For the partial rate our result yields a sizable correction to the latter and also to the local result. The variation between the different models is negligible.

\subsubsection{Lepton Energy Spectrum}

Finally, we come to the lepton energy spectrum. Integrating Eq.~\eqref{dG3_wq} over $P_+$ and $P_-$ or Eq.~\eqref{dG2_Elq2} over $q^2$ we find
\begin{equation}
\label{dG_El}
\begin{split}
\frac{\df\G^u}{\df E_\ell}
&= 4\G_0^u \theta(E_\ell) \int\df\w \theta(\D_\w) M_\w \biggl\{
4E_\ell^2 (M_\w + 2\D_\w) (F_0 - K_0)(\w)
\\ & \quad
+ 12M_\w\D_\w \bigl(2E_\ell + \D_\w\ln(\D_\w/M_\w) \bigr)K_0(\w)
\\ & \quad
- 6\Bigl(2E_\ell(E_\ell + 2\D_\w + \w) - \bigl\{M_\w\D_\w^2\ln(\D_\w/M_\w)\bigr\}' \Bigr) (G_5 - H_5)(\w)
\\ & \quad
- 6 M_\w \D_\w \bigl\{\D_\w\ln(\D_\w/M_\w)\bigr\}' (G_5 + H_5)(\w)
\\ & \quad
-12\D_\w \bigl(2E_\ell - \bigl\{M_\w\D_\w\ln(\D_\w/M_\w)\bigr\}' \bigr) (R_4 + L_3)(\w)
\\ & \quad
- 3\D_\w^2 \bigl\{M_\w \ln(\D_\w/M_\w)\bigr\}'' (G_8 - H_8)(\w)
\\ & \quad
- 3 \frac{\D_\w}{M_\w}\Bigl(4E_\ell - \bigl\{M_\w^2\D_\w\ln(\D_\w/M_\w)\bigr\}''\Bigr) (G_8 + H_8)(\w)
\\ & \quad
+ 6\D_\w \bigl\{M_\w\D_\w\ln(\D_\w/M_\w)\bigr\}'' [R_{10} - \la (R_4 + L_3)](\w)
\biggr\}
,\end{split}
\end{equation}
where $G_8(\w)$, $H_8(\w)$, and $L_3(\w)$ are given in Eq.~\eqref{GH8} above, and we use the notation
\begin{equation}
\bigl\{f(\w)\bigr\}' = \frac{f(\w) - f(0)}{\w}
\q
\bigl\{f(\w)\bigr\}'' = 2\frac{f(\w) - f(0) - \w f'(0)}{\w^2}
.\end{equation}
Expanding Eq.~\eqref{dG_El} to subleading twist reproduces the result in Ref.~\cite{Lee:2004ja}. It also agrees with Ref.~\cite{Bosch:2004cb}. However, for some reason, the authors there divide their result by an additional factor $M_B - \w$ and subtract a compensating term $(\w - \la)F_0(\w)$.

\begin{figure}%
\centering
\includegraphics[width=0.495\columnwidth]{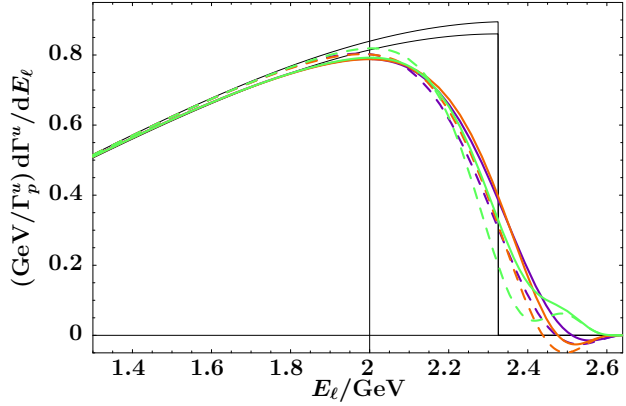}%
\caption{\label{fig:dGEl1}(color online) Lepton energy spectrum spectrum. The solid lines show the result from Eq.~\eqref{dG_El}, and the dashed ones that of Ref.~\cite{Mannel:2004as}. The upper black line shows the partonic spectrum, and the lower one the nonsingular part of the local spectrum to $\ord{\La^2}$.}
\end{figure}
\begin{figure}%
\centering
\includegraphics[width=0.495\columnwidth]{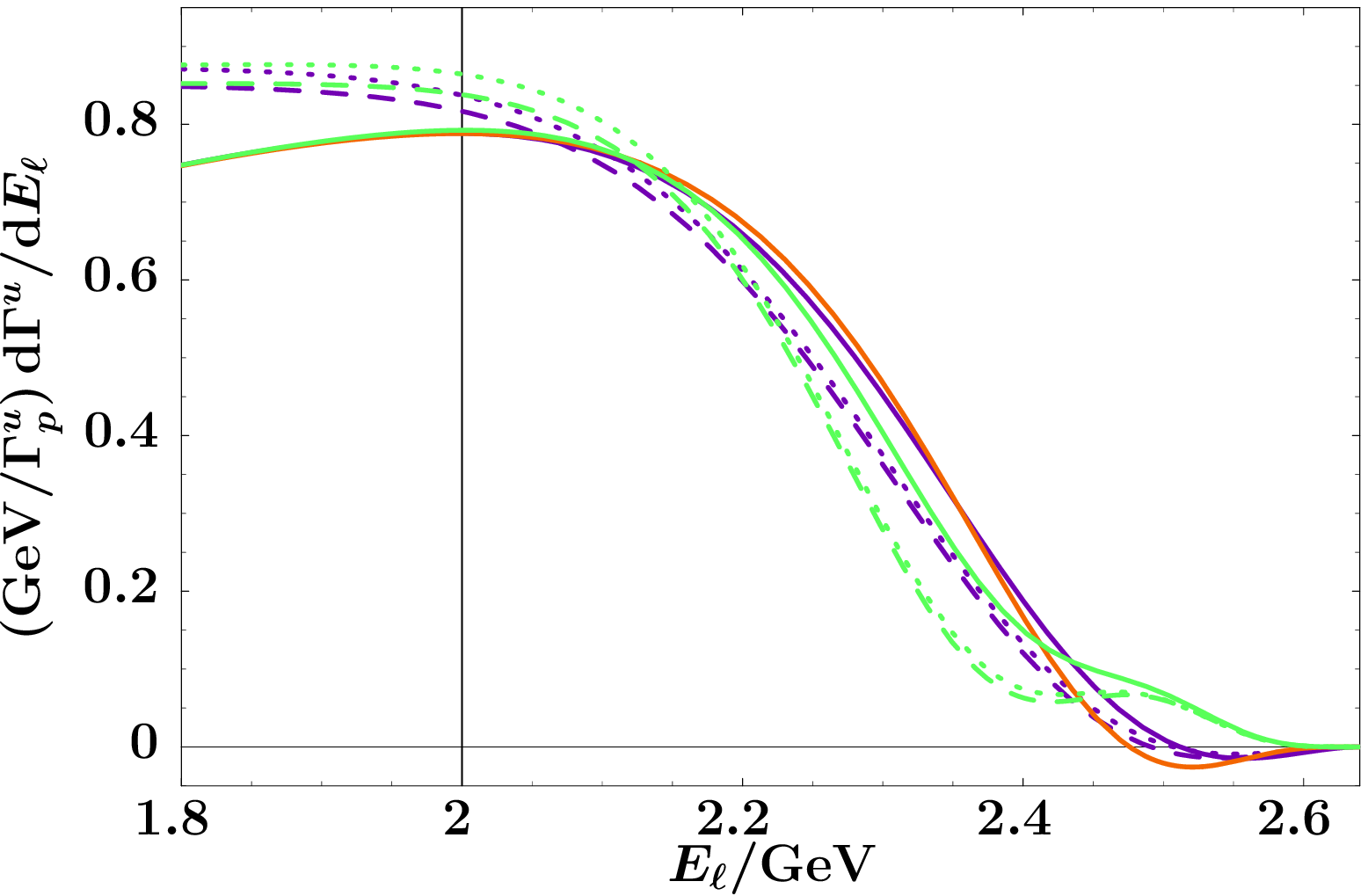}\hfill
\includegraphics[width=0.495\columnwidth]{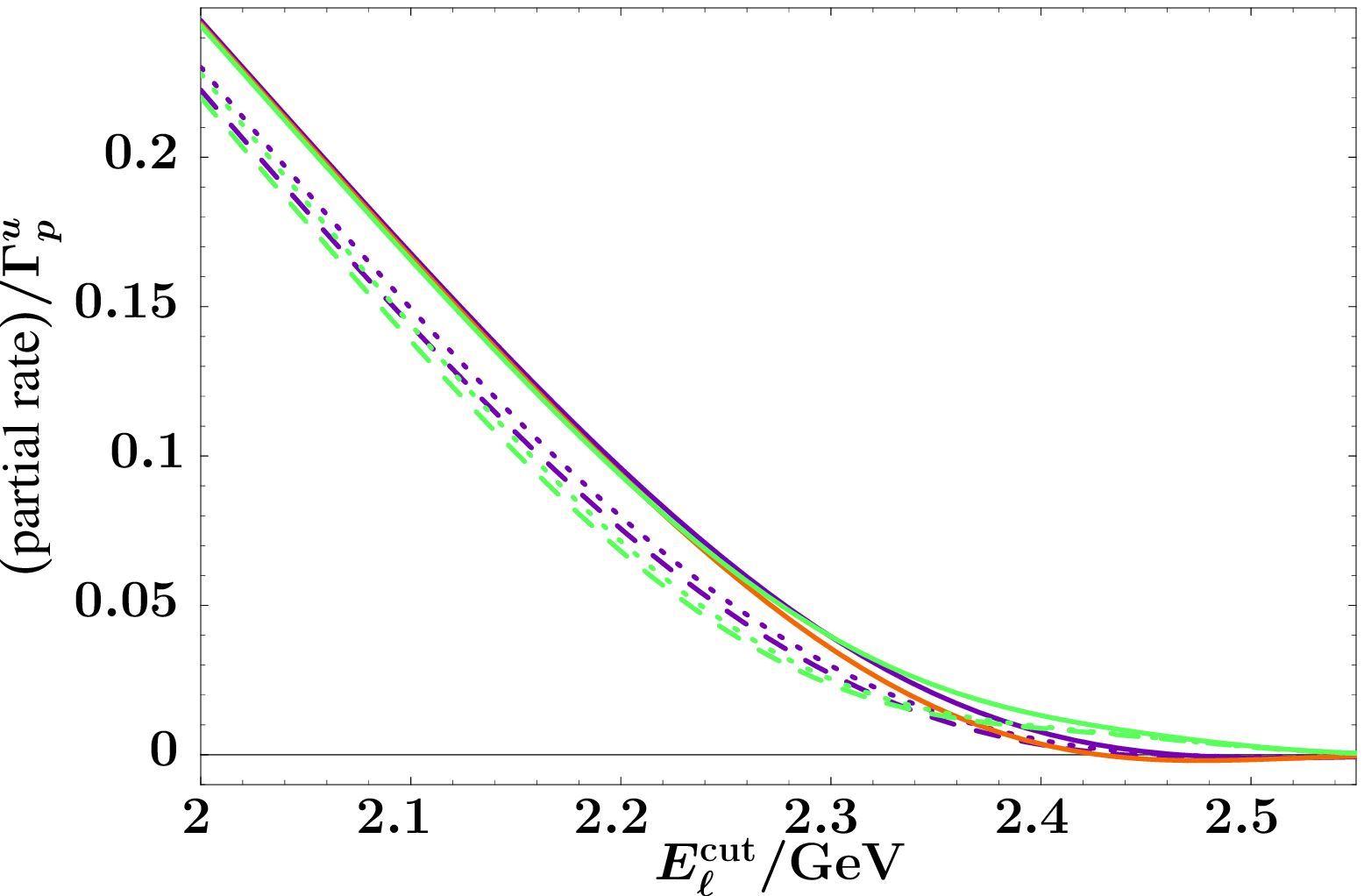}%
\caption{\label{fig:dGEl2}(color online) Lepton energy spectrum (left) and partial rate (right).
The solid lines show the full result, the dashed lines the subleading twist result, and the dotted lines the result of Ref.~\cite{Bosch:2004cb}. The vertical line denotes the BaBar cut $E_\ell = 2\GeV$ \cite{Aubert:2004bv}.}
\end{figure}

Eq.~\eqref{dG_El} and the result derived in Ref.~\cite{Mannel:2004as} agree when both are expanded to subleading twist. However, we cannot expect them to be identical, since the light-cone directions in Ref.~\cite{Mannel:2004as} and in the present case are different. The modified expansion thus retains different higher order twist corrections in each case. The direct computation in Ref.~\cite{Mannel:2004as} yields a much more concise result, because it uses a light-cone direction natural to the lepton energy spectrum. On the other hand one could say that the present choice includes more higher order twist corrections, since it keeps them already at the level of the triple differential rate. However, to make this statement precise one would need to compute all subsubleading twist contributions. The two results are shown in Fig.~\ref{fig:dGEl1}. Below $E_\ell \sim 1.8\GeV$ they are dominated by the local spectrum. Above $E_\ell \sim 2.2\GeV$ they become quite different, which indicates that the higher order kinematic twist corrections are important.

In Fig.~\ref{fig:dGEl2} we compare our result (solid) with the subleading twist result (dashed and dotted). We note two things. First, the spectrum on the left shows that the twist expansion breaks down around $E_\ell \sim 2.1\GeV$, and cannot be trusted for smaller energies. Secondly, we see again that the higher order twist corrections are important. Interestingly, the variations between the different models are significantly larger for the subleading twist result than for our result. We already observed a similar but smaller effect in the $s_H$ spectrum. This suggests that the endpoint spectrum is to a large extent determined by higher order kinematic twist corrections. If this is the case, it would be promising to the $\abs{V_{ub}}$ extraction from the lepton energy endpoint spectrum, where the shape-function dependence is a limiting factor in the achievable accuracy. To confirm this one certainly needs to consider a wider spread of model functions. One could also compute the true subsubleading twist corrections to see if they have an equally large effect on the spectrum or not.

\section{Conclusions}
\label{sec:conclusions}

We have studied the $\LQCD/M_B$ corrections to inclusive $B$-meson decays, with attention to the radiative decay $B\to X_s \g$ and the semileptonic decay $B \to X_u \ell \bar{\nu}_\ell$.

Usually, the twist expansion is valid in the shape-function region and the local expansion in the rest of phase space. Following Ref.~\cite{Mannel:2004as} we used a modification of the twist expansion which avoids the restriction to the shape-function region and yields an expansion applicable over the full phase space, except for the resonance region. This effectively provides a smooth interpolation between the otherwise separate regimes of local and twist expansion. So far, we only worked at tree level. It would certainly be interesting to see how much of our approach can be carried over to include radiative corrections, at least for the contributions proportional to the leading shape function, since the $\alpha_s$ corrections to the leading twist result are known \cite{Bauer:2003pi,Bosch:2004th}. To extend the matching calculation presented here to order $\alpha_s$ one needs to study the renormalization of the relevant light-cone operators in QCD. The renormalization properties of the shape functions will be different when they are defined via QCD rather then HQET operators. Similarly, the $\alpha_s$ corrections to the relations between shape-function moments and the standard HQET parameters will differ between QCD and HQET shape functions.

We performed the expansion directly in QCD light-cone operators and gave a discussion of the general operator basis appearing at tree level, including its parametrization in terms of QCD shape functions. We used reparametrization invariance under rotations of the light-cone direction to reduce the number of independent shape functions, and showed that the different lepton energy spectra obtained in Refs.~\cite{Mannel:2004as,Lee:2004ja,Bosch:2004cb} are in agreement.

The results for the various decay distributions are presented in Sec.~\ref{sec:decayrates}. The photon spectrum for $B \to X_s \g$ is given in Eq.~\eqref{dG_Eg}. The triple differential decay rate for $B\to X_u \ell \bar{\nu}_\ell$ is given in terms of $E_\ell$, $P_+$, and $P_-$ in Eq.~\eqref{dG3_wq} and in terms of $E_\ell$, $s_H$, and $q^2$ in Eq.~\eqref{dG3_sHq2}. Except for the resonance region, the decay rates are valid over the entire phase space, being exact to order $\LQCD/M_B$ in the region of hadronic masses $s_H \sim \ord{\LQCD M_B}$, and to order $\LQCD^2/M_B^2$ away from it.

Employing different shape-function models our results allow to quantify the impact of shape-function effects on decay distributions and partial rates for any desired kinematic cuts. We stress again, that the shape-function models we used only differ in the third and higher moments of the shape functions, but still give quite different shapes. The observed variations in the results provide a direct measure of the true shape-function sensitivity of a quantity, i.e., its sensitivity to the unknown specific form or higher order moments of the shape functions. For the total uncertainty, one has to additionally vary the local parameters, most notably $\la_1$ and $m_b$, as well, the effect of which should be regarded separately.

An application is to study the transition between the local and twist expansion. The primary example is the lepton energy spectrum, and we saw that the usual twist expansion cannot be trusted below $E_\ell \sim 2.1\GeV$.

We are free to choose any kinematic variables, and discussed several examples of interest.
In particular, we can study decay rates which for given values of the kinematic variables receive contributions from the phase space regions of both local and twist expansion, such as the hadronic energy, or the variable $s_h^\mathrm{max}$ used in the $q^2$-$E_\ell$ analysis. The hadronic energy spectrum has not received much attention so far. We point out that, with or without an additional cut on $s_H$, it represents a viable alternative to the existing $s_H$-$q^2$ analyses to extract $\abs{V_{ub}}$.

\begin{acknowledgments}
I like to thank Zoltan Ligeti for many helpful discussions and comments on the manuscript. I also thank Thomas Mannel, Bj\"orn Lange, and Matthias Neubert for discussions related to this subject.
\end{acknowledgments}

\appendix*

\section{Parametrization of Light-Cone Operators}
\label{app:param-of-lc-ops}

Here we collect the results related to the parametrization the light-cone operators from the body of the paper and add some further details. First, the number of independent Lorentz structures in the operator basis in Eq.~\eqref{lc-ops_basis} can be reduced employing the relations
\begin{equation}
\label{nO-rel_full}
\begin{aligned}
n_\mu \cO_1^{\G\mu}(\w) &= \bigl[\w \cO_0^\G(\w)\bigr]'
,\\
n_\mu \cO_3^{\G\mu}(\w) &= -\w \cO_0^\G(\w)
\q &
n_\mu \cO_4^{\G\mu}(\w) &= 0
,\\
n_\nu \cO_2^{\G(\mu\nu)}(\w)
&= \tfrac{1}{2} \bigl[\w\cO_1^{\G\mu}(\w) \bigr]' + \tfrac{1}{2} \cO_1^{\G\mu}(\w)
,\\
n_\nu \cO_5^{\G(\mu\nu)}(\w) &= - \w \cO_1^{\G\mu}(\w) + \cO_3^{\G\mu}(\w)
\q &
n_\nu \cO_5^{\G[\mu\nu]}(\w) &= - \cO_4^{\G\mu}(\w)
,\\
n_\nu \cO_8^{\G(\mu\nu)}(\w) &= - \w \cO_3^{\G\mu}(\w)
\q &
n_\nu \cO_8^{\G[\mu\nu]}(\w) &= - \w \cO_4^{\G\mu}(\w)
,\\
n_\nu \cO_6^{\G\mu\nu}(\w) &= \bigl[\w \cO_3^{\G\mu}(\w)\bigr]'
\q &
n_\mu \cO_6^{\G\mu\nu}(\w) &= \cO_3^{\G\nu} - \w \cO_1^{\G\nu}
,\\
n_\nu \cO_7^{\G\mu\nu}(\w) &= \bigl[\w \cO_4^{\G\mu}(\w)\bigr]'
\q &
n_\mu \cO_7^{\G\mu\nu}(\w) &= \cO_4^{\G\nu}
,\\
n_\nu \cO_9^{\G\mu\nu}(\w) &= -\w \cO_3^{\G\mu}(\w)
,\\
n_\nu \cO_{10}^{\G\mu\nu}(\w) &= -\w \cO_4^{\G\mu}(\w)
,\end{aligned}
\end{equation}
which hold for any Dirac structure $\G$. Note that there is no relation for $n_\mu \cO_{9,10}^{\G\mu\nu}(\w)$. They also imply
\begin{equation}
\begin{aligned}
n_\mu n_\nu \cO_2^{\G\mu\nu}(\w) &= \tfrac{1}{2} \bigl[\w^2 \cO_0^\G(\w)\bigr]''
,\\
n_\mu n_\nu \cO_5^{\G\mu\nu}(\w) &= n_\mu n_\nu \cO_6^{\G\mu\nu}(\w) = -\bigl[\w^2 \cO_0^\G(\w)\bigr]'
,\\
n_\mu n_\nu \cO_8^{\G\mu\nu}(\w) &= n_\mu n_\nu \cO_9^{\G\mu\nu}(\w) = \w^2 \cO_0^\G(\w)
,\\
n_\mu n_\nu \cO_7^{\G\mu\nu}(\w) &= n_\mu n_\nu \cO_{10}^{\G\mu\nu}(\w) = 0
.\end{aligned}
\end{equation}

For completeness we repeat the parametrization of the leading operator, Eq.~\eqref{FK0_def},
\begin{equation}
\vevB{\cO_0^\alpha(\w)}
= F_0(\w) v^\alpha + K_0(\w) (n - v)^\alpha
,\end{equation}
and the $\ord{\eps}$ operators, Eqs.~\eqref{FK1_def},
\begin{equation}
\begin{aligned}
\vevB{v_\mu\cO_1^{\alpha\mu}(\w)} &= - (F_1 - \la F_0)'(\w) v^\alpha - (K_1 - \la K_0)'(\w) (n - v)^\alpha
,\\
\vevB{\eta_{\perp\alpha\mu}\cO_1^{\alpha\mu}(\w)} &= - L_1'(\w)
,\\
\vevB{v_\mu\cO_3^{\alpha\mu}(\w)} &= (F_3 - \la F_0)(\w) v^\alpha + (K_3 - \la K_0)(\w) (n - v)^\alpha
,\\
\vevB{\eta_{\perp\alpha\mu} \cO_3^{\alpha\mu}(\w)} &=L_3(\w)
,\\
\vevB{\img\eps_{\perp\alpha\mu}\cP_4^{\alpha\mu}(\w)} &= R_4(\w)
.\end{aligned}
\end{equation}
The set of all nonzero matrix elements for the $\ord{\eps^2}$ operators is given in Eq.~\eqref{OP_e2_nonzero}. We need%
\begin{equation}
\begin{aligned}
\vevB{\eta_{\perp\mu\nu}\cO_{5,8}^{\alpha\mu\nu}(\w)}
&= G_{5,8}(\w) v^\alpha + M_{5,8}(\w) (n-v)^\alpha
,\\
\vevB{\img \eps_{\perp\mu\nu}\cP_{5,8}^{\alpha[\mu\nu]}(\w)}
&= H_{5,8}(\w) (n - v)^\alpha + N_{5,8}(\w) v^\alpha
,\\
\vevB{\bn_\mu \eta_{\perp\alpha\nu} \cO_9^{\alpha\mu\nu}(\w)}
&= (L_9 - \la L_3)(\w)
,\\
\vevB{\bn_\mu \img\eps_{\perp\alpha\nu} \cP_{10}^{\alpha\mu\nu}(\w)}
&= (R_{10} - \la R_4)(\w)
.\end{aligned}
\end{equation}
The four-quark operators, defined in Eq.~\eqref{cQi_def} give rise to the shape functions
\begin{equation}
\label{GH4q_def}
\begin{split}
\vevB{\cQ_1^{f\alpha}(\w)} &= G^f_1(\w) v^\alpha + M^f_1(\w) (n - v)^\alpha
,\\
\vevB{\cQ_2^{f\alpha}(\w)} &= H^f_2(\w) (n - v)^\alpha + N^f_2(\w) v^\alpha
.\end{split}
\end{equation}
They depend on the flavor of the final-state quark, and are thus different for $B\to X_s\g$ and $B\to X_u\ell \bar{\nu}_\ell$. In addition, they also differ for charged and neutral $B$-mesons.

The RPI constraints in Eqs.~\eqref{deR_Os} require
\begin{equation}
\label{O12_RPI}
\begin{aligned}
L_1'(\w) &= 2K_0(\w)
\q &
L_2'(\w) &= 2K_1(\w)
,\\
G_2(\w) &= -2[(\w - \la) F_0 + F_1](\w)
\q &
M_2'(\w) &= -2[(\w - \la) K_0 + K_1]'(\w) - 2K_0(\w)
,\end{aligned}
\end{equation}
where the functions $G_2(\w)$, $L_2(\w)$, and $M_2(\w)$ are defined by
\begin{equation}
\begin{split}
\vevB{\eta_{\perp\mu\nu}\cO_2^{\alpha(\mu\nu)}(\w)}
&= - \tfrac{1}{2} G_2'(\w) v^\alpha - \tfrac{1}{2} M_2'(\w) (n-v)^\alpha
,\\
\vevB{\eta_{\perp\alpha(\mu} v_{\nu)} \cO_2^{\alpha(\mu\nu)}(\w)}
&= \tfrac{1}{2} (L_2 - \la L_1)''(\w)
.\end{split}
\end{equation}
They also require
\begin{equation}
\label{O67_RPI}
\begin{split}
\vevB{\eta_{\perp\mu\nu}\cO_6^{\alpha\mu\nu}(\w)}
&= -2 [(\w - \la) F_0 + F_3](\w) v^\alpha - [2(\w - \la) K_0 + 2K_3 + L_3](\w)(n-v)^\alpha
,\\
\vevB{v_\mu \eta_{\perp\alpha\nu} \cO_6^{\alpha\mu\nu}(\w)} &= -(K_3 - \la K_0)(\w)
,\\
\vevB{\img \eps_{\perp\mu\nu}\cP_7^{\alpha\mu\nu}(\w)}
&= R_4(\w) (n - v)^\alpha + 0 v^\alpha
,\\
\vevB{v_\mu \img\eps_{\perp\alpha\nu} \cP_7^{\alpha\mu\nu}(\w)} &= 0
.\end{split}
\raisetag{7ex}
\end{equation}

In the remainder of this appendix we use $\La^n$ to denote $\LQCD^n$ divided by an appropriate power of $m_b$. The functions $M_{5,8}(\w)$, $N_{5,8}(\w)$, and $L_9(\w)$ are suppressed by $1/m_b$. Hence, they are effectively twist $\ord{\La^3}$ and can be neglected. The HQET equations of motion imply%
\begin{equation}
\label{eom1}
\begin{split}
F_3(\w) &= \ord{\La^2}
,\\
G_8(\w) &= -2 (\w - \la)^2 F_0(\w) + \ord{\La^3}
,\\
H_8(\w) &= (\w - \la) R_4(\w) + \ord{\La^3}
,\end{split}
\end{equation}
where the neglected terms are of higher twist order, and the relations for $G_8(\w)$ and $H_8(\w)$ follow from the form of their n$^\mathrm{th}$ moments. In addition, we may neglect also all twist $\ord{\La^2}$ shape functions without moments of local $\ord{\La^2}$. The RPI relations \eqref{O12_RPI} and \eqref{O67_RPI} and the HQET equations of motions then imply
\begin{equation}
\label{eom2}
\begin{aligned}
F_1(\w) &= \ord{\La^2}
\q &
\w F_1'(\w) + F_3(\w) &= \ord{\La^3}
,\\
K_1(\w) &= \ord{\La^3}
\q &
K_3(\w) &= \ord{\La^3}
,\\
L_1'(\w) &= 2 K_0(\w)
\q &
L_3(\w) &= -2(\w - \la) K_0(\w) + \ord{\La^3}
.\end{aligned}
\end{equation}
The neglected terms are now only of higher order in the local power counting. The relation for $L_1(\w)$ is exact. There is no formal relation fixing $R_{10}(\w)$, but we may model it as
\begin{equation}
\label{R10_rel}
R_{10}(\w) = - (\w - \la) H_5(\w) + \ord{\La^4}
\end{equation}
which correctly reproduces its first two moments, see below.

Putting everything together by employing the above shape-function definitions and Eqs.~\eqref{nO-rel_full}, \eqref{eom1}, and \eqref{eom2}, the parametrization of the operators \eqref{lc-ops_alpha} in the OPE are%
\begin{equation}
\label{Oi_param}
\begin{split}
\vevB{\cO_0^\alpha(\w)} &= F_0(\w) v^\alpha + K_0(\w) (n-v)^\alpha
,\\
\vevB{\cO_{5\perp}^\alpha(\w)} &= G_5(\w) v^\alpha + \dotsb
,\\
\vevB{\cP_{5\perp}^\alpha(\w)} &= H_5(\w) (n - v)^\alpha + \dotsb
,\\
\vevB{\cO_{8\perp}^\alpha(\w)} &= -2(\w - \la)^2 F_0(\w) v^\alpha + \dotsb
,\\
\vevB{\cP_{8\perp}^\alpha(\w)} &= (\w - \la) R_4(\w) (n - v)^\alpha + \dotsb
,\\
\vevB{\cR_{4\perp}^{\alpha\mu}(\w)}
&= \tfrac{1}{2} [R_4 - 2(\w - \la)K_0](\w) \eta_\perp^{\alpha\mu} +\dotsb
,\\
\vevB{\cR_{10\perp}^{\alpha\mu}(\w)}
&= \tfrac{1}{2} [R_{10} - \la R_4 + 2\la(\w - \la)K_0](\w) \eta_\perp^{\alpha\mu} + \dotsb
.\end{split}
\end{equation}

At last, we look at the moment expansions of the shape functions. For $F_0(\w)$ and $K_0(\w)$ they were given in Eqs.~\eqref{FK0_exp},
\begin{equation}
\label{sf0_exp}
\begin{split}
F_0(\w)
&= \delta(\w - \la) - \frac{\la_0}{2m_b} \delta'(\w - \la) - \frac{\la_1 + \tau_1/m_b}{6} \delta''(\w - \la)
- \frac{\rho_1}{18} \delta'''(\w - \la) + \dotsb
,\\
K_0(\w)
&= \frac{2\la_0 - \rho_0/m_b}{6m_b} \delta'(\w - \la) + \frac{\rho_0}{6m_b} \delta''(\w - \la) + \dotsb
.\end{split}
\end{equation}
For the shape functions arising from the $\ord{\eps}$ operators we have
\begin{equation}
\label{sf1_exp}
\begin{split}
F_1(\w)
&= - \frac{\la_0}{2m_b} \delta(\w-\la) + \ord{\La^4}\delta'(\w-\la)
 - \frac{\rho_1}{18}\delta''(\w-\la) + \dotsb
,\\
F_3(\w)
&= - \frac{\la_0}{2m_b} \delta(\w-\la) + \ord{\La^4}\delta'(\w-\la) + \ord{\La^4}\delta''(\w-\la) + \dotsb
,\\
K_{1,3}(\w) &= \frac{\rho_0}{6m_b} \delta'(\w-\la) + \dotsb
,\\
L_{1,3}(\w) &= \frac{2\la_0 - \rho_0/m_b}{3m_b} \delta(\w-\la) + \frac{\rho_0}{3m_b}\delta'(\w-\la) + \dotsb
,\\
R_4(\w) &= - (\la_2 + \tau_2/m_b) \delta'(\w - \la) - \frac{\rho_2}{2} \delta''(\w - \la) + \dotsb
.\end{split}
\end{equation}
Finally, the shape functions arising from the $\ord{\eps^2}$ operators we need obey the expansion
\begin{equation}
\label{sf2_exp}
\begin{split}
G_5(\w) &= -\frac{2}{3}(\la_1 + \tau_1/m_b) \delta'(\w-\la) + \ord{\La^4} \delta''(\w-\la) + \dotsb
,\\
H_5(\w) &= - (\la_2 + \tau_2/m_b) \delta'(\w-\la) + \ord{\La^4} \delta''(\w-\la) + \dotsb
,\\
G_8(\w) &= \frac{2}{3}(\la_1 + \tau_1/m_b) \delta(\w - \la)
+ \frac{2\rho_1}{3} \delta'(\w - \la) + \dotsb
,\\
H_8(\w) &= (\la_2 + \tau_2/m_b) \delta(\w - \la) + \rho_2 \delta'(\w - \la) + \dotsb
,\\
L_9(\w) &= \frac{\rho_0}{3m_b} \delta(\w - \la) + \dotsb
,\\
R_{10}(\w) &= - (\la_2 + \tau_2/m_b) \delta(\w-\la) + \ord{\La^4} \delta'(\w-\la) + \dotsb
.\end{split}
\end{equation}
All moments of $M_{5,8}(\w)$ and $N_{5,8}(\w)$ are $\ord{\La^4}$ and higher.

\providecommand{\href}[2]{#2}


\begin{thebibliography}{10}

\bibitem{Aubert:2004aw}
BABAR Collaboration, B.~Aubert {\em et~al.},
\newblock Phys. Rev. Lett. {\bf 93}, 011803 (2004),
  [\href{http://www.arXiv.org/abs/hep-ex/0404017}{hep-ex/0404017}].

\bibitem{Luth:2004gs}
BABAR Collaboration, V.~G. Luth,
\newblock \href{http://www.arXiv.org/abs/hep-ex/0411047}{hep-ex/0411047}.

\bibitem{Bauer:2004ve}
C.~W. Bauer, Z.~Ligeti, M.~Luke, A.~V. Manohar, and M.~Trott,
\newblock Phys. Rev. D {\bf 70}, 094017 (2004),
  [\href{http://www.arXiv.org/abs/hep-ph/0408002}{hep-ph/0408002}].

\bibitem{Gibbons:2004dg}
CLEO Collaboration, L.~Gibbons,
\newblock AIP Conf. Proc. {\bf 722}, 156 (2004),
  [\href{http://www.arXiv.org/abs/hep-ex/0402009}{hep-ex/0402009}].

\bibitem{Battaglia:2004ti}
M.~Battaglia and L.~Gibbons,
\newblock \href{http://www.arXiv.org/abs/hep-ph/0402095}{hep-ph/0402095}.

\bibitem{Aubert:2004tw}
BABAR Collaboration, B.~Aubert {\em et~al.},
\newblock \href{http://www.arXiv.org/abs/hep-ex/0408045}{hep-ex/0408045}.

\bibitem{Aubert:2004bq}
BABAR Collaboration, B.~Aubert {\em et~al.},
\newblock \href{http://www.arXiv.org/abs/hep-ex/0408068}{hep-ex/0408068}.

\bibitem{Aubert:2004bv}
BABAR Collaboration, B.~Aubert {\em et~al.},
\newblock \href{http://www.arXiv.org/abs/hep-ex/0408075}{hep-ex/0408075}.

\bibitem{Abe:2004sc}
BELLE Collaboration, K.~Abe {\em et~al.},
\newblock \href{http://www.arXiv.org/abs/hep-ex/0408115}{hep-ex/0408115}.

\bibitem{Chay:1990da}
J.~Chay, H.~Georgi, and B.~Grinstein,
\newblock Phys. Lett. B {\bf 247}, 399 (1990).

\bibitem{Bigi:1993fe}
I.~I. Bigi, M.~A. Shifman, N.~G. Uraltsev, and A.~Vainshtein,
\newblock Phys. Rev. Lett. {\bf 71}, 496 (1993),
  [\href{http://www.arXiv.org/abs/hep-ph/9304225}{hep-ph/9304225}].

\bibitem{Blok:1994va}
B.~Blok, L.~Koyrakh, M.~Shifman, and A.~I. Vainshtein,
\newblock Phys. Rev. D {\bf 49}, 3356 (1994),
  [\href{http://www.arXiv.org/abs/hep-ph/9307247}{hep-ph/9307247}].

\bibitem{Manohar:1994qn}
A.~V. Manohar and M.~B. Wise,
\newblock Phys. Rev. D {\bf 49}, 1310 (1994),
  [\href{http://www.arXiv.org/abs/hep-ph/9308246}{hep-ph/9308246}].

\bibitem{Mannel:1994su}
T.~Mannel,
\newblock Nucl. Phys. B {\bf 413}, 396 (1994),
  [\href{http://www.arXiv.org/abs/hep-ph/9308262}{hep-ph/9308262}].

\bibitem{Falk:1994dh}
A.~F. Falk, M.~Luke, and M.~J. Savage,
\newblock Phys. Rev. D {\bf 49}, 3367 (1994),
  [\href{http://www.arXiv.org/abs/hep-ph/9308288}{hep-ph/9308288}].

\bibitem{Bauer:2001mh}
C.~W. Bauer, M.~Luke, and T.~Mannel,
\newblock Phys. Rev. D {\bf 68}, 094001 (2003),
  [\href{http://www.arXiv.org/abs/hep-ph/0102089}{hep-ph/0102089}].

\bibitem{Neubert:1994ch}
M.~Neubert,
\newblock Phys. Rev. D {\bf 49}, 3392 (1994),
  [\href{http://www.arXiv.org/abs/hep-ph/9311325}{hep-ph/9311325}].

\bibitem{Neubert:1994um}
M.~Neubert,
\newblock Phys. Rev. D {\bf 49}, 4623 (1994),
  [\href{http://www.arXiv.org/abs/hep-ph/9312311}{hep-ph/9312311}].

\bibitem{Bigi:1994ex}
I.~I.~Y. Bigi, M.~A. Shifman, N.~G. Uraltsev, and A.~I. Vainshtein,
\newblock Int. J. Mod. Phys. A {\bf 9}, 2467 (1994),
  [\href{http://www.arXiv.org/abs/hep-ph/9312359}{hep-ph/9312359}].

\bibitem{Mannel:1994pm}
T.~Mannel and M.~Neubert,
\newblock Phys. Rev. D {\bf 50}, 2037 (1994),
  [\href{http://www.arXiv.org/abs/hep-ph/9402288}{hep-ph/9402288}].

\bibitem{Korchemsky:1994jb}
G.~P. Korchemsky and G.~Sterman,
\newblock Phys. Lett. B {\bf 340}, 96 (1994),
  [\href{http://www.arXiv.org/abs/hep-ph/9407344}{hep-ph/9407344}].

\bibitem{Leibovich:2002ys}
A.~K. Leibovich, Z.~Ligeti, and M.~B. Wise,
\newblock Phys. Lett. B {\bf 539}, 242 (2002),
  [\href{http://www.arXiv.org/abs/hep-ph/0205148}{hep-ph/0205148}].

\bibitem{Bauer:2002yu}
C.~W. Bauer, M.~Luke, and T.~Mannel,
\newblock Phys. Lett. B {\bf 543}, 261 (2002),
  [\href{http://www.arXiv.org/abs/hep-ph/0205150}{hep-ph/0205150}].

\bibitem{Burrell:2003cf}
C.~N. Burrell, M.~E. Luke, and A.~R. Williamson,
\newblock Phys. Rev. D {\bf 69}, 074015 (2004),
  [\href{http://www.arXiv.org/abs/hep-ph/0312366}{hep-ph/0312366}].

\bibitem{Kraetz:2002rv}
M.~Kraetz and T.~Mannel,
\newblock Phys. Lett. B {\bf 556}, 163 (2003),
  [\href{http://www.arXiv.org/abs/hep-ph/0210248}{hep-ph/0210248}].

\bibitem{Mannel:2004as}
T.~Mannel and F.~J. Tackmann,
\newblock Phys. Rev. D {\bf 71}, 034017 (2005),
  [\href{http://www.arXiv.org/abs/hep-ph/0408273}{hep-ph/0408273}].

\bibitem{Lee:2004ja}
K.~S.~M. Lee and I.~W. Stewart,
\newblock Nucl. Phys. B {\bf 721}, 325 (2005),
  [\href{http://www.arXiv.org/abs/hep-ph/0409045}{hep-ph/0409045}].

\bibitem{Bosch:2004cb}
S.~W. Bosch, M.~Neubert, and G.~Paz,
\newblock JHEP {\bf 11}, 073 (2004),
  [\href{http://www.arXiv.org/abs/hep-ph/0409115}{hep-ph/0409115}].

\bibitem{Beneke:2004in}
M.~Beneke, F.~Campanario, T.~Mannel, and B.~D. Pecjak,
\newblock JHEP {\bf 06}, 071 (2005),
  [\href{http://www.arXiv.org/abs/hep-ph/0411395}{hep-ph/0411395}].

\bibitem{manoharwise}
A.~V. Manohar and M.~B. Wise,
\newblock {\em Heavy Quark Physics}, Cambridge monographs on particle physics,
  nuclear physics, and cosmology No. ~10 (Cambridge University Press, 2000).

\bibitem{Falk:1992fm}
A.~F. Falk, M.~Neubert, and M.~E. Luke,
\newblock Nucl. Phys. B {\bf 388}, 363 (1992),
  [\href{http://www.arXiv.org/abs/hep-ph/9204229}{hep-ph/9204229}].

\bibitem{Luke:1992cs}
M.~E. Luke and A.~V. Manohar,
\newblock Phys. Lett. B {\bf 286}, 348 (1992),
  [\href{http://www.arXiv.org/abs/hep-ph/9205228}{hep-ph/9205228}].

\bibitem{Campanario:2002fy}
F.~Campanario and T.~Mannel,
\newblock Phys. Rev. D {\bf 65}, 094017 (2002),
  [\href{http://www.arXiv.org/abs/hep-ph/0201136}{hep-ph/0201136}].

\bibitem{Chay:2002vy}
J.~Chay and C.~Kim,
\newblock Phys. Rev. D {\bf 65}, 114016 (2002),
  [\href{http://www.arXiv.org/abs/hep-ph/0201197}{hep-ph/0201197}].

\bibitem{Manohar:2002fd}
A.~V. Manohar, T.~Mehen, D.~Pirjol, and I.~W. Stewart,
\newblock Phys. Lett. B {\bf 539}, 59 (2002),
  [\href{http://www.arXiv.org/abs/hep-ph/0204229}{hep-ph/0204229}].

\bibitem{Pirjol:2002km}
D.~Pirjol and I.~W. Stewart,
\newblock Phys. Rev. D {\bf 67}, 094005 (2003),
  [\href{http://www.arXiv.org/abs/hep-ph/0211251}{hep-ph/0211251}],
\newblock [Erratum-ibid.\ D {\bf 69}, 019903 (2004)].

\bibitem{Neubert:2004cu}
M.~Neubert,
\newblock \href{http://www.arXiv.org/abs/hep-ph/0411027}{hep-ph/0411027}.

\bibitem{Gremm:1997df}
M.~Gremm and A.~Kapustin,
\newblock Phys. Rev. D {\bf 55}, 6924 (1997),
  [\href{http://www.arXiv.org/abs/hep-ph/9603448}{hep-ph/9603448}].

\bibitem{Mannel:1999gs}
T.~Mannel and S.~Recksiegel,
\newblock Phys. Rev. D {\bf 60}, 114040 (1999),
  [\href{http://www.arXiv.org/abs/hep-ph/9904475}{hep-ph/9904475}].

\bibitem{Barger:1990tz}
V.~D. Barger, C.~S. Kim, and R.~J.~N. Phillips,
\newblock Phys. Lett. B {\bf 251}, 629 (1990).

\bibitem{Falk:1997gj}
A.~F. Falk, Z.~Ligeti, and M.~B. Wise,
\newblock Phys. Lett. B {\bf 406}, 225 (1997),
  [\href{http://www.arXiv.org/abs/hep-ph/9705235}{hep-ph/9705235}].

\bibitem{Bigi:1997dn}
I.~I.~Y. Bigi, R.~D. Dikeman, and N.~Uraltsev,
\newblock Eur. Phys. J. C {\bf 4}, 453 (1998),
  [\href{http://www.arXiv.org/abs/hep-ph/9706520}{hep-ph/9706520}].

\bibitem{DeFazio:1999sv}
F.~De~Fazio and M.~Neubert,
\newblock JHEP {\bf 06}, 017 (1999),
  [\href{http://www.arXiv.org/abs/hep-ph/9905351}{hep-ph/9905351}].

\bibitem{Bauer:2001rc}
C.~W. Bauer, Z.~Ligeti, and M.~Luke,
\newblock Phys. Rev. D {\bf 64}, 113004 (2001),
  [\href{http://www.arXiv.org/abs/hep-ph/0107074}{hep-ph/0107074}].

\bibitem{Bouzas:1994cr}
A.~O. Bouzas and D.~Zappala,
\newblock Phys. Lett. B {\bf 333}, 215 (1994),
  [\href{http://www.arXiv.org/abs/hep-ph/9403313}{hep-ph/9403313}].

\bibitem{Greub:1996ed}
C.~Greub and S.-J. Rey,
\newblock Phys. Rev. D {\bf 56}, 4250 (1997),
  [\href{http://www.arXiv.org/abs/hep-ph/9608247}{hep-ph/9608247}].

\bibitem{Bauer:2000xf}
C.~W. Bauer, Z.~Ligeti, and M.~E. Luke,
\newblock Phys. Lett. B {\bf 479}, 395 (2000),
  [\href{http://www.arXiv.org/abs/hep-ph/0002161}{hep-ph/0002161}].

\bibitem{Kowalewski:2002qs}
R.~V. Kowalewski and S.~Menke,
\newblock Phys. Lett. B {\bf 541}, 29 (2002),
  [\href{http://www.arXiv.org/abs/hep-ex/0205038}{hep-ex/0205038}].

\bibitem{Bauer:2003pi}
C.~W. Bauer and A.~V. Manohar,
\newblock Phys. Rev. D {\bf 70}, 034024 (2004),
  [\href{http://www.arXiv.org/abs/hep-ph/0312109}{hep-ph/0312109}].

\bibitem{Bosch:2004th}
S.~W. Bosch, B.~O. Lange, M.~Neubert, and G.~Paz,
\newblock Nucl. Phys. B {\bf 699}, 335 (2004),
  [\href{http://www.arXiv.org/abs/hep-ph/0402094}{hep-ph/0402094}].

\end{thebibliography}
\end{document}